\documentclass[prc,showkeys,longbibliography]{revtex4-1}
\usepackage{epsfig,graphicx,float,color}
\usepackage{amssymb}
\usepackage{amsmath}
\usepackage{longtable}
\usepackage{setspace}
\usepackage[ulem=normalem,defaultcolor=red]{changes}
\begin{document}
\title{At the borderline of shape coexistence: Mo and Ru}
\author{E.~Maya-Barbecho$^{1}$, S.~Baid$^{2}$, J.M.~Arias$^{2,3}$ and J.E.~Garc\'{\i}a-Ramos$^{1,3}$}
\affiliation{
  $^1$Departamento de  Ciencias Integradas y Centro de Estudios Avanzados en F\'isica, Matem\'atica y Computaci\'on, Universidad de Huelva, 21071 Huelva, Spain\\
  $^2$Departamento de F\'isica At\'omica, Molecular y Nuclear, Facultad de F\'isica\text{,} Universidad de Sevilla, Apartado 1065, E-41080 Sevilla, Spain\\
  $^3$Instituto Carlos I de F\'{\i}sica Te\'orica y Computacional,  Universidad de Granada, Fuentenueva s/n, 18071 Granada, Spain
} 

\begin{abstract}
\begin{description}
\item [Background]
Even-even isotopes of Mo ($Z=42$) and Ru ($Z=44$) are nuclei close to the subshell closure at $Z=40$, where shape coexistence plays a significant role. As a result, their spectroscopic properties are expected to resemble those of Sr ($Z=38$)  and Zr ($Z=40$). Exploring the evolution of these properties as they move away from the subshell closure is of great interest.
\item [Purpose]
The purpose of this study is to reproduce the spectroscopic properties of even-even 
$^{96-110}_{\phantom{961-}42}$Mo and $^{98-114}_{\phantom{961-}44}$Ru isotopes and to determine the influence of shape coexistence.
\item [Method]
We have employed the interacting boson model with configuration mixing as the framework to calculate all the observables for Mo and Ru isotopes. We have considered two types of configurations: 0-particle-0-hole and 2-particle-2-hole excitations. The model parameters have been determined using a least-squares fitting to match the excitation energies and the $B(E2)$ transition rates.
\item [Results]
We have obtained the excitation energies, $B(E2)$ values, two-neutron separation energies, nuclear radii, and isotope shifts for the entire chain of isotopes. Our theoretical results have shown good agreement with experimental data. Furthermore, we have conducted a detailed analysis of the wave functions and obtained the mean-field energy surfaces and the nuclear deformation parameter, $\beta$, for all considered isotopes.
\item [Conclusions]
Our findings reveal that shape coexistence plays a significant role in Mo isotopes, with the crossing of intruder and regular configurations occurring at neutron number $60$ ($A=102$), which induces a quantum phase transition. In contrast, in Ru isotopes, the intruder states have minimal influence, remaining at higher energies. However, at neutron number $60$, also a quantum phase transition occurs in Ru isotopes.
\end{description}
\end{abstract}

\date{\today}
\maketitle

\section{Introduction}
\label{sec-intro}
The subshell closure at $Z=40$ is known to exhibit rapid onset of deformation, particularly around neutron number $60$, resulting from the filling of the neutron $1g_{7/2}$ orbit, which interacts with the proton $1g_{9/2}$ one. This phenomenon was well explained in \cite{Fede77,Fede79a,Fede79b}, and more recently in \cite{Togashi16}, providing insights into the origin of deformation in this mass region. Generally, the shape of a nucleus arises from a delicate balance between pairing and quadrupole nuclear interactions, favoring spherical and deformed shapes, respectively. It is important not to overlook the role of monopole interaction, which is responsible for the evolution of single-particle energies.

The region around $Z=40$ is also known for the presence of states with different shapes within a narrow energy range. This situation, known as shape coexistence, was initially proposed in nuclear physics by Morinaga \cite{morinaga56} to explain the nature of the first excited state ($0^+$) in $^{16}$O, which was assumed to be deformed while the ground state is obviously spherical due to its doubly magic nature. These experimental findings were theoretically confirmed in \cite{Brown64,Brown66a,Brown66b}, where particle-hole excitations across the energy shell gap were allowed. A deformed band, originated at low energy due to the quadrupole part of the nucleon-nucleon interactions, emerged on top of the first excited $0^+$ state. The presence of additional effective valence nucleons is crucial in explaining the characteristic dropping of this deformed band, often referred to as the intruder band. Another notable example of shape coexistence is observed experimentally in $^{40}$Ca and is effectively described through shell model calculations involving multi-particle multi-hole excitations \cite{Caur07,Poves16,Poves18}.

The presence of shape coexistence is clearly manifested through experimental observations of nuclear radii and isotope shifts. The phenomenon was first identified in the odd-even staggering of the radii of mercury isotopes, which indicated the coexistence of states with significantly different degrees of deformation. Subsequent studies, such as the measurements of mercury radii in the neutron-deficient region well beyond the mid-shell \cite{Marsh2018}, have confirmed the role of shape coexistence in explaining nuclear structure in various mass regions. This phenomenon is particularly prominent near shell or subshell closures in protons (neutrons) with neutrons (protons) around the mid-shell. Shape coexistence has been observed in light, medium, and heavy nuclei \cite{hey83,wood92,heyde11, Garr22}.

From a theoretical perspective, shape coexistence is described using two complementary approaches: self-consistent methods based on Hartree-Fock (HF) or Hartree-Fock-Bogoliubov (HFB) theories, and the nuclear shell model. In the region around $Z=40$, notable studies have been conducted using the relativistic interaction PC-PK1, focusing on Kr, Sr, Zr, and Mo isotopes near neutron number $60$ \cite{Xian12}. These studies revealed a rapid evolution in Sr and Zr isotopes, while a more moderate evolution was observed in Mo and Kr isotopes. Prolate-oblate shape coexistence was observed in $^{98}$Sr and $^{100}$Zr. In another study \cite{Nomu16}, even-even Ru, Mo, Zr, and Sr isotopes were investigated using the HFB approach with a Gogny-D1M interaction. The spectroscopic properties were obtained by mapping the energy density functional into an interacting boson model with configuration mixing (IBM-CM) energy surface. The Ru isotopes exhibited a smooth shape evolution with no evidence of intruder bands. The Mo isotopes required two different particle-hole configurations, resulting in a good reproduction of the yrast band but with some deficiencies in describing the non-yrast band and inter-band transitions. State-of-the-art calculations within the HFB framework were carried out by \citet{Rodr10}, allowing for the treatment of axial and triaxial degrees of freedom on an equal footing. These calculations, applied to Sr, Zr, and Mo isotopes, revealed an oblate shape for Mo isotopes at neutron number $58$, gradually transitioning to a triaxial shape as the neutron number increased. An island of triaxiality was evident from neutron number $60$ up to $68$. Another study \cite{Abus17a} investigated Mo and Ru isotopes using the relativistic-Hartree-Bogoliubov formalism with density-dependent zero- and finite-range nucleon-nucleon interactions, as well as a separable pairing. The results were in agreement with other mean-field calculations. Additionally, the study \cite{Thak21} employed the density-dependent meson exchange model DD-ME2 and density-dependent point coupling models DD-PC1 and DD-PCX to explore the shape evolution of Zr, Mo, and Ru isotopes, considering only axial situations. The predictions indicated a spherical shape for the lightest Ru isotopes, nearly degenerate prolate and oblate minima for $^{96-102}$Ru, a prolate and an oblate degenerate minima in $^{104}$Ru, and an oblate shape for the heaviest Ru isotopes. In the case of Mo isotopes, the lightest isotopes were predicted to have spherical shapes, with nearly degenerate prolate and oblate minima for $^{94-100}$Mo.

Moving into the shell-model framework, it is important to note that the description of the region around $Z=40$ is influenced by the simultaneous occupation of neutron and proton spin-orbit partners. When the neutron $1g_{7/2}$ orbital begins to be filled, the interaction with the proton $1g_{9/2}$ orbital favors the existence of a deformed region in Zr and Sr nuclei with a neutron number larger than $58$, and likely in Mo and Ru nuclei as well.
This concept has been recently applied to the Zr region \cite{Togashi16} using the Monte Carlo Shell Model \citet{Otsuka01,Shimizu12,Shimizu17}, which has the capability to handle open shells. However, the idea was first introduced in the seminal works of Federman and Pittel \cite{Fede77,Fede79a,Fede79b} where the simultaneous occupation of neutron-proton spin-orbit partners was emphasized as crucial. Federman and their colleagues extensively explored this mass region using a reduced model space consisting of the $3s_{1/2}$, $2d_{3/2}$, and $1g_{7/2}$ neutron orbits, and the $2p_{1/2}$, $1g_{9/2}$, and $2d_{5/2}$ proton orbits \cite{Fede84,Heyd88,Etch89,Pitt93}. More recently, large-scale shell-model calculations have been performed for the same mass region using more realistic valence spaces, as demonstrated in studies such as \cite{Holt00} and \cite{Siej09}. In \cite{Cora2022}, a shell-model calculation was conducted for $^{100}$Mo and $^{100}$Ru, starting from a realistic nucleon-nucleon potential and deriving the effective shell-model Hamiltonian and decay operators within many-body perturbation theory, with a focus on studying neutrinoless double-$\beta$ decay. In \cite{Baks22}, the multi-quasiparticle triaxial projected shell model was used to investigate the band structures of $^{98-106}$Ru isotopes, providing a consistent description. In \cite{Jian16}, the odd-even and even-even isotopes of $^{95-102}$Ru were studied using the nucleon pair approximation with a phenomenological pairing plus quadrupole interaction, yielding good agreement with experimental data.

Other works that provide insights into the nature of this mass region include the following studies. In \cite{Gian16,Gian17}, the authors analyzed $^{98}$Ru within the framework of the IBM-2 \cite{Arima77} and concluded that a clear vibrational pattern is present. In \cite{Kisy16}, even-even and even-odd Ru isotopes were investigated using the IBM \cite{iach87}, revealing a transitional behavior. The g-factors of Ru and Pd nuclei were calculated using the IBM-2 in \cite{Gian2013}. The even-even $^{98-110}$Ru isotopes were studied using the affine $\widehat{SU(1,1)}$ Lie Algebra in \cite{Jafa16}. In \cite{Boyu10}, the $A=100$ region was described using the IBM-1 with a single Hamiltonian featuring constant parameters. Although this work captured the overall trends well, it could not reproduce the fine details of the spectra, especially the rapid shape evolution observed around neutron number $60$. In \cite{Leva21}, the even-even Mo isotopes were investigated using a Bohr Hamiltonian with a sextic potential in the $\beta$ direction, without dependence on $\gamma$. This approach provided a good description of the spectra across the entire chain of isotopes, and it was concluded that $^{104}$Ru is the closest nucleus to the critical point symmetry E(5) \cite{Iach00}. In \cite{Kern1995}, a large set of isotopes were studied to identify good candidates for vibrational-like behavior, i.e., U(5) nuclei. Among others, $^{100}$Mo and $^{98-104}$Ru were identified as suitable candidates. It is important to note that this work is relatively old, and with the present experimental knowledge, the conclusions may have evolved.

The present work extends our previous analysis of the $Z\approx 40$ and $A\approx 100$ region \cite{Garc03, Garc19, Garc20, Maya2022, Maya2023u}, particularly focusing on the even-even Mo and Ru isotopes. Mo is an excellent candidate for studying the influence of intruder states on the onset of deformation for $N\approx 60$ due to the rapid lowering of the energy of the $2_1^+$ state, a significant increase in the ratio $E(4_1^+)/E(2_1^+)$, or a sudden increase of the radius. However, Ru exhibits a smoother trend in these observables, and the influence of intruder states seems to be minimal. Nevertheless, the onset of deformation at $N\approx 60$, specifically in $^{104}$Ru, has been suggested to generate a relatively flat energy surface, reminiscent of the concept of critical point symmetry E(5) \cite{Fran2001}.

The paper is organized as follows. In Section \ref{sec-exp}, the present experimental knowledge on Mo and Ru nuclei is reviewed. In Section \ref{sec-ibm-cm}, the theoretical framework used in this work is presented, namely the IBM-CM (interacting boson model with configuration mixing). The procedure for obtaining the fitting parameters of the model will also be discussed. In Section \ref{sec-corr_energy}, the correlation energy gain is studied. It plays a crucial role in understanding the nuclear structure and the interaction between nucleons. In Section \ref{sec-comp}, a detailed comparison between theory and experiment for excitation energies and E2 transition rates is presented. This analysis will provide insights into the agreement between the theoretical predictions of the IBM-CM and the experimental data. In Section \ref{sec-wf}, the wave functions of the nuclear states are analyzed. The structure and configuration mixing of these states will be examined, allowing for a deeper understanding of the underlying nuclear dynamics. In Section \ref{sec-other}, additional observables, including radii, isotopic shifts, and two-neutron separation energies, are studied. These quantities provide valuable information about the nuclear shapes, deformations, and binding energies. In Section \ref{sec-deformation}, a calculation of the IBM mean-field energy surfaces and an investigation of the deformations in Mo and Ru nuclei is presented. %This analysis will shed light on the interplay between shape coexistence, intruder states, and the possible existence of a quantum phase transition. 
In Section \ref{sec-quest}, an analysis of the possible existence of a quantum phase transition in the studied nuclei is discussed. Finally, in Section \ref{sec-conclu}, the summary and the conclusions of the paper are presented.

\section{Experimental data in the even-even Mo and Ru nuclei }
\label{sec-exp}
\begin{figure}[hbt]
\centering
\includegraphics[width=0.95\linewidth]{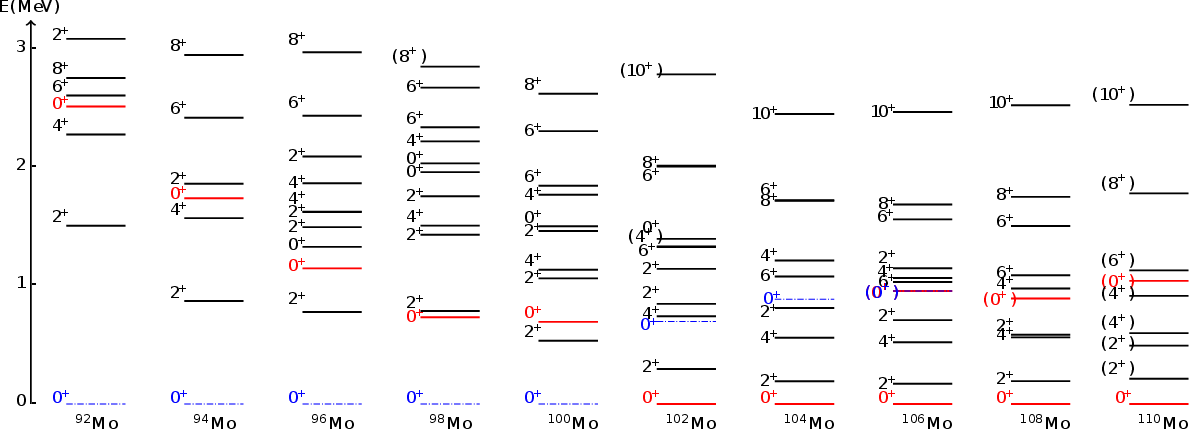}
\caption{The experimental energy level systematics of low-lying positive parity states for the Mo isotopes are displayed, showing levels up to approximately 3 MeV in energy. Levels with dashed blue lines likely correspond to spherical shapes, while those in red represent deformed shapes (for more details refer to the information in Section \ref{sec-exp}).}
\label{fig-e-systematics-mo}
\end{figure}

The energy systematics of isotopes $^{92-110}$Mo below 3 MeV are presented in Fig.\ \ref{fig-e-systematics-mo}, which illustrates a significant increase in the density of states in the lower part of the spectrum as the mass increases. Additionally, a transformation from a vibrational-like pattern to a rotational one is observed, beginning at $^{102}$Mo and continuing for heavier isotopes. Another important observation is the decrease in energy of the first excited $0^+$ state, with the minimum also occurring at $^{102}$Mo. Throughout the entire chain, the energies of the $4_1^+$ and $2_2^+$ states are relatively close, almost degenerate in the case of $^{108}$Mo, indicating the presence of a certain degree of $\gamma$ softness in the heavier isotopes.

\begin{figure}[hbt]
\centering
\includegraphics[width=0.95\linewidth]{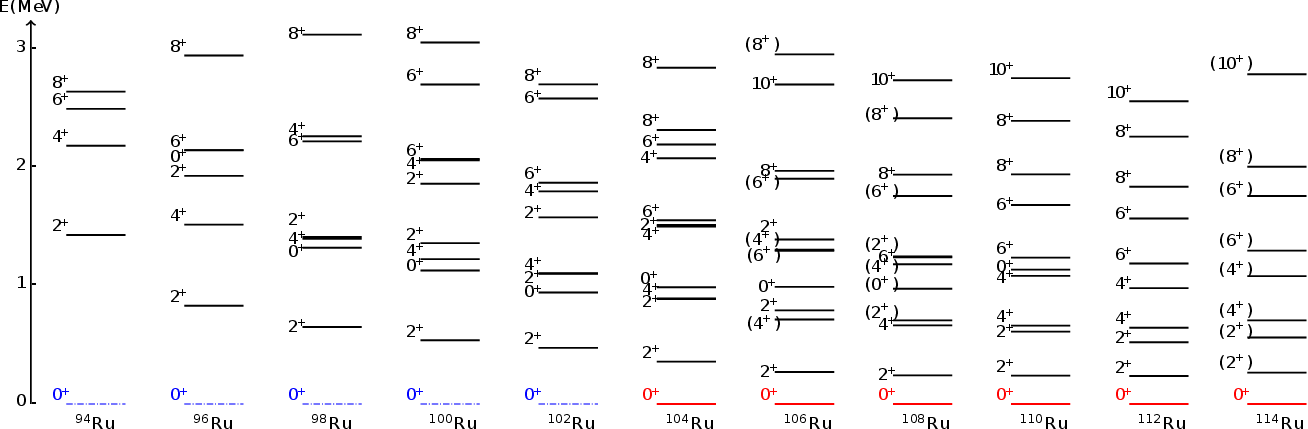}
\caption{The same as Fig.\ \ref{fig-e-systematics-mo} but for Ru isotopes.}
\label{fig-e-systematics-ru}
\end{figure}

Fig.\ \ref{fig-e-systematics-ru} depicts the energy systematics of $^{94-114}$Ru isotopes. A clear vibrational pattern is evident for isotopes $^{94-102}$Ru, which later evolves into a rotational structure with some degree of $\gamma$ softness, as indicated by the close proximity of the $4_1^+$ and $2_2^+$ states. Similarly, a dropping $0^+$ state is observed, with a minimum energy at $^{102}$Ru.
  
For the comparison with theoretical calculations, we will consider the evaluated experimental data from Nuclear Data Sheets publications for specific isotopes: $A=96$ \cite{Abri08}, $A=98$ \cite{Sing03,Chen20}, $A=100$ \cite{Sing08}, $A=102$ \cite{Defr09}, $A=104$ \cite{Blac07}, $A=106$ \cite{Defr2008}, $A=108$ \cite{Blac08}, $A=110$ \cite{Defr09}, $A=112$ \cite{Lalk2015}, $A=114$ \cite{Blac2012}. In addition to these sources, we have incorporated the most up-to-date references for certain isotopes as described below.

An excellent experimental overview of this mass region, including Mo and Ru isotopes, can be found in \cite{Garr22} with updated references. In \cite{Thom2016}, $\gamma\gamma$ angular correlation experiments were conducted to study the low-lying states of $^{96-98}$Mo, allowing the determination of angular momenta and multiple mixing ratios. Detailed Coulomb-excitation studies of $^{98}$Mo and $^{100}$Mo were performed in \cite{Ziel2002} and \cite{Wrzo12} respectively.
In \cite{Ha20}, $^{106-108-110}$Mo nuclei were investigated through $\beta$-delayed $\gamma$-ray spectroscopy, and for the first time, the $0_2^+$ band in $^{108-110}$Mo was measured. The authors of \cite{Free2017} measured neutron and proton occupancy in $^{98}$Mo and $^{100}$Mo, revealing a clear change in the filling of the proton g$_{9/2}$ shell between the two isotopes.

For Ru isotopes, in \cite{Garr20}, the $0_2^+$ and $\gamma$ bands in $^{98}$Ru were observed using $\gamma$-ray spectroscopy following the $\beta$-decay of $^{98}$Rh, as well as via the $^{100}$Ru(p,~t) reaction. The $0_2^+$ state is suggested to be an intruder state rather than a two-phonon vibrational state, although the mean-field calculation presented in the same work does not fully support this hypothesis. The lifetimes of states $2_1^+$, $2_2^+$, and $4_1^+$ in $^{98}$Ru were measured in \cite{Rade12} in an attempt to resolve discrepancies observed in the literature regarding the lifetime of the $4_1^+$ state.
In \cite{Urba2013}, the $0_2^+$ band in $^{102}$Ru was studied, along with an analysis of the mixing between the $0_1^+$ and $0_2^+$ states to understand the deformation evolution of the ground state in even-even Ru isotopes. Measurement of $28$ E2 and $3$ M1 matrix elements involving $17$ low-lying excited states in $^{104}$Ru was conducted using Coulomb excitation in \cite{Srebrny06}. The $g$-factor of the $2_1^+$ state in $^{96-104}$Ru was obtained in \cite{Tayl11}. More recently, in \cite{Garr22b} a Coulomb excitation experiment was conducted for $^{102}$Ru obtaining a little larger E2 matrix elements than the evaluated ones for the transitions from $2_1^+$ and $2^+_2$ to the $0_1^+$ state, and measuring for the first time the E2 matrix element of the transition $2_3^+ \rightarrow 0_1^+$. They concluded that the $0_2^+$ state, with $\beta\approx 0.18$ is a little less deformed than the ground state, with $\beta\approx 0.24$.    

\section{The Interacting Boson Model with configuration mixing formalism}
\label{sec-ibm-cm}
\subsection{The formalism}
\label{sec-formalism}
The framework used in this work is the IBM-CM (interacting boson model with configuration mixing). This model is an extension of the original IBM (interacting boson model) proposed by Arima and Iachello \cite{iach87}. The IBM-CM allows for the simultaneous treatment of multiple boson configurations corresponding to particle-hole excitations across a shell or subshell closure \cite{duval81,duval82}.

In this version of the model, known as IBM-1, no distinction is made between proton and neutron bosons or particles and holes. For the study of Mo and Ru nuclei, we consider the closure for protons of the $Z=40$ subshell, where the regular states correspond to a 0h-2p (0 holes and 2 particles) proton configuration for Mo and to a 0h-4p proton configuration for Ru, and the intruder states correspond to a 2h-4p proton configuration for Mo and to a 2h-6p proton configuration for Ru. The number of valence neutrons is determined considering a neutron closed shell at $N=50$. Hence, the number of valence bosons, denoted as $N$, will be half of the sum of the valence protons, which is $2$ for Mo and $4$ for Ru, plus half the number of valence neutrons. The intruder configuration will have additionally $2$ bosons. Therefore, the regular and intruder spaces will form a $[N] \oplus [N+2]$ Hilbert space.

The Hamiltonian of the system consists of two sectors: one corresponding to the regular part, $[N]$, and another corresponding to the intruder part, $[N+2]$. The total Hamiltonian is written as follows:
\begin{equation}
  \hat{H}=\hat{P}^{\dag}_{N}\hat{H}^N_{\rm ecqf}\hat{P}_{N}+
  \hat{P}^{\dag}_{N+2}\left(\hat{H}^{N+2}_{\rm ecqf}+
    \Delta^{N+2}\right)\hat{P}_{N+2}\
  +\hat{V}_{\rm mix}^{N,N+2}~,
\label{eq:ibmhamiltonian}
\end{equation}
where $\hat{P}_{N}$ and $\hat{P}_{N+2}$ are projection operators onto the $[N]$ and the $[N+2]$ boson subspaces, respectively,
\begin{equation}
  \hat{H}^i_{\rm ecqf}=\varepsilon_i \hat{n}_d+\kappa'_i
  \hat{L}\cdot\hat{L}+
  \kappa_i
  \hat{Q}(\chi_i)\cdot\hat{Q}(\chi_i)
  \label{eq:cqfhamiltonian}
\end{equation}
is the Hamiltonian of the extended consistent-Q formalism (ECQF), \cite{warner83,lipas85} with $i=N,N+2$, $\hat{n}_d$ is the $d$ boson number operator, $\hat{L}=\sqrt{10}\left[d^\dag \times \tilde d  \right]^{(1)}$ is the angular momentum, and 
$\hat{Q}(\chi)=\left[s^\dag \times \tilde d + d^\dag \times s \right]^{(2)}+ \chi \left[d^\dag \times \tilde d  \right]^{(2)}$ is
the quadrupole operator. Note that the ECQF corresponds to a simplified version of the general IBM Hamiltonian. The parameter $\Delta^{N+2}$ represents the energy needed to excite two proton particles across the $Z=40$ subshell gap, resulting in 2p-2h excitations. The operator $\hat{V}_{\rm mix}^{N,N+2}$ is the mixing between the $N$ and the $N+2$ configurations and is given by
\begin{equation}
  \hat{V}_{\rm mix}^{N,N+2}=\omega_0^{N,N+2}(s^\dag\times s^\dag + s\times
  s)+\omega_2^{N,N+2} (d^\dag\times d^\dag+\tilde{d}\times \tilde{d})^{(0)}.
\label{eq:vmix}
\end{equation}
In this study, we assume that $\omega_0^{N,N+2}=\omega_2^{N,N+2}=\omega$, where $\omega$ is a constant parameter.

The $E2$ transition operator is built with the same quadrupole operator that appears in the Hamiltonian (\ref{eq:cqfhamiltonian}). It is defined as the sum of two contributions that act separately in the regular and the intruder sectors without crossed contributions, 
\begin{equation}
  \hat{T}(E2)_\mu=\sum_{i=N,N+2} e_i \hat{P}_i^\dag\hat{Q}_\mu(\chi_i)\hat{P}_i~.
  \label{eq:e2operator}
\end{equation}
The $e_i$ ($i=N,N+2$) are the effective boson charges and the parameters $\chi_i$ take the same values that in the Hamiltonian (\ref{eq:cqfhamiltonian}). Note that the operator cannot connect the regular with the intruder sector or viceversa. 

The free parameters associated with the above operators need to be determined in order to reproduce a set of excitation energies and transition rates, as described in Section \ref{sec-fit-procedure}.

This approach has been successfully employed in recent studies for Sr \cite{Maya2022}, Zr \cite{Garc19,Garc20}, Pt \cite{Garc09,Garc11}, Hg \cite{Garc14b,Garc15b} and Po isotopes \cite{Garc15,Garc15c}.

\subsection{The fitting procedure: energy spectra and absolute $B(E2)$ reduced transition probabilities}
\label{sec-fit-procedure}

\begin{table}
\caption{Hamiltonian and $\hat{T}(E2)$ parameters resulting from the study of Mo isotopes in the present work. All quantities have dimensions of energy (given in keV), except $\chi_N$ and $\chi_{N+2}$, which are dimensionless, and $e_N$ and $e_{N+2}$, which are given in units of $\sqrt{\text{W.u.}}$. It should be noted that the value of $\Delta^{N+2}=1500$ keV is fixed for all isotopes.}
\begin{center}
\begin{ruledtabular}
\begin{tabular}{cccccccccccc}
Nucleus&$\varepsilon_N$&$\kappa_N$&$\chi_{N}$&$\kappa'_N$&$\varepsilon_{N+2}$& $\kappa_{N+2}$&$\chi_{N+2}$&$\kappa'_{N+2}$&$\omega$&$e_{N}$&$e_{N+2}$\\
  \hline
$^{96}$Mo &  695.84&     0.00&  1.50&   15.00&   191.56&    -9.96&  1.12&   13.69&  45.0&  2.48&   4.00 \\
$^{98}$Mo &  813.16&     0.00& -1.70&   -5.00&   873.21&   -25.08&  1.50&   -5.00&  45.0&  2.73&  -0.87 \\
$^{100}$Mo&  517.29&    -2.00& -1.49&   10.00&   408.52&   -21.98&  0.05&    4.34&  45.0&  2.24&   2.88 \\
$^{102}$Mo&  470.23&    -4.93& -1.50&   -5.00&   446.78&   -35.00&  0.14&    1.08&  15.0&  4.00&  -2.15 \\
$^{104}$Mo&  150.00&   -10.00& -1.34&    7.02&   450.04&   -35.00&  0.49&   -0.63&  15.0&  2.11&  -2.05 \\
$^{106}$Mo&  294.38&   -15.00& -0.58&    0.02&   263.83&   -29.90&  0.37&    7.68&  15.0&  1.90&   1.93 \\
$^{108}$Mo&  294.38\footnotemark[1]&   -15.00\footnotemark[1]& -0.58\footnotemark[1]&    0.02\footnotemark[1]&   203.58&   -28.85&  0.12&    8.98&  15.0&  1.90\footnotemark[2]&   2.12 \\
$^{110}$Mo&  294.38\footnotemark[1]&   -15.00\footnotemark[1]& -0.58\footnotemark[2]&    0.02\footnotemark[1]&   200.00&   -31.79&  0.01&    7.58&  15.0&  1.90\footnotemark[2]&   2.12\footnotemark[3] \\
\end{tabular}
\end{ruledtabular}
\end{center}
\footnotetext[1]{Hamiltonian parameters for the regular sector taken from $^{106}$Mo.}
\footnotetext[2]{$\hat{T}(E2)$ regular parameter taken from $^{106}$Mo.}
\footnotetext[3]{$\hat{T}(E2)$ intruder parameter taken from $^{108}$Mo.}
\label{tab-fit-par-mix-Mo}
\end{table}

\begin{table}
\caption{Hamiltonian and $\hat{T}(E2)$ parameters resulting from the study of Ru isotopes in the present work. All quantities have dimensions of energy (given in keV), except $e_N$ and $e_{N+2}$, which are given in units of $\sqrt{\text{W.u.}}$. It should be noted that the values $\chi_{N}=\chi_{N+2}=0$, $\kappa'_{N+2}=0$ keV, $\omega=15$ keV and $\Delta^{N+2}=2200$ keV were fixed for all isotopes.}
\label{tab-fit-par-mix-Ru}
\begin{center}
\begin{ruledtabular}
\begin{tabular}{ccccccccc}
Nucleus&$\varepsilon_N$&$\kappa_N$&$\kappa'_N$&$\varepsilon_{N+2}$& $\kappa_{N+2}$&$e_{N}$&$e_{N+2}$\\
  \hline
  $^{98}$Ru  & 683.60&   -15.79&   -1.28&   410.72\footnotemark[1]&   -24.08\footnotemark[1]& 2.55 &   4.00\footnotemark[1]\\ 
  $^{100}$Ru & 546.85&   -19.89&    8.60&   410.72&   -24.08& 2.34&   4.00\\
  $^{102}$Ru & 535.28&   -20.46&    4.75&   410.72\footnotemark[1]&   -24.08\footnotemark[1]& 2.27&   4.00\footnotemark[1]\\
  $^{104}$Ru & 412.75&   -24.34&    7.03&   410.72\footnotemark[1]&   -24.08\footnotemark[1]& 2.11&   4.00\footnotemark[1]\\
  $^{106}$Ru & 360.60&   -26.37&    5.67&   410.72\footnotemark[1]&   -24.08\footnotemark[1]& 1.96&   4.00\footnotemark[1]\\
  $^{108}$Ru & 298.53&   -29.19&    7.60&   410.72\footnotemark[1]&   -24.08\footnotemark[1]& 1.67&   4.00\footnotemark[1]\\
  $^{110}$Ru & 250.00&   -30.00&    9.55&   410.72\footnotemark[1]&   -24.08\footnotemark[1]& 1.55&   4.00\footnotemark[1]\\
  $^{112}$Ru & 250.00&   -30.00&    6.43&   410.72\footnotemark[1]&   -24.08\footnotemark[1]& 1.76&   4.00\footnotemark[1]\\
  $^{114}$Ru & 299.33&   -34.25&    5.55&   410.72\footnotemark[1]&   -24.08\footnotemark[1]& 1.80&   4.00\footnotemark[1]\\
\end{tabular}
\end{ruledtabular}
\end{center}
\footnotetext[1]{Hamiltonian and $\hat{T}(E2)$ parameters for the intruder sector taken from $^{100}$Ru.}
\end{table}

In this section, we describe how the parameters of the Hamiltonian (\ref{eq:ibmhamiltonian}), (\ref{eq:cqfhamiltonian}), and (\ref{eq:vmix}), as well as the effective charges of the $\hat{T}(E2)$ transition operator (\ref{eq:e2operator}), were determined.

We focus on studying the even-even isotopes $^{96-110}$Mo and $^{98-114}$Ru, covering a large portion of the neutron shell $50-82$. We exclude nuclei very close to the neutron shell closure due to the limited reliability of IBM results for those cases.

The goal of the fitting procedure is to achieve a satisfactory overall agreement with the available excitation energies and $B(E2)$ reduced transition probabilities. A standard $\chi^2$ method is used to determine the values of the parameters appearing in the Hamiltonian and the $\hat{T}(E2)$ operator, following the approach described in \cite{Garc09,Garc14b,Garc15,Garc19}. In general, there are $13$ parameters involved, but the number may be smaller for most nuclei. We impose the constraint that the parameters should vary smoothly from one isotope to another. Additionally, we strive to keep as many parameters as possible at constant values, particularly the parameters $\Delta^{N+2}$ and $\omega$.

The resulting parameter values for the IBM-CM Hamiltonian and $\hat{T}(E2)$ operator are presented in Tables \ref{tab-fit-par-mix-Mo} and \ref{tab-fit-par-mix-Ru} for Mo and Ru, respectively. In these tables, certain parameters could not be determined unambiguously from the available experimental information. Specifically, the parameters corresponding to the intruder sector of most Ru isotopes and the parameters of the regular sector of $^{108-110}$Mo. For Ru, only a few $0^+$ states in $^{100}$Ru could be identified as intruder members, but there is no strong evidence of other intruder states in the rest of the isotope chain. Therefore, we have to assume the same intruder parameters obtained for $^{100}$Ru for the entire Ru chain. As a result, the description of Ru intruder states should be considered only as approximate. In the case of $^{108-110}$Mo, no evidence of regular states exists, so the regular parameters of $^{106}$Mo are used for those isotopes.

It is worth noting the smooth variation or constancy of certain parameters, such as $\chi_{N}=\chi_{N+2}=0$ and $\kappa'_{N+2}=0$ keV in the Ru chain. Thus, both configurations in Ru exhibit $\gamma$-unstability. The value of $\omega$ is constant in Ru and in the majority of Mo isotopes, with $\omega=15$ keV, except for $^{96-100}$Mo where $\omega=45$ keV. A value of $\Delta^{N+2}=1500$ keV is employed for the entire Mo chain, which is compatible with the values used for Zr ($\Delta^{N+2}=820-3200$ keV) and Sr ($\Delta^{N+2}=1360-1900$ keV). For the entire Ru chain, a value of $\Delta^{N+2}=2200$ keV is considered, but it should be noted that this value is only constrained by the experimental information of $^{100}$Ru.

\begin{figure}
  \centering
  \begin{tabular}{cc} \includegraphics[width=0.418\linewidth]{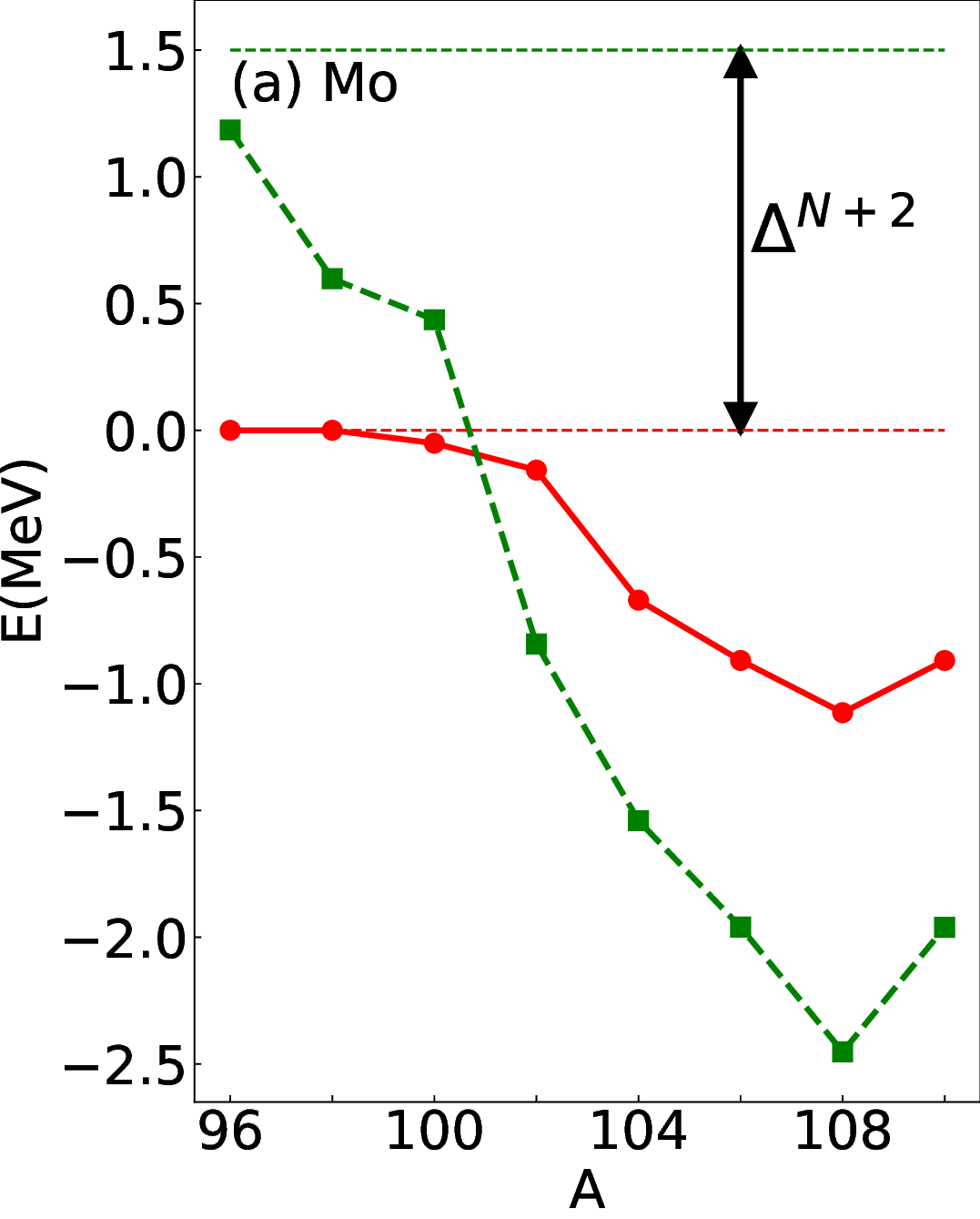}&                                                                      \includegraphics[width=0.42\linewidth]{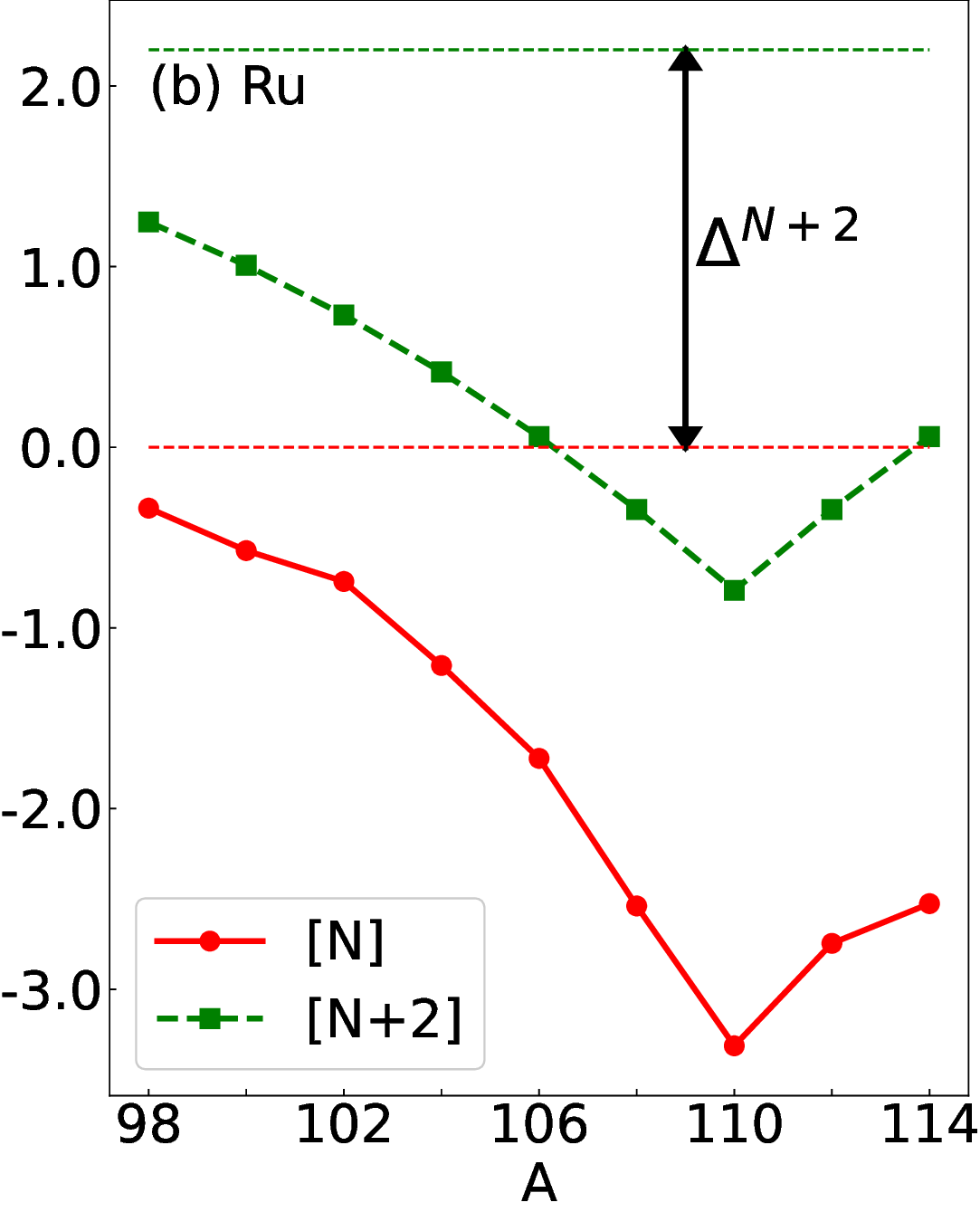}
  \end{tabular}
  \caption{Absolute energy of the lowest unperturbed regular (red) and intruder (green) 0$^+_1$ states for $^{96-110}$Mo (a) and $^{98-114}$Ru (b). Thin dashed lines correspond to the reference lines for regular (red) and intruder (green) configurations (see text). %The arrows correspond to the correlation energies in the N and N+2 subspaces (see also the text for a more detailed discussion).
  }
  \label{fig-energ-corr}
\end{figure}

\section{Correlation energy and unperturbed energy spectra}
\label{sec-corr_energy}
The position of intruder states is generally expected to be at higher energy compared to regular states due to the creation of a 2p-2h excitation across the shell gap. The parameter $\Delta^{N+2}=1500$ keV for Mo and $\Delta^{N+2}=2200$ keV for Ru represents the energy needed for this excitation. However, in practice, the energy of the intruder states is corrected by the pairing energy gain resulting from the formation of two extra $0^+$ pairs \cite{hey85, Hey87}.

The presence of extra bosons leads to a reduction in energy for the considered configuration. As a result, the lowest energies are expected to appear around the mid-shell, and the reduction in energy will be more significant for the intruder configuration due to the larger number of bosons (two units larger). Therefore, the actual energies of intruder states may differ from the initial expectation based solely on the shell gap energy.

\begin{figure}[hbt]
  \centering
  \begin{tabular}{cc}
    \includegraphics[width=.5\linewidth]{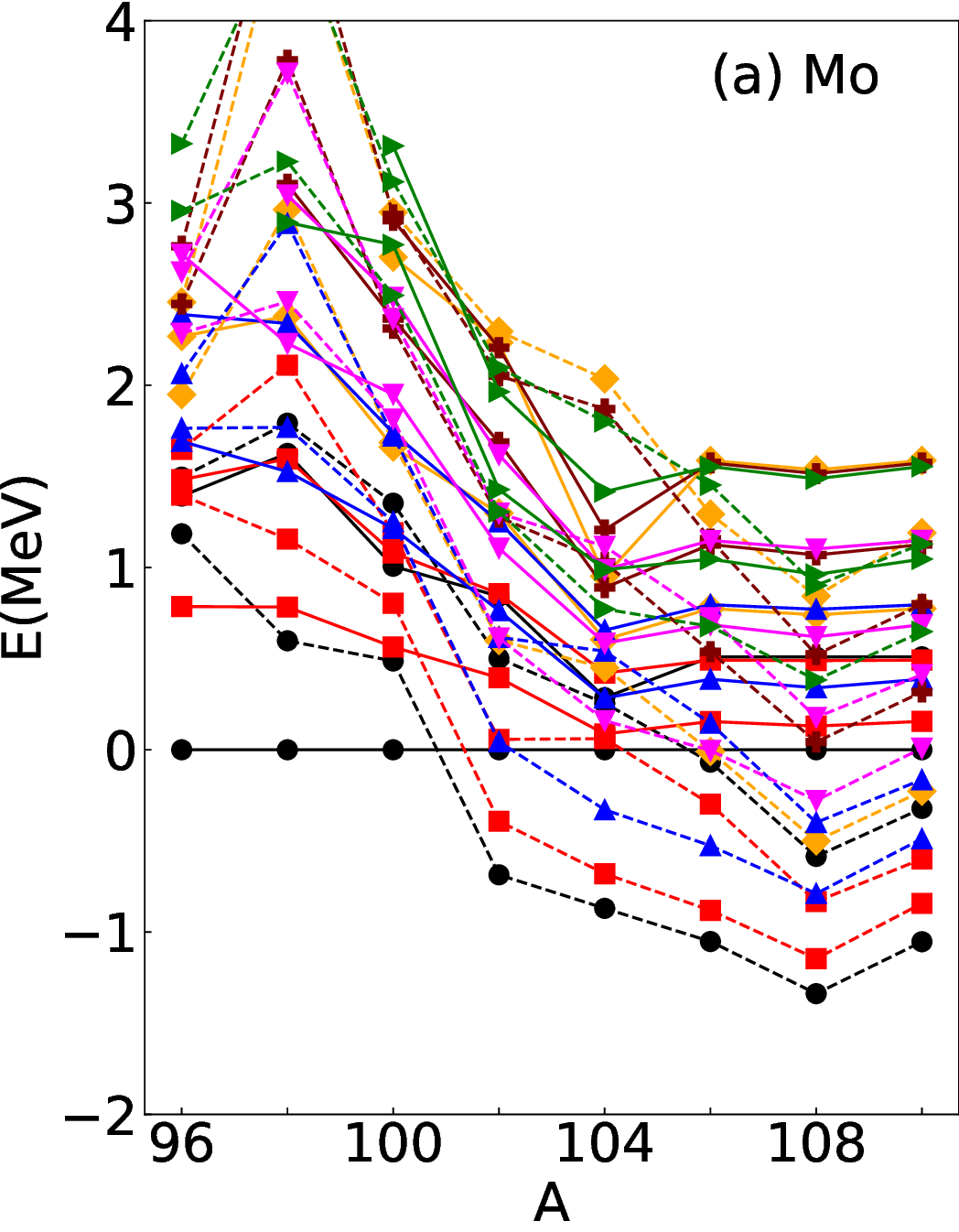}& \includegraphics[width=.5\linewidth]{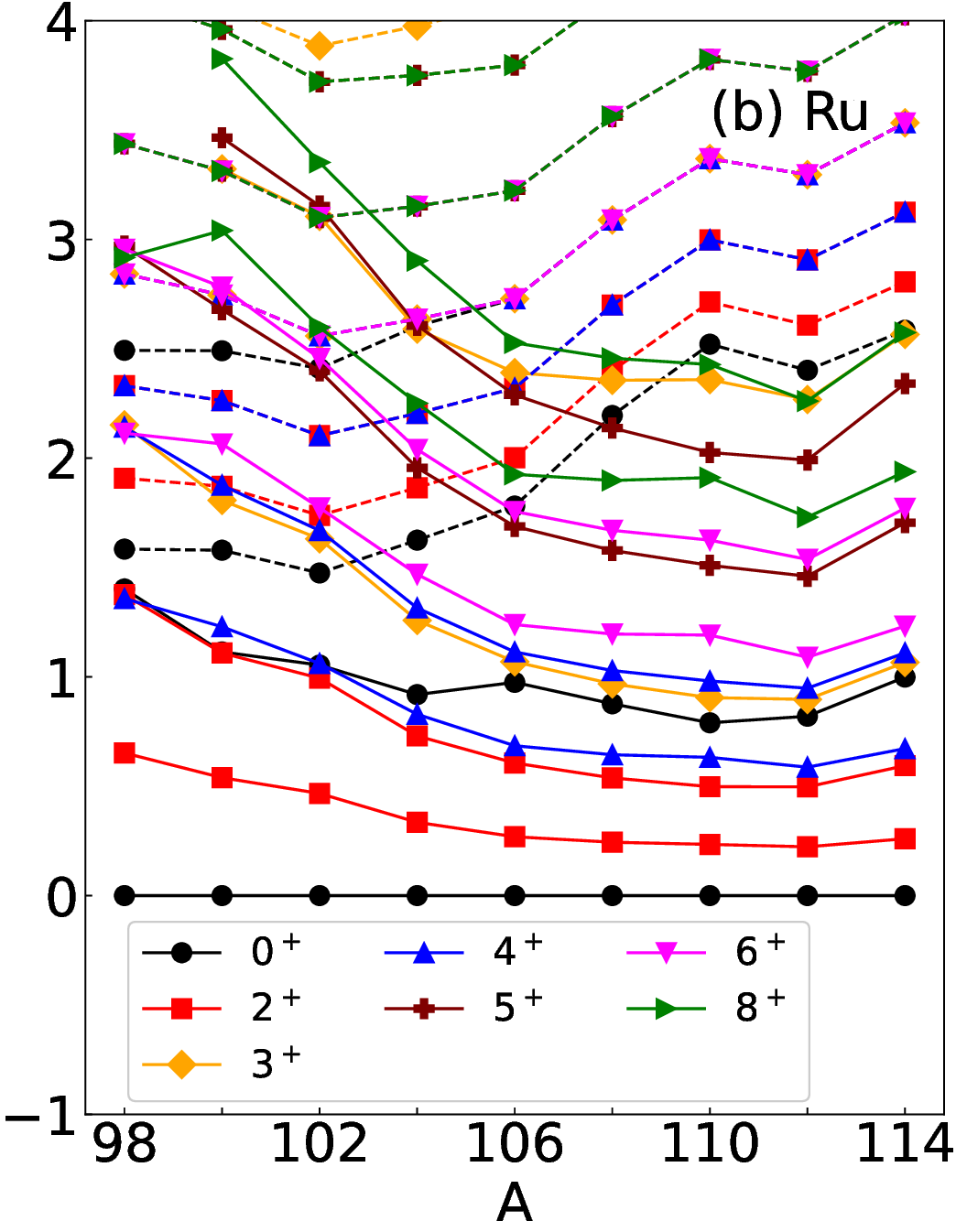}
  \end{tabular}
  \caption{
  Energy spectra for the Mo isotopes (panel (a)) and Ru isotopes (panel (b)), obtained from the IBM-CM Hamiltonian presented in Table \ref{tab-fit-par-mix-Mo} and \ref{tab-fit-par-mix-Ru}, respectively. For these calculations, the mixing term in the Hamiltonian has been switched off. For each angular momentum, the energy levels of the two lowest-lying regular and intruder states are displayed. The regular states are represented by full,  while the intruder states are shown with dashed lines.}
  \label{fig-ener-nomix}
\end{figure}

To gain a better understanding of the energy systematics in the regular and intruder configurations, we can examine the absolute energies of the lowest regular and intruder $0^+$ states by suppressing the mixing term in the IBM-CM Hamiltonian. In this analysis, we consider pure states where the reference regular energy corresponds to $0$, while the intruder energy corresponds to $\Delta^{N+2}$.
In Fig.\ \ref{fig-energ-corr}, panel (a) depicts the energy curves for Mo isotopes, while panel (b) shows the energy curves for Ru isotopes. In the case of Mo isotopes, a notable observation is the crossing of the regular and intruder energies between $A=100$ and $A=102$, corresponding to a neutron number of $60$. From $A=102$ onwards, the intruder configuration becomes the ground state. The correlation energy of the regular configuration increases gradually, while the intruder configuration exhibits a much more rapid change. This behavior can be attributed to the larger number of bosons in the intruder configuration and to have $|\kappa_{N+2}|>|\kappa_{N}|$. The minima of both curves occur precisely at the mid-shell region.

Turning to the case of Ru isotopes, we observe that the regular configuration remains the lowest energy state throughout the isotopic chain. Notably, the gain in correlation energy is more significant for the regular state. However, it is important to note that the Hamiltonian parameters for the intruder sector are kept constant and are the same as those used for $^{100}$Ru. Therefore, these results should be interpreted with caution, considering the potential limitations of using the same Hamiltonian parameters in the intruder sector for all Ru isotopes.

Overall, these energy correlation plots provide insights into the relative energies of the regular and intruder configurations in Mo and Ru isotopes, highlighting their different behaviors and the impact of correlation energy on their ordering.

To gain an initial understanding of the distribution of intruder and regular states in the spectrum, we can examine the unperturbed spectra of Mo and Ru isotopes, with the regular ground state as the reference. Fig.\ \ref{fig-ener-nomix} illustrates these spectra, with panel (a) representing Mo isotopes and panel (b) representing Ru isotopes. Full lines for the regular states and dashed lines for the intruder ones.

In the case of Mo isotopes (panel (a)), we observe that the regular configuration exhibits a more vibrational character for the lighter isotopes. However, as the mass increases, it switches to a more rotational or O(6) behavior. This transition is evident from the reduction in excitation energies of the $2^+$ and $4^+$ states. On the other hand, the intruder configuration demonstrates a more prominent rotational character throughout the isotopic chain, which becomes more evident for isotopes with $A>100$.

In the case of Ru isotopes (panel (b)), the intruder configuration is considerably higher in energy, with a minimum occurring at $A=98-102$. Beyond this region, the energy of the intruder configuration increases rapidly with increasing mass. The regular configuration, on the other hand, displays a clear vibrational pattern in the lighter isotopes. It is possible to easily identify the two-phonon triplet ($4^+$, $2^+$, $0^+$) or the three-phonon quintuplet ($6^+$, $4^+$, $3^+$, $2^+$, $0^+$). However, in the heavier isotopes, the presence of a $\gamma$-unstable structure becomes evident. The clear manifestation of this structure is seen in the seniority two doublet ($2^+$ and $4^+$) and the seniority three quartet ($6^+$, $4^+$, $3^+$, and $0^+$). Notably, at $A=104$ (neutron number $60$), there is a transition point where the system switches from one structure to the other.

Overall, these unperturbed spectra provide valuable insights into the nature of intruder and regular states in Mo and Ru isotopes, highlighting the vibrational and rotational characteristics and the presence of clear structural patterns in the heavier isotopes.
\begin{figure}[hbt]
  \centering
  \begin{tabular}{cc}
    \includegraphics[width=0.521\linewidth]{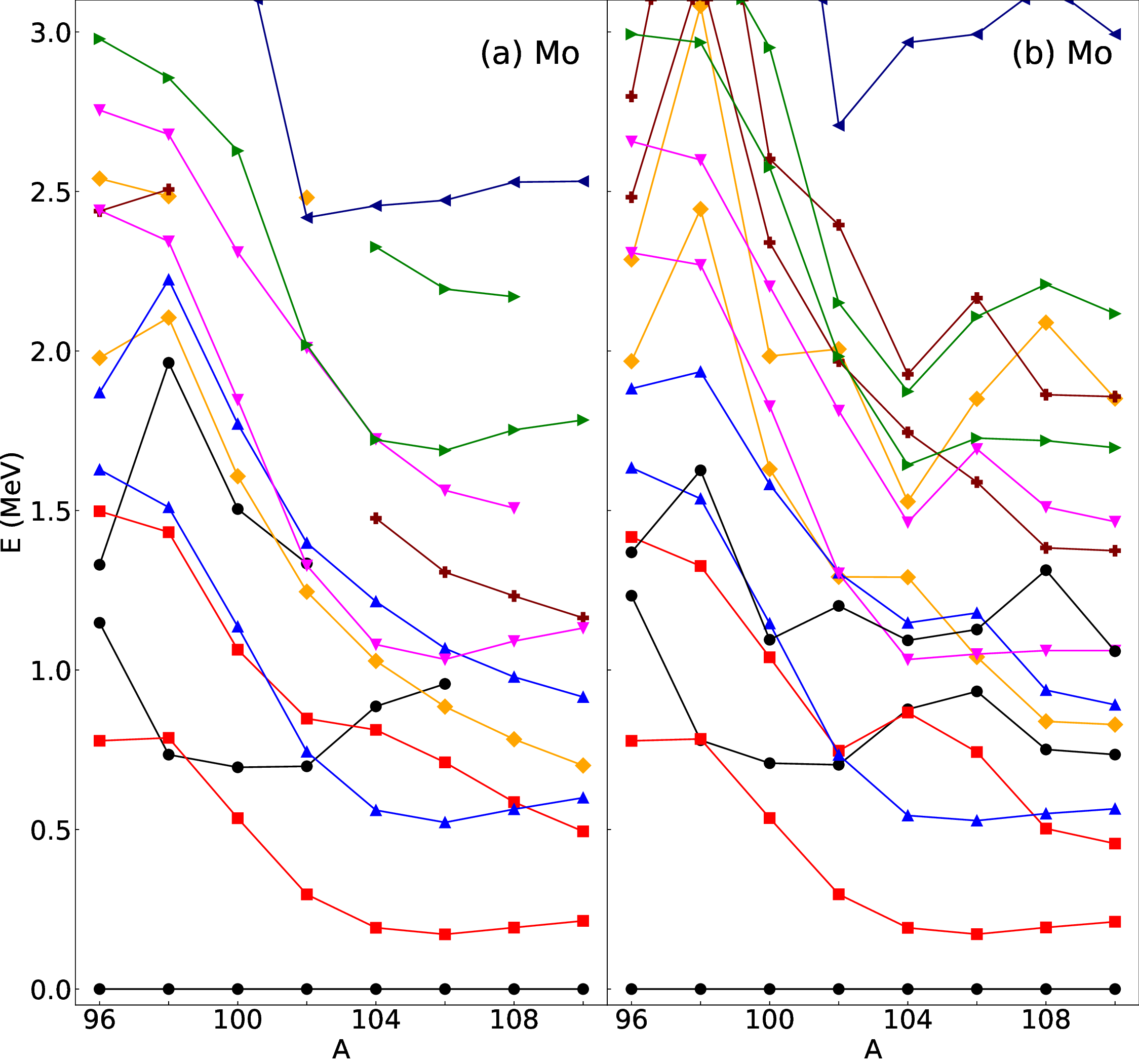} &
    \includegraphics[width=0.55\linewidth]{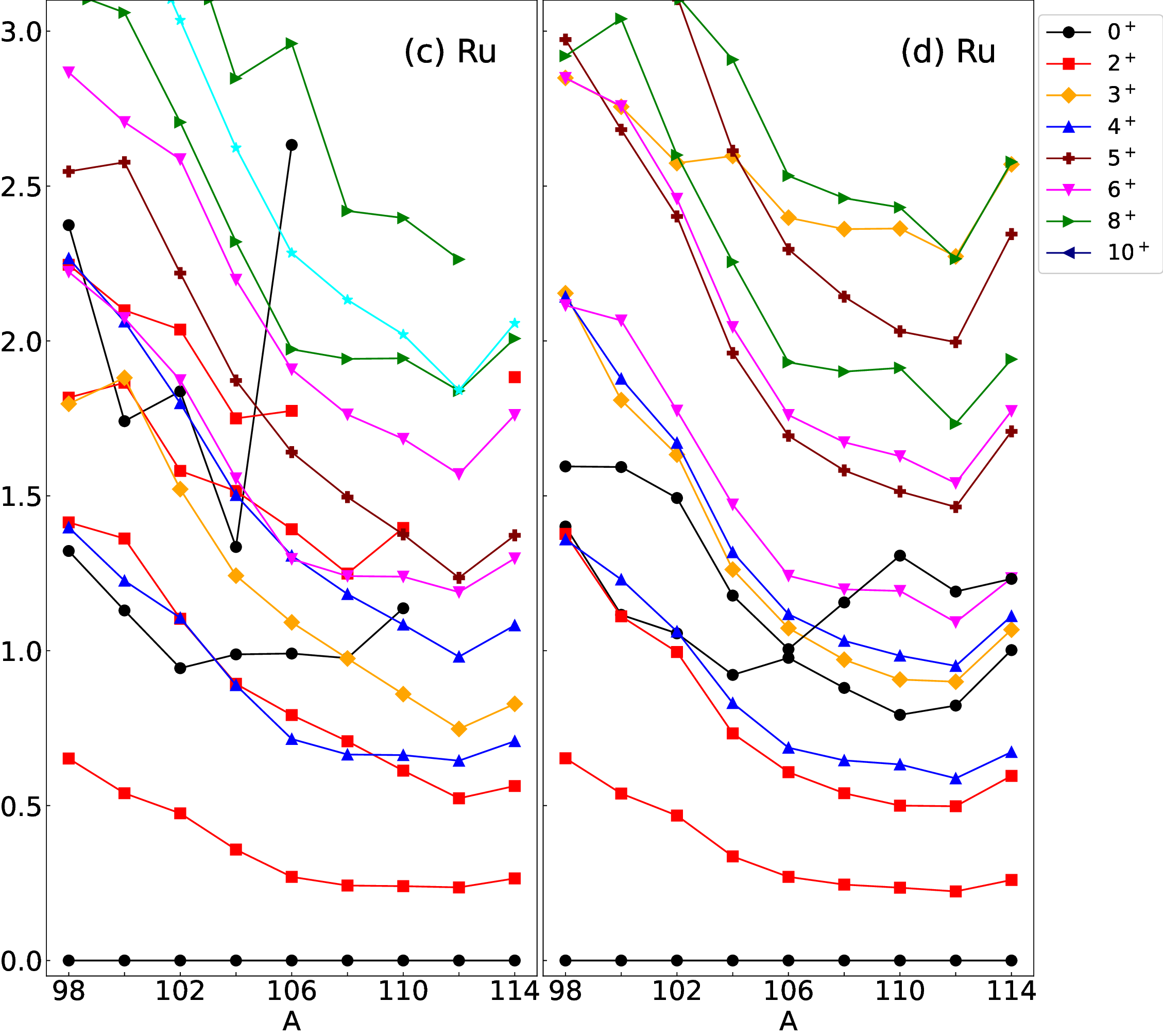} 
  \end{tabular}                 
  \caption{Excitation energies (up to $E \approx 3.0$  MeV) for Mo in panel (a) with the experimental data and in panel (b) with the IBM-CM theoretical results. Same information for Ru isotopes, in panel (c) experimental data and in panel (d) IBM-CM theoretical results. Only two excited states (if known experimentally) per angular momentum are plotted.} 
  \label{fig-energ-comp-combined}
\end{figure}

\section{Detailed comparison for energy spectra and $B(E2)$ transition rates}
\label{sec-comp}

\begin{figure}[hbt]
\centering
\begin{tabular}{cc}
  \includegraphics[width=0.503\linewidth]{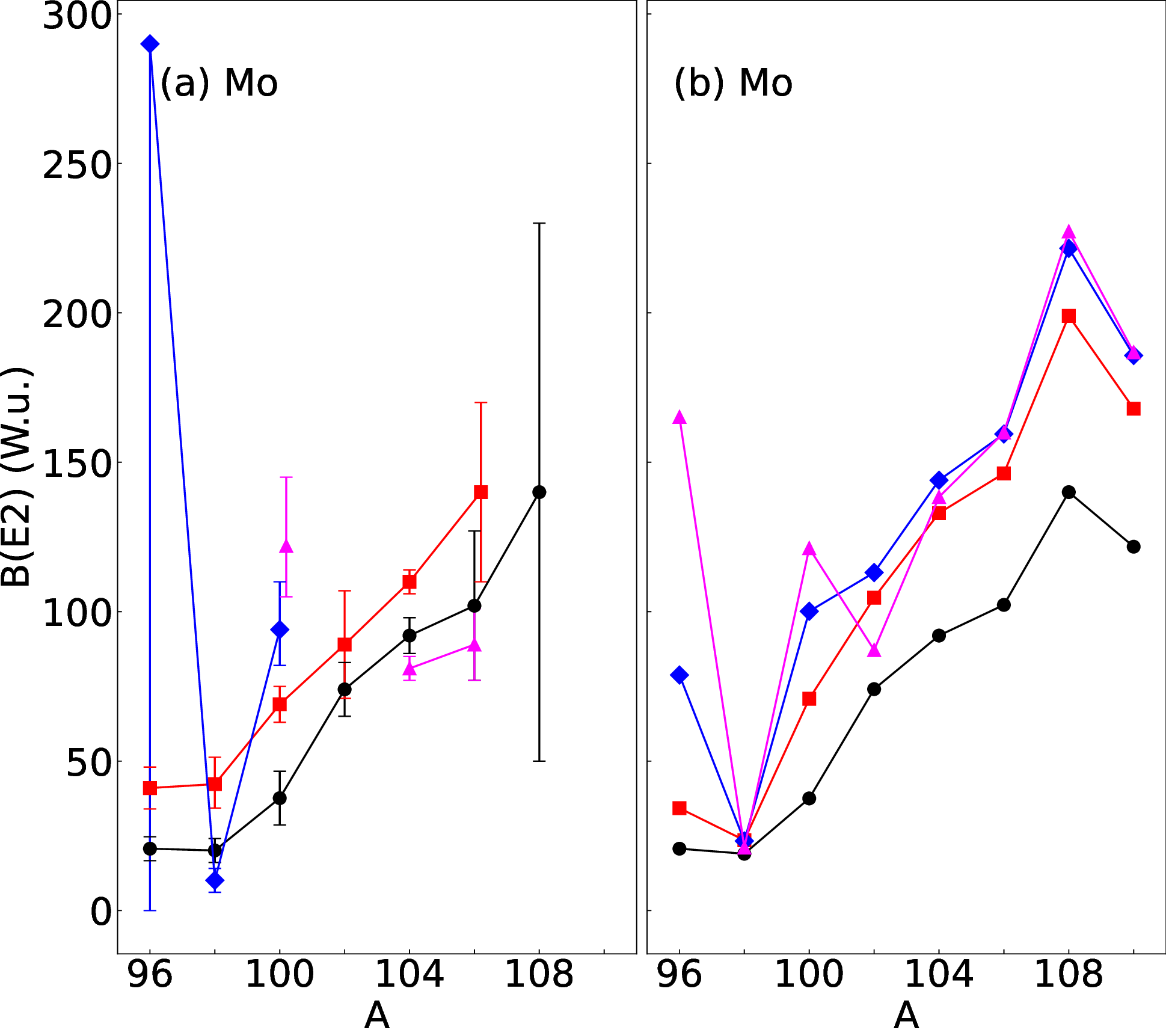}&
    \includegraphics[width=0.487\linewidth]{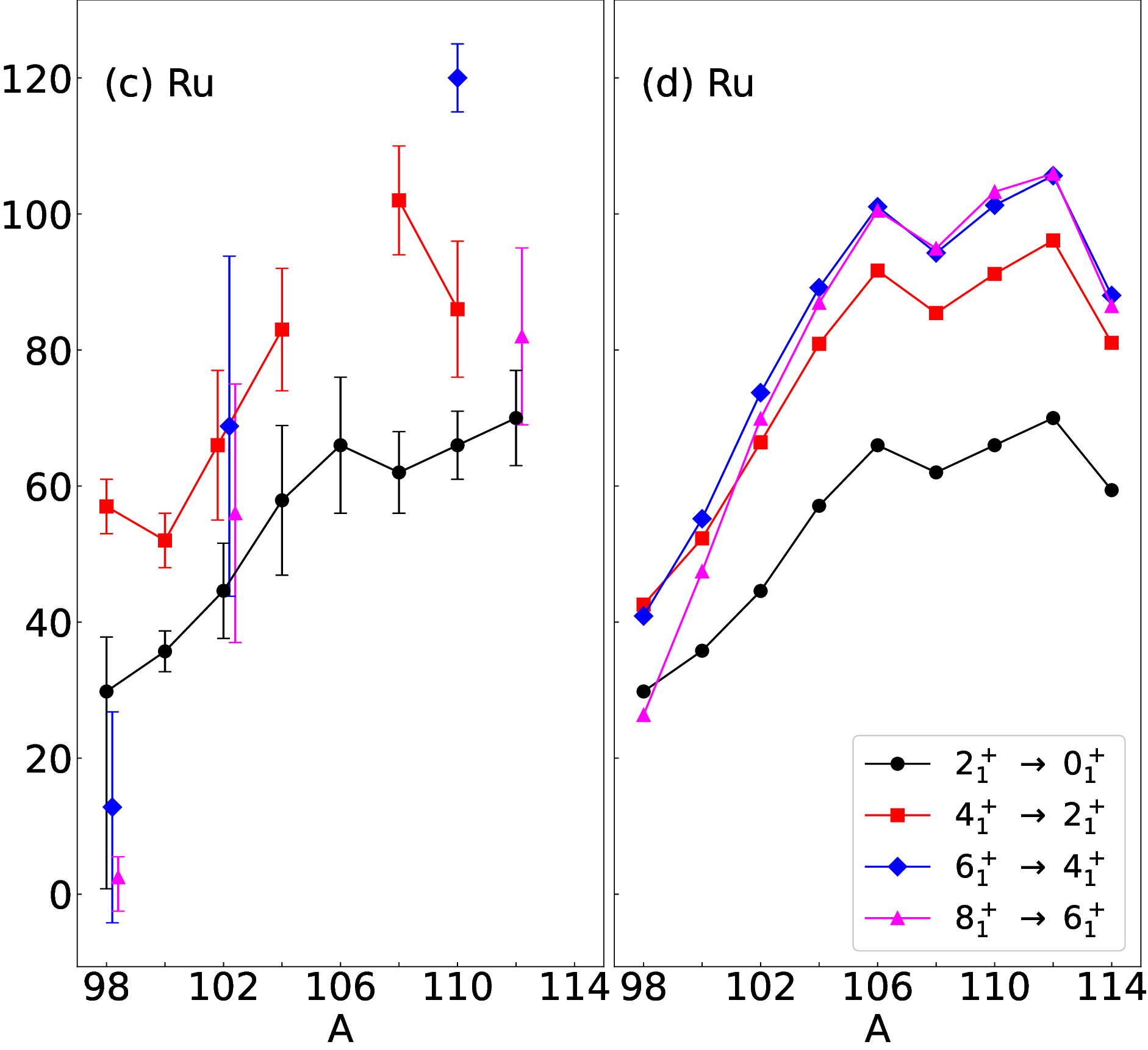}
\end{tabular}
\caption{Comparison of the absolute $B(E2)$ transition probabilities along the yrast band, given in W.u. Panels (a) and (c) corresponds to known experimental data for Mo and Ru, respectively and panels (b) and (d) to the theoretical IBM-CM results, also for Mo and Ru, respectively.} 
  \label{fig-be2-1-combined}
\end{figure}

In this section, we compare the theoretical calculations with the experimental data up to an excitation energy of approximately $3$ MeV. Fig.\ \ref{fig-energ-comp-combined} presents a detailed comparison between the experimental excitation energies of Mo (panel (a)) and their corresponding theoretical values (panel (b)). Similarly, it compares the experimental excitation energies of Ru (panel (c)) with their theoretical counterparts (panel (d)). It is important to note that, throughout the entire chain, the theoretical $2_1^+$ excitation energy closely matches the experimental data. Consequently, we utilize this experimental energy as a reference to normalize the theoretical spectra (refer to Section 3.2 of Ref.\ \cite{Garc09} for more details).

Beginning with the Mo isotopes, the spectra of the lighter isotopes exhibit the expected characteristics for nuclei near the neutron number $50$ shell closure. Additionally, there is a sudden increase in the excitation energies of certain states attributed to the presence of the neutron number $56$ subshell closure. Subsequently, the spectra evolve smoothly towards a more collective behavior, characterized by a compressed spectrum. The IBM-CM accurately reproduces all of these features, particularly the position of the $0_2^+$ state along the entire chain.

In the case of Ru isotopes, the theoretical energies quantitatively reproduce the experimental ones, clearly depicting a transition from a purely vibrational-like spectrum (where the one-, two-, and three-phonon states are easily identified) to an O(6) spectrum. However, it should be noted that the experimental levels are more scattered compared to the theoretical ones. There is a notable discrepancy observed for the $0_3^+$ state in $^{106}$Ru, which appears much higher in energy than predicted by the theoretical model. It is worth mentioning that this particular state was not included in the fitting procedure. Additionally, a slight discrepancy is observed for the $0_2^+$ state in the case of $^{108-110}$Ru, where its energy is slightly higher than the corresponding experimental value.
\begin{figure}[hbt]
  \centering
  \begin{tabular}{cc}
  \includegraphics[width=0.503\linewidth]{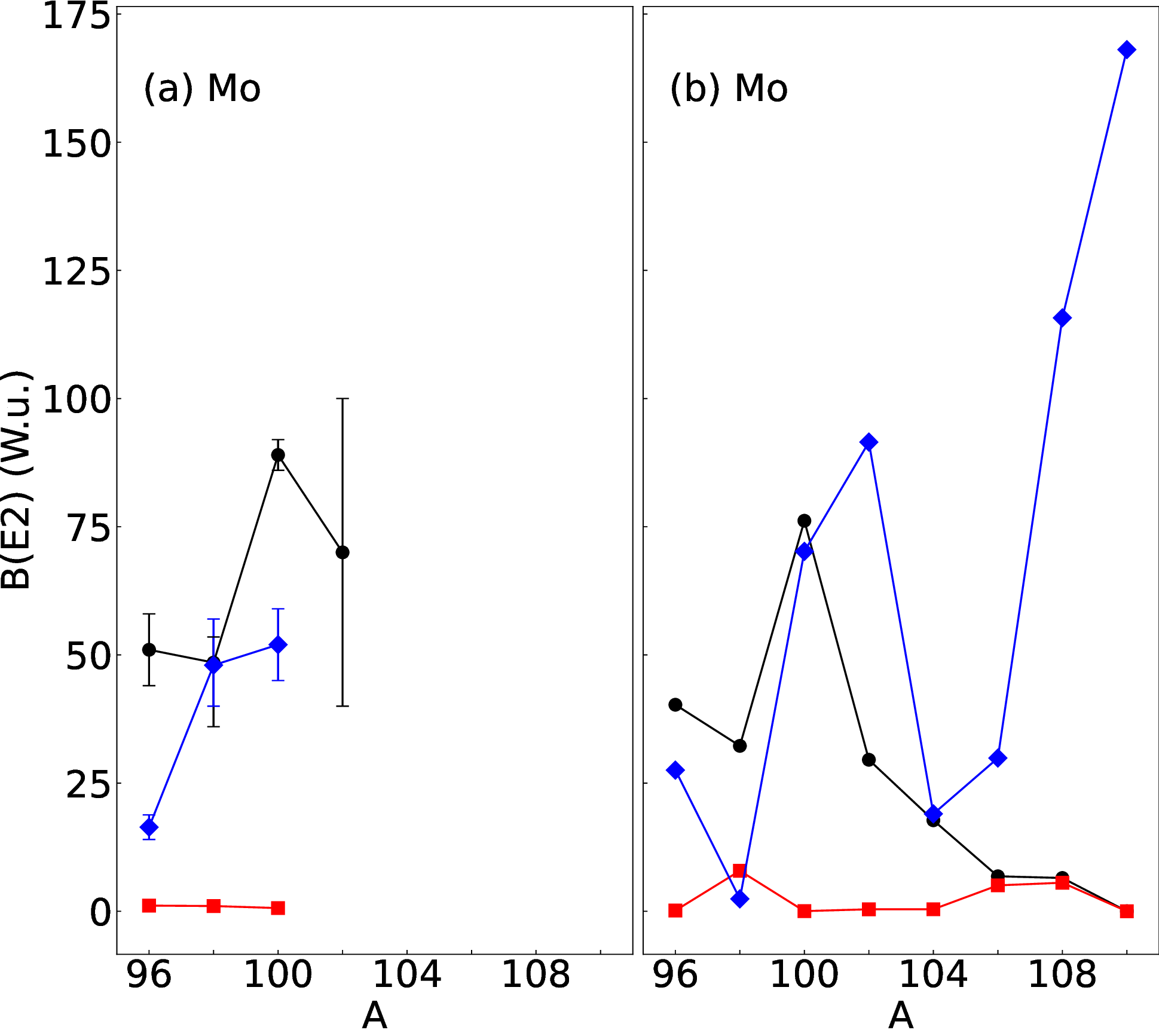}& \includegraphics[width=0.487\linewidth]{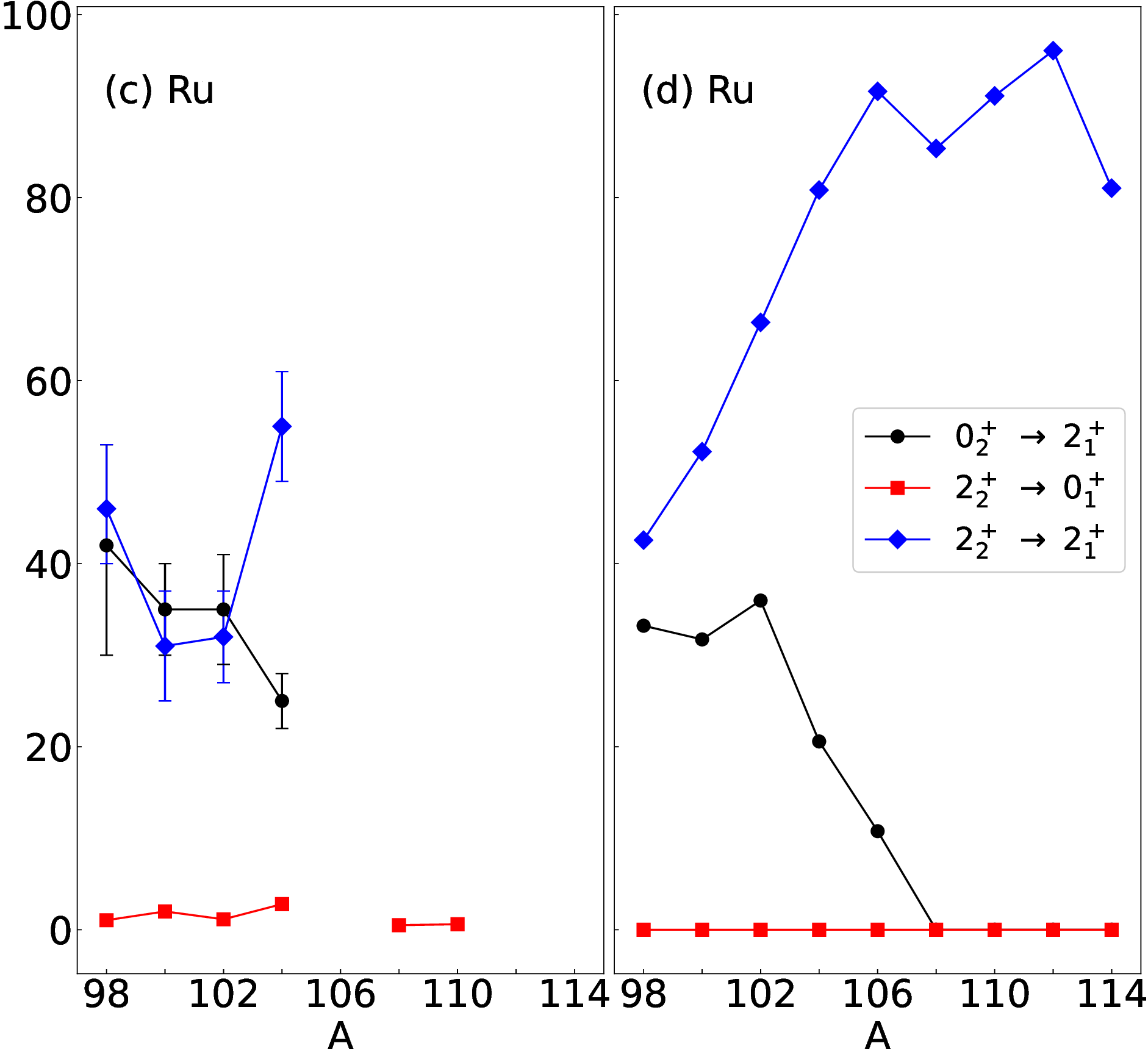}
  \end{tabular}
  \caption{Comparison of the few non-yrast intraband  absolute $B(E2)$ transition probabilities in Mo and Ru, given in W.u. Panels (a) and (c) correspond to the known experimental data for Mo and Ru, respectively, panel (b) and (d) to the theoretical IBM-CM results, also for Mo and Ru, respectively.} 
  \label{fig-be2-2-combined}
\end{figure}

The information regarding $B(E2)$ transitions is presented through a series of tables and figures. Tables \ref{tab-be2-Mo} and \ref{tab-be2-Ru} provide a detailed comparison between the known experimental values and the corresponding theoretical ones for Mo and Ru, respectively. Additionally, figures \ref{fig-be2-1-combined} and \ref{fig-be2-2-combined} illustrate selected intra- and interband transitions, respectively, highlighting the results for both Mo and Ru.

Overall, there is a satisfactory agreement between the theoretical predictions and experimental data, with only a few specific discrepancies. Notably, in $^{98}$Mo, the model predicts a smaller $B(E2; 2_2^+\rightarrow 2_1^+)$ value than observed experimentally. Additionally, it is worth mentioning the nearly equal $B(E2)$ values for transitions within the yrast band. Generally, in both chains of isotopes, the natural increase of $B(E2)$ values with angular momentum, reaching a maximum at mid-shell, is accurately reproduced by the model.

The reliable reproduction of the transition rates serves as a stringent test for the model. Therefore, the agreement obtained between theory and experiment demonstrates the reliability of the presented calculations, particularly in cases where experimental information is more abundant.

\begin{figure}[hbt]
  \centering
  \includegraphics[width=1\linewidth]{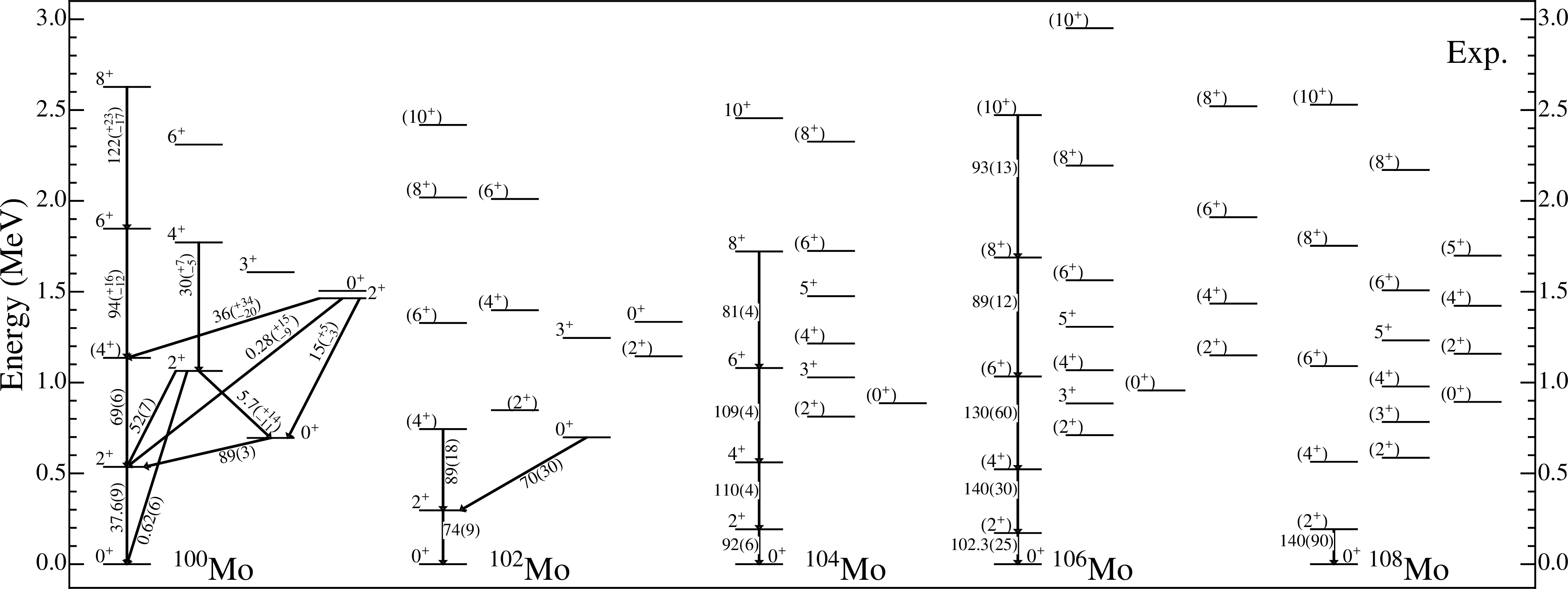}
  \caption{Experimental excitation energies and absolute $B(E2)$ transition rates (given in W.u.) for selected states in $^{100-108}$Mo.} 
  \label{fig-exp-Mo}
\end{figure}

\begin{figure}[hbt]
  \centering
  \includegraphics[width=1\linewidth]{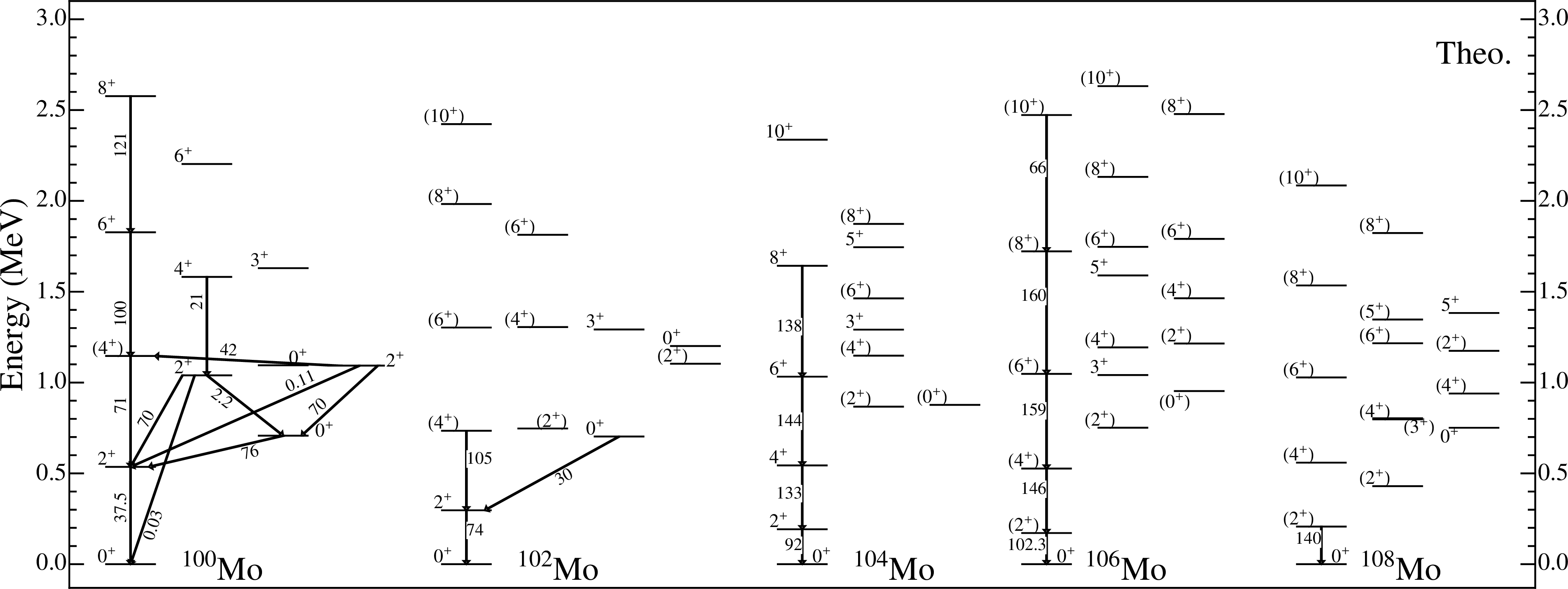} 
  \caption{The same as Fig.\ \ref{fig-exp-Mo} but for the theoretical IBM-CM results. } 
  \label{fig-theo-Mo}
\end{figure}

%\vspace*{-1cm}
%\begingroup
%\squeezetable
%\setlength{\LTcapwidth}{\linewidth}
%\renewcommand{\arraystretch}{3.5}
  \begin{longtable}{p{3.5cm}p{3.5cm}p{3.5cm}p{3.5cm}}
  \caption{Comparison of the experimental absolute $B(E2)$ values (given in W.u.) with the IBM-CM Hamiltonian results for Mo isotopes.
    Data are taken from the Nuclear Data Sheets \cite{Abri08, Sing03,Chen20, Sing08, Defr09, Blac07, Defr2008, Blac08, Defr09, Lalk2015, Blac2012}
    complemented with references presented in section \ref{sec-exp}. }  \\
  \label{tab-be2-Mo}
%   \vspace{-.5cm}
 %  \begin{center}
%\begin{ruledtabular}
% \begin{tabular}{cccc}
% Isotope~~~~~   &Transition~~~~~             &Experiment~~~~~ &IBM-CM~~~~~ \\
 Isotope  &Transition   &Experiment &IBM-CM \\
 \hline
 \endhead
   $^{96}$Mo&$2_1^+\rightarrow 0_1^+$& 20.7(4)        & 20.7\\
           &$0_2^+\rightarrow 2_1^+$& 51(7)       & 40\\   
           &$2_2^+\rightarrow 2_1^+$& 16.4(24)       & 27.5 \\   
           &$2_2^+\rightarrow 0_1^+$& 1.10(11)       & 0.14 \\   
           &$2_3^+\rightarrow 2_1^+$& $<28$       & 5 \\   
           &$2_3^+\rightarrow 0_1^+$& $<0.18$       & 0.32 \\   
           &$2_4^+\rightarrow 2_1^+$& 0.43(20)       & 0.07 \\   
           &$2_4^+\rightarrow 0_1^+$& 0.080(11)       & 0.022 \\   
           &$4_1^+\rightarrow 2_1^+$& 41(7)      & 34 \\   
           &$4_2^+\rightarrow 4_1^+$& 0.18$(^{+9}_{-10})$       & 14 \\   
           &$4_2^+\rightarrow 2_2^+$& 22.0$(^{+6}_{-10})$       & 56.4\\   
           &$4_2^+\rightarrow 2_1^+$& 1.9$(^{+5}_{-9})$        & 0.0 \\   
           &$4_3^+\rightarrow 2_2^+$& $<72$       & 7\\   
           &$4_3^+\rightarrow 2_1^+$& $<0.72$        & 0.08 \\   
           &$2_5^+\rightarrow 3_1^+$& 140$(^{+30}_{-40})$        & 115\\   
           &$2_5^+\rightarrow 2_3^+$& 4$(^{+8}_{-4})$        & 12\\   
           &$6_1^+\rightarrow 4_1^+$& $<290$       & 79\\   
           &$6_2^+\rightarrow 4_3^+$& $<52$        & 96\\   
           &$6_2^+\rightarrow 4_2^+$& $<1.1$        & 1.8\\   
           &$6_2^+\rightarrow 4_1^+$& $<47$        & 0.5\\   
           &$6_2^+\rightarrow 6_1^+$& $<65$        & 17\\   
           &$3_1^+\rightarrow 2_2^+$& $<1.8$        & 25\\   
           &$3_1^+\rightarrow 2_1^+$& $<1.3$        & 0.06\\   
           &$5_1^+\rightarrow 4_3^+$& $<1700$        & 20\\   
           &$5_1^+\rightarrow 3_1^+$& $<2000$        & 69\\   
           &$5_1^+\rightarrow 4_2^+$& $<100$       & 3\\   
           &$3_3^+\rightarrow 4_1^+$& 2.7$(^{+1.7}_{-2.7})$       & 0.1\\   
           &$3_3^+\rightarrow 2_3^+$& 4.1$(^{+2.4}_{-4.1})$        & 0.1\\   
           &$3_3^+\rightarrow 2_1^+$& 0.19$(^{+10}_{-19})$         & 0.00\\   
\hline
   $^{98}$Mo&$2_1^+\rightarrow 0_1^+$& 20.1(4)         &19    \\  
           &$2_1^+\rightarrow 0_2^+$& 9.7$(^{+10}_{-25})$       & 6.5\\   
           &$4_1^+\rightarrow 2_1^+$& 42.3$(^{+9}_{-8})$    & 23.5\\ 
           &$6_1^+\rightarrow 4_1^+$& 10.1(4)       & 23.3\\   
           &$2_2^+\rightarrow 0_1^+$& 1.02$(^{+15}_{-12})$       & 7.90 \\   
           &$2_2^+\rightarrow 0_2^+$& 2.3$(^{+5}_{-4})$       & 2.3 \\   
           &$2_2^+\rightarrow 2_1^+$& 48$(^{+9}_{-8})$       & 2 \\   
           &$4_1^+\rightarrow 2_2^+$& 15.2$(^{+33}_{-30})$       & 16.6 \\   
           &$2_3^+\rightarrow 4_1^+$& 14(4)       & 5 \\   
           &$2_3^+\rightarrow 2_2^+$& $<22$       & 3.5 \\   
           &$2_3^+\rightarrow 2_1^+$& 3.0(7)       & 30.9 \\   
           &$2_3^+\rightarrow 0_2^+$& 7.5$(^{+6}_{-5})$       & 0.0 \\   
           &$2_3^+\rightarrow 0_1^+$& 0.032$(^{+7}_{-6})$       & 0.228 \\   
 \hline
   $^{100}$Mo&$2_1^+\rightarrow 0_1^+$& 37.6(9)   & 37.5   \\  
           &$0_2^+\rightarrow 2_1^+$& 89(3)  &  76 \\    
           &$4_1^+\rightarrow 2_1^+$& 69(6) &  71  \\ 
           &$6_1^+\rightarrow 4_1^+$& 94$(^{+16}_{-12})$       & 100 \\  
           &$2_2^+\rightarrow 0_1^+$& 0.62(6)      & 0.03 \\   
           &$2_2^+\rightarrow 0_2^+$& 5.7$(^{+14}_{-11})$       & 2.2 \\   
           &$2_2^+\rightarrow 2_1^+$& 52(7)       & 70 \\   
           &$2_3^+\rightarrow 4_1^+$& 36$(^{+34}_{-20})$       & 42 \\   
           &$2_3^+\rightarrow 0_2^+$& 15$(^{+5}_{-3})$       & 70 \\   
           &$2_3^+\rightarrow 2_1^+$& 0.28$(^{+15}_{-9})$      & 0.11 \\   
           &$4_2^+\rightarrow 2_2^+$& 30$(^{+7}_{-5})$       & 21 \\   
           &$8_1^+\rightarrow 6_1^+$& 122$(^{+23}_{-17})$      & 121 \\   
   \hline
   $^{102}$Mo&$2_1^+\rightarrow 0_1^+$& 74(9)        & 74    \\  
           &$0_2^+\rightarrow 2_1^+$& 70(30)       & 30 \\   
           &$4_1^+\rightarrow 2_1^+$& 89(18)    & 105   \\ 
   \hline
   $^{104}$Mo&$2_1^+\rightarrow 0_1^+$& 92(6)         & 92    \\ 
           &$4_1^+\rightarrow 2_1^+$& 110(4)      & 133 \\   
           &$6_1^+\rightarrow 4_1^+$& 109(4)       & 144 \\   
           &$8_1^+\rightarrow 6_1^+$& 81(4)       & 138 \\   
   \hline
   $^{106}$Mo&$2_1^+\rightarrow 0_1^+$& 102.3(25)         & 102.3    \\ 
           &$4_1^+\rightarrow 2_1^+$& 140 (30)     & 146 \\   
           &$6_1^+\rightarrow 4_1^+$& 130(60)       & 159 \\   
           &$8_1^+\rightarrow 6_1^+$& 89(12)       & 160 \\ 
           &$10_1^+\rightarrow 8_1^+$& 93(13)       & 66 \\ 
   \hline
   $^{108}$Mo&$2_1^+\rightarrow 0_1^+$& 140(90)        & 140   \\ 
   \hline
 %\end{tabular}
 %\end{ruledtabular}
 %\end{center}
% \vspace*{-.2cm}
% \footnotetext[1]{Data taken from Ref.\ \cite{Regi17}.}
\end{longtable}
%\endgroup

%\begingroup
%\squeezetable
\begin{longtable}{p{3.5cm}p{3.5cm}p{3.5cm}p{3.5cm}}
  \caption{Same as Table \ref{tab-be2-Mo} but for Ru isotopes.}  \\
  \label{tab-be2-Ru}
%   \vspace{-.5cm}
   %\begin{center}
% \begin{ruledtabular}
% \begin{tabular}{cccc} 
  Isotope~~~~~   &Transition~~~~~             &Experiment~~~~~ &IBM-CM~~~~~ \\
 \hline
 \endhead 
   $^{98}$Ru&$2_1^+\rightarrow 0_1^+$& 29.8(10)         & 29.8\\ 
            &$0_2^+\rightarrow 2_1^+$& 42 $(^{+12}_{-11})$       & 33\\  
            &$4_1^+\rightarrow 2_1^+$& 57(4)         & 43\\   
            &$6_1^+\rightarrow 4_1^+$& 12.8$(^{+17}_{-14})$         & 40.9\\      
           &$2_2^+\rightarrow 0_1^+$& 1.04$(^{+17}_{-14})$      & 0.00 \\   
           &$2_2^+\rightarrow 2_1^+$& 46$(^{+7}_{-6})$      & 43 \\   
           &$8_1^+\rightarrow 6_1^+$& $2.5(^{+5}_{-3})$       & 26 \\   
           &$10_1^+\rightarrow 8_1^+$& 1.27$(^{+31}_{-23})$        & 0.43 \\   
 \hline
   $^{100}$Ru&$2_1^+\rightarrow 0_1^+$& 35.7(3)         &35.8    \\  
           &$0_2^+\rightarrow 2_1^+$& 35(5)   & 32 \\   
           &$4_1^+\rightarrow 2_1^+$& 52(4)  & 52 \\     
           &$2_2^+\rightarrow 2_1^+$& 31(6)    & 52 \\     
           &$2_2^+\rightarrow 0_1^+$& 2.0(4)   & 0.0 \\     
           &$2_3^+\rightarrow 4_1^+$& 17(5)    & 11 \\     
           &$2_3^+\rightarrow 0_2^+$& 37$(^{+8}_{-9})$    & 23 \\     
           &$2_3^+\rightarrow 2_1^+$& 1.24$(^{+0.43}_{-0.53})$    & 0.00 \\     
           &$2_3^+\rightarrow 0_1^+$& 0.43(10)    & 0.02 \\     
           &$2_4^+\rightarrow 0_3^+$& 270$(^{+60}_{-50})$    & 196 \\   
           &$2_4^+\rightarrow 4_1^+$& 1.9$(^{+0.6}_{-0.5})$    & 0.2 \\ 
           &$2_4^+\rightarrow 0_2^+$& 1.9(5)    & 0.0 \\              
           &$4_2^+\rightarrow 2_2^+$& 41$(^{+27}_{-21})$    & 29 \\     
           &$4_2^+\rightarrow 4_1^+$& 27$(^{+18}_{-14})$    & 26 \\     
           &$4_2^+\rightarrow 2_1^+$& 1.9$(^{+13}_{-10})$   & 0.0 \\     
           &$4_3^+\rightarrow 2_3^+$& 77$(^{+32}_{-29})$   & 7 \\     
           &$4_3^+\rightarrow 4_1^+$& 1.8$(^{+9}_{-8})$   & 0 \\     
           &$4_3^+\rightarrow 2_1^+$& 0.9(4)    & 0.3 \\     
           &$3_1^+\rightarrow 2_2^+$& 9.6$(^{+46}_{-41})$    & 39.2 \\   
           &$3_1^+\rightarrow 4_1^+$& 15(5)    & 16 \\     
           &$3_1^+\rightarrow 2_1^+$& 3.9$(^{+13}_{-12})$    & 0 \\     
 \hline
   $^{102}$Ru&$2_1^+\rightarrow 0_1^+$& 44.6(7)   & 44.6    \\  
           &$0_2^+\rightarrow 2_1^+$& 35(6)  &  36  \\   
           &$4_1^+\rightarrow 2_1^+$& 66(11) &  66  \\
           &$6_1^+\rightarrow 4_1^+$& 68(25)  &  74  \\  
           &$2_2^+\rightarrow 0_1^+$& 1.14(15)  &  0.00  \\ 
           &$2_2^+\rightarrow 2_1^+$& $32(5)$  &  66  \\  
           &$8_1^+\rightarrow 6_1^+$& 56(19)  &  70  \\  
           &$10_1^+\rightarrow 8_1^+$& 57(21)  &  58  \\  
             %reordenar los valores
   \hline
   $^{104}$Ru&$2_1^+\rightarrow 0_1^+$& 57.9(11)         & 57.1    \\  
           &$0_2^+\rightarrow 2_1^+$& 25(3)      & 21 \\   
           &$4_1^+\rightarrow 2_1^+$& 83(9)    & 81   \\  
           &$2_2^+\rightarrow 0_1^+$& 2.8(5)       & 0.0 \\   
           &$2_2^+\rightarrow 2_1^+$& 55(6)       & 81 \\   
   \hline 
   $^{106}$Ru&$2_1^+\rightarrow 0_1^+$& 66(10)         & 66    \\   
   \hline
   $^{108}$Ru&$2_1^+\rightarrow 0_1^+$& 62(6)         & 62    \\  
           &$4_1^+\rightarrow 2_1^+$& 102(8)    & 85   \\  
           &$2_2^+\rightarrow 0_1^+$& 0.5(4)    & 0.0   \\  
           &$2_3^+\rightarrow 4_1^+$& 0.08(7)    & 0.00   \\  
           &$2_3^+\rightarrow 0_1^+$& 0.005(3)    & 0.000   \\  
   \hline
   $^{110}$Ru&$2_1^+\rightarrow 0_1^+$&      66(5)   & 66    \\  
           &$4_1^+\rightarrow 2_1^+$& 86(10)    & 91   \\  
           &$6_1^+\rightarrow 4_1^+$& 120(50)       & 101 \\
           &$2_2^+\rightarrow 0_1^+$& 0.6(3)       & 0.0 \\
   \hline
   $^{112}$Ru&$2_1^+\rightarrow 0_1^+$& 70(7)         & 70    \\  
           &$8_1^+\rightarrow 6_1^+$& 82(13)      & 106 \\   
           &$10_1^+\rightarrow 8_1^+$& 85(13)    & 100   \\  
           &$7_1^+\rightarrow 5_1^+$& 83(12)       & 68 \\
           \hline
 %\end{tabular}
 %\end{ruledtabular}
 %\end{center}
% \vspace*{-.2cm}
% \footnotetext[1]{Data taken from Ref.\ \cite{Regi17}.}
\end{longtable}
    %     \endgroup

Figs.\ \ref{fig-exp-Mo} and \ref{fig-theo-Mo} display the detailed excitation energies up to approximately $3$ MeV and the $B(E2)$ transition rates for both experimental data and theoretical results. These figures focus on a selected set of Mo isotopes, namely $^{100-108}$Mo, where the coexistence of two bands is most evident. The separation into bands has been performed by first considering the yrast band and then grouping the remaining levels around the $0^+$ or $2^+$ bandheads, or according to an O(6) scheme.

In the case of $^{100}$Mo, the yrast band exhibits a predominantly vibrational character, which is accurately reproduced by the IBM-CM calculation. However, the remaining states cannot be easily grouped in any obvious manner. The main features of the $B(E2)$ transition rates are generally well reproduced, except for the $2_3^+\rightarrow 0_2^+$ transition, where the theoretical model predicts a larger value than observed experimentally.
Moving on to $^{102}$Mo, the yrast band begins to deviate from the harmonic behavior in terms of both energies and $B(E2)$ values. Nevertheless, the energies and $B(E2)$ values are correctly reproduced by the theoretical calculations.
In the cases of $^{104}$Mo and $^{106}$Mo, the spectra appear quite similar, and the IBM-CM calculations accurately reproduce them. However, based on the analysis presented in Section \ref{sec-wf}, the $0_2^+$ state is regular in the former case while is intruder in the latter, and the $2_2^+$ state is fully mixed in the former case but is intruder in the latter.
Lastly, in the isotope $^{108}$Mo, part of the spectra exhibit an O(6) character, although there are no known experimental $B(E2)$ values available for comparison. Table \ref{tab-be2-Mo} provides a comparison between the relevant $B(E2)$ values and their corresponding theoretical predictions, which complements the information provided in the former figures.

\begin{figure}[hbt]
  \centering
  \includegraphics[width=1\linewidth]{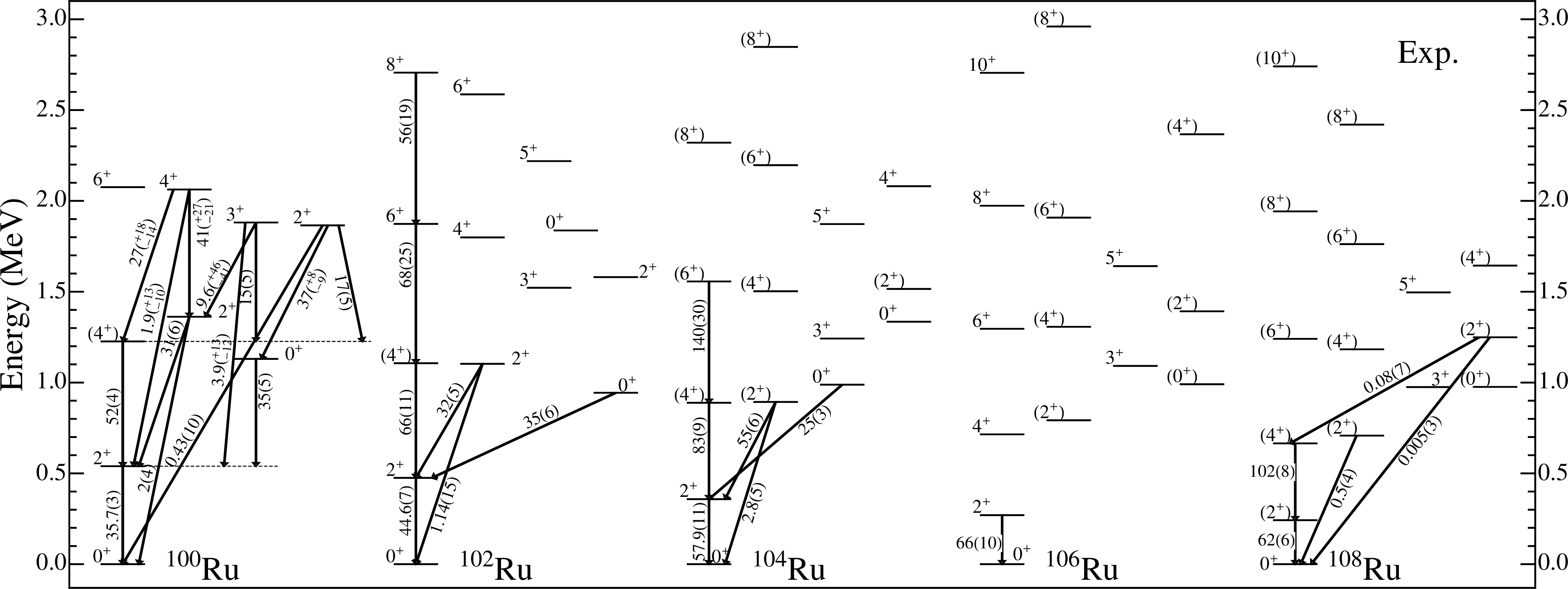}
  \caption{Experimental excitation energies and absolute $B(E2)$ transition rates (given in W.u.) for selected states in $^{100-108}$Ru.} 
  \label{fig-exp-Ru}
\end{figure}

\begin{figure}[hbt]
  \centering
  \includegraphics[width=1\linewidth]{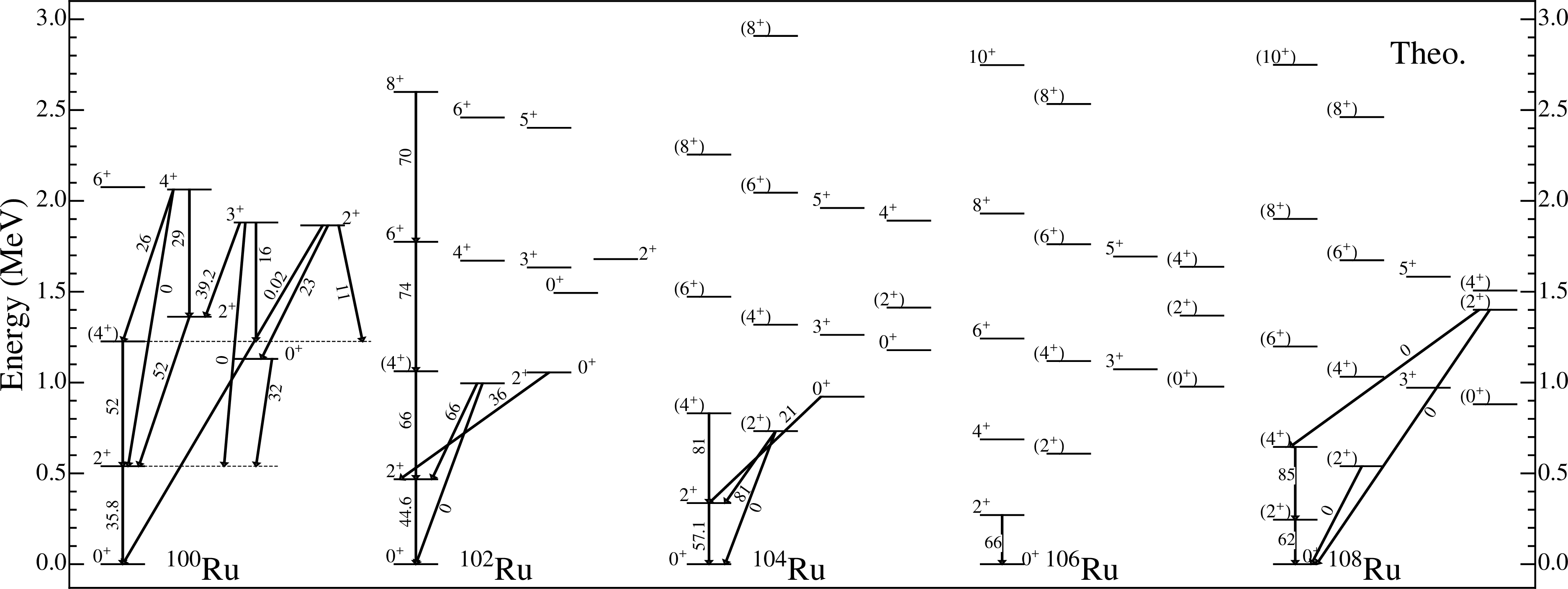} 
  \caption{The same as Fig.\ \ref{fig-exp-Ru} but for theoretical results.} 
  \label{fig-theo-Ru}
\end{figure}

Figs.\ \ref{fig-exp-Ru} and \ref{fig-theo-Ru} illustrate the experimental and theoretical spectra for the range of $^{100-108}$Ru isotopes. The overall agreement between experiment and theory is remarkable, with no significant discrepancies observed. As demonstrated in Fig.\ \ref{fig-energ-comp-combined}, nuclei up to $^{104}$Ru exhibit a vibrational-like structure, allowing for the identification of different members of one-, two-, three-, and even four-phonon multiplets. Starting from $^{106}$Ru, a transition towards an O(6)-like structure becomes evident, with $^{104}$Ru being the critical point of this transition.
Another noteworthy feature is the absence of intruder states, except for the $0_4^+$ and $0_5^+$ states in $^{100}$Ru (not shown in the figure). This highlights the consistency of the theoretical model in reproducing the experimental spectra.
Table \ref{tab-be2-Ru} provides a comparison between the relevant $B(E2)$ values and their corresponding theoretical predictions for Ru isotopes,  which complements the information provided in the former figures.

\section{Wave function structure}
\label{sec-wf}
In this section, we investigate the structure of wave functions, specifically focusing on the fraction lying in the regular sector, denoted as $[N]$.
The wave function, within the IBM-CM, can be written as
\begin{eqnarray}
  \Psi(k,JM) &=& \sum_{i} a^{k}_i(J;N) \psi((sd)^{N}_{i};JM) 
                  \nonumber\\
              &+& \
                  \sum_{j} b^{k}_j(J;N+2)\psi((sd)^{N+2}_{j};JM)~,
                  \label{eq:wf:U5}
\end{eqnarray}
where $k$ refers to the different states with a given $J$, while $i$, and $j$ run over the bases of the $[N]$ and $[N+2]$ sectors, respectively. The  weight of the wave function contained within the $[N]$-boson subspace, can then be defined as the sum of the squared amplitudes,
  \begin{equation} 
    w^k(J,N) \equiv \sum_{i}\mid a^{k}_i(J;N)\mid ^2.
    \label{wk}
  \end{equation} 

\begin{figure}[hbt]
  \centering
  \begin{tabular}{cc}
  \includegraphics[width=.5\linewidth]{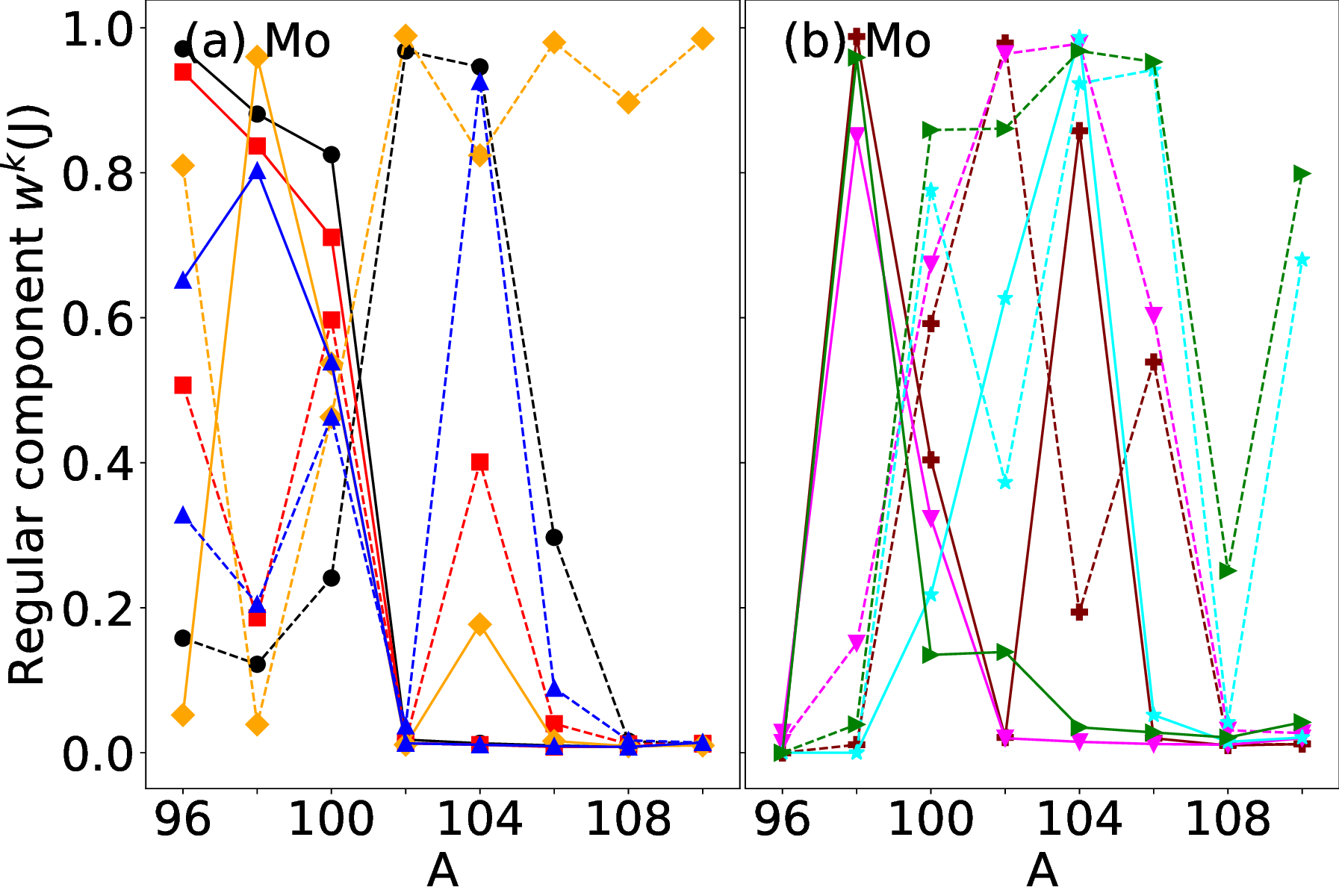}&
  \includegraphics[width=.5\linewidth]{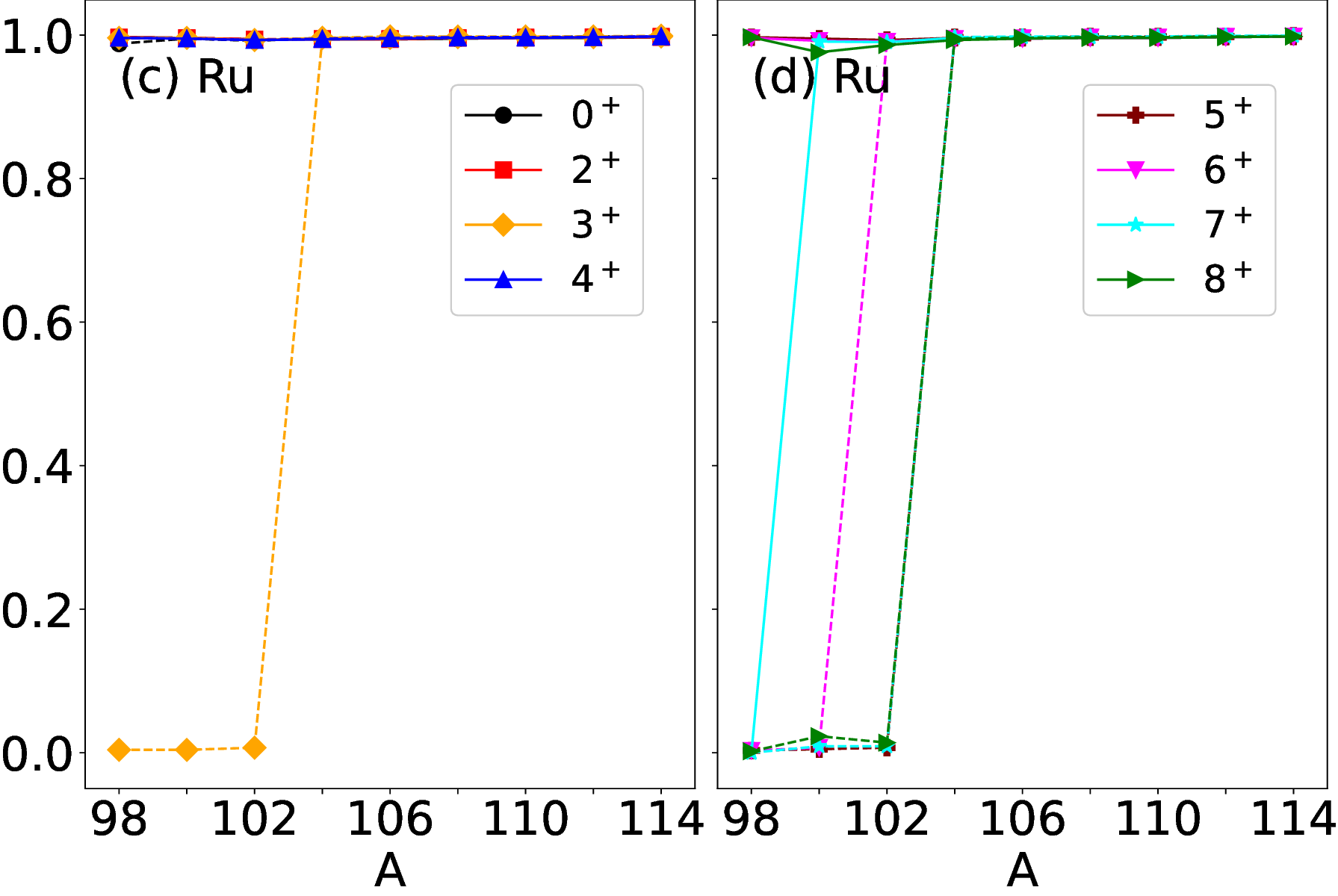}
  \end{tabular}
  \caption{Regular content of the Mo (panels (a) and (b)) and Ru (panels (c) and (d)) for the two lowest-lying states for each $J$ value (full lines with closed symbols correspond with the first state while dashed lines correspond with the second state) resulting from the IBM-CM  calculation.
  }
  \label{fig-wf-combined}
\end{figure}

Fig.\ \ref{fig-wf-combined} illustrates $w^k(J)$ for the first two states of each angular momentum, with the full line representing the first state and the dashed line representing the second state.

For Mo isotopes, the ground state undergoes a rapid transition from an almost entirely regular structure up to $A=100$, to a fully intruder one from $A=102$ onwards. This trend is also observed for the $2_1^+$, $4_1^+$, $6_1^+$, and $8_1^+$ states, with the exception of $A=96$, where the $6_1^+$ and $8_1^+$ states correspond to intruder configurations. No clear trend can be observed for odd angular momenta, both for the first and second members. The second $0^+$ state predominantly exhibits an intruder character in most cases, except for $A=102-104$, where it is fully regular. On the other hand, the second $2^+$ state shows significant mixing for $A=96-100$ and $104$, while being almost purely intruder for the remaining cases.

In the case of Ru isotopes, the observed trend is relatively straightforward. For the first member of all angular momentum states a regular wave function is predominant. However, for the second member, in some cases it undergoes a transition from an intruder character to a regular one starting in the majority of cases from $A=104$ and onwards.
\begin{figure}[hbt]
  \centering
  \begin{tabular}{cc}
  \includegraphics[width=.5\linewidth]{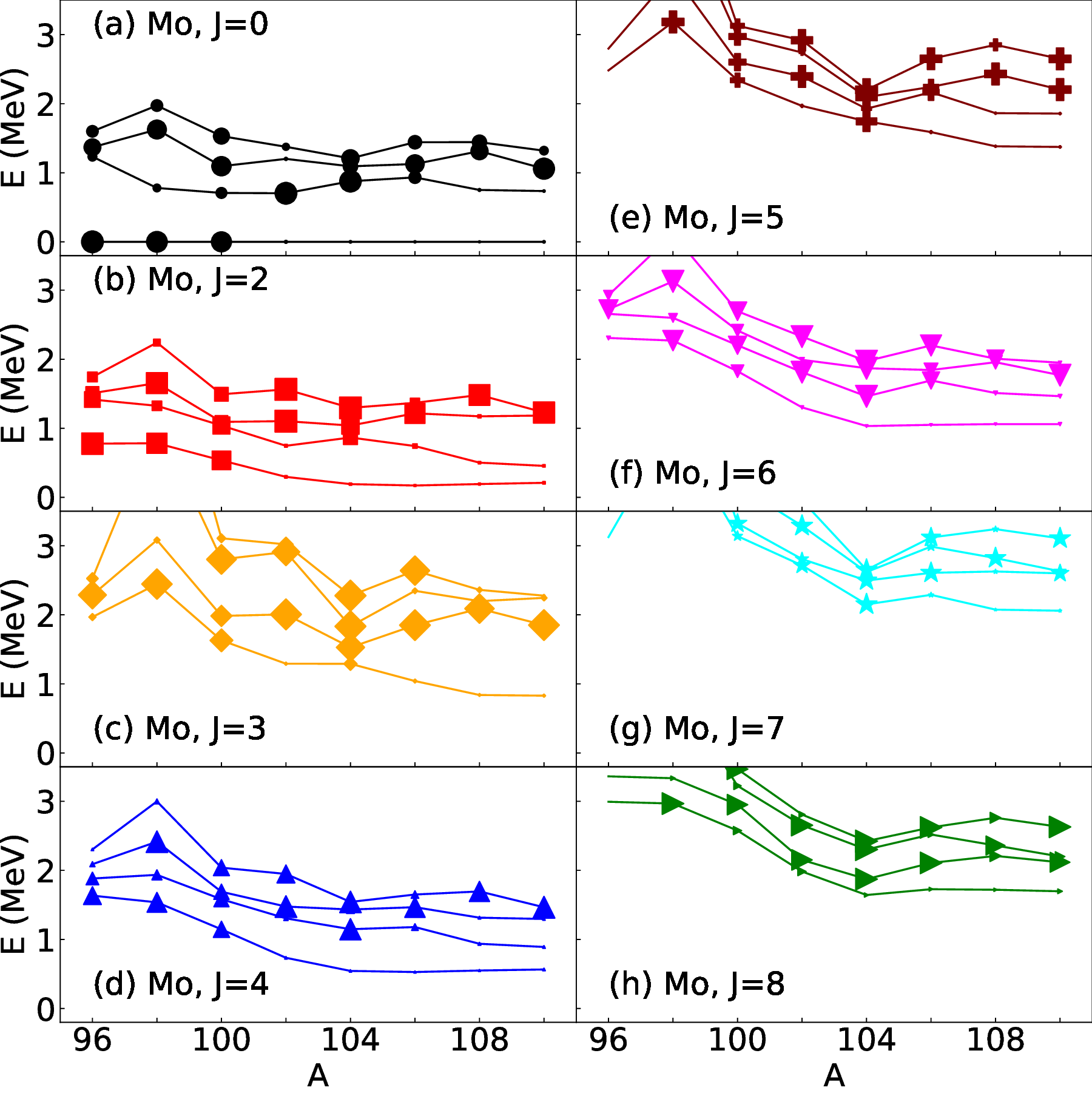}&
    \hspace{-0cm}\includegraphics[width=.5\linewidth]{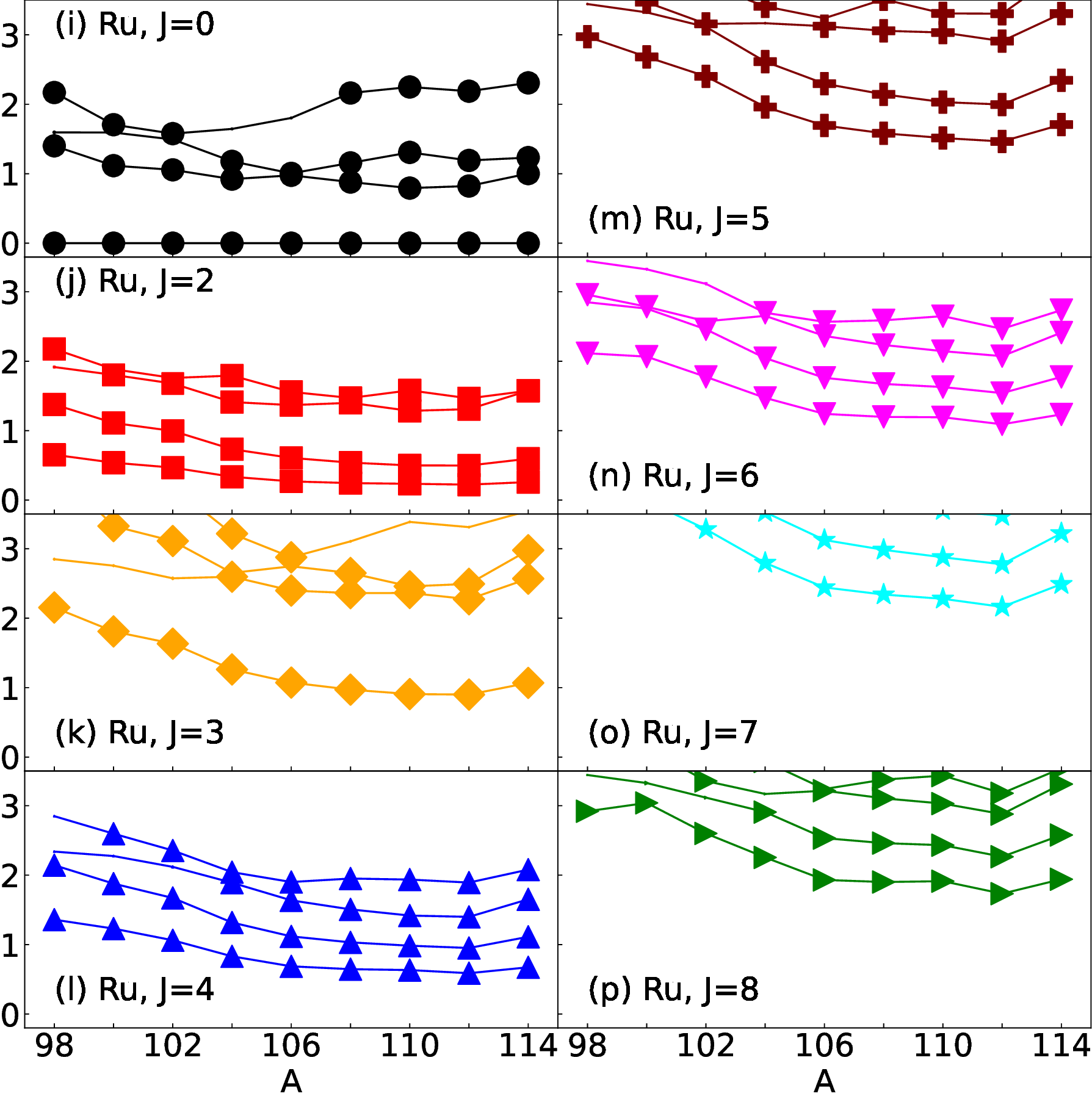}
  \end{tabular}
  \caption{The energy systematics of the four lowest states below $4$ MeV for both Mo and Ru isotopes. The panels (a)-(h) represent Mo, while panels (i)-(p) correspond to Ru. Each panel corresponds to a specific angular momentum: (a) and (i) for $J=0$, (b) and (j) for $J=2$, (c) and (k) for $J=3$, (d) and (l) for $J=4$, (e) and (m) for $J=5$, (f) and (n) for $J=6$, (g) and (o) for $J=7$, and (h) and (p) for $J=8$. The size of each symbol is proportional to the fraction of the wave function lying in the regular sector. The dot for the state $0_1^+$ in $^{98}$Ru corresponds to $100\%$.}
  \label{fig-wf-ener-combined}
\end{figure}

The representation of Fig.\ \ref{fig-wf-combined} can become cumbersome, especially in cases like panel (b) where the levels cross each other. To provide a clearer visualization, it is more effective to combine the fraction of the wave function within the regular sector with the excitation energy of the states. In Fig.\ \ref{fig-wf-ener-combined}, we present the regular content of the first four states per angular momentum along with their corresponding excitation energies. The size of each dot associated with a state is proportional to the regular content of its wave function.
To provide a reference point, the size of the dot for the $0_1^+$ states in $^{98}$Ru (panel (i)) corresponds to $100\%$ of regular content. In the case of Mo isotopes, it is evident how the regular content of the $0_1^+$ state transitions from the ground state in $^{96-100}$Mo to the first and second excited $0^+$ states in $^{102-104}$Mo and $^{106-110}$Mo, respectively. Similarly, for the $2^+$ and $4^+$ states, the regular content transitions from the first member in $^{96-102}$Mo to the second or third member in $^{104-110}$Mo. For angular momenta $6^+$ and $8^+$, the regular content is mainly concentrated in the second and third members throughout the isotopic chain.
In contrast, the situation in Ru is much simpler, with the majority of states belonging to the regular sector, as indicated by their significant regular content.

\section{Study of other observables: radii, isotopic shifts, and two-neutron separation energies}
\label{sec-other}

\begin{figure}[hbt] 
  \centering
  \begin{tabular}{cc}
  \includegraphics[width=0.31\linewidth]{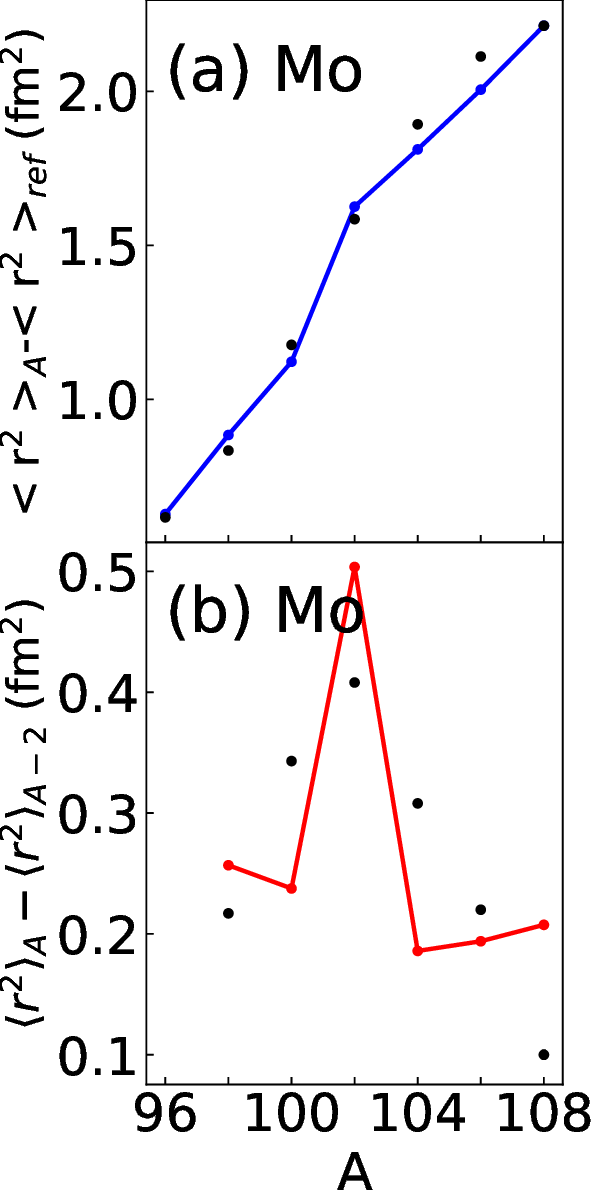}&
  \includegraphics[width=0.301\linewidth]{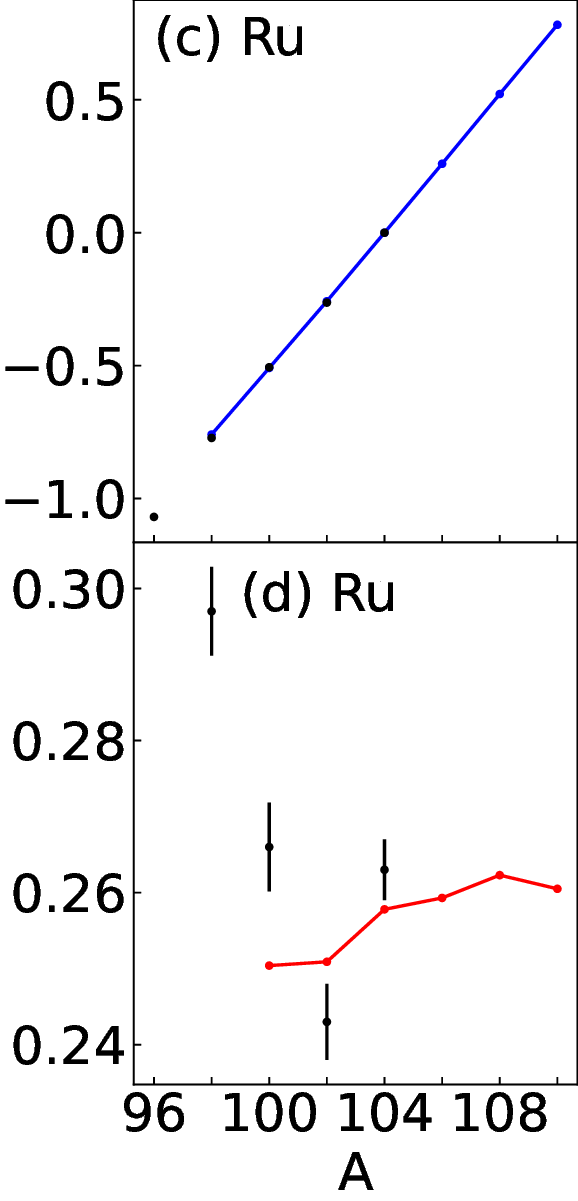}
  \end{tabular}
  \caption{Charge mean-square radii for the Mo (a) and Ru nuclei (c) and isotopic shift for the Mo (b) and Ru nuclei (d). The data are taken from \cite{Ange13}. Lines with dots for theoretical values and dots with error bars for experimental data. For Mo, $\langle r^2\rangle_{ref}$ is fixed to reproduce the value of $\langle r^2\rangle$ in $^{96}$Mo, while for Ru to reproduce $^{96}$Ru.}
  \label{fig-iso-shift-combined}
\end{figure}

\subsection{Radii and isotopic shifts}
\label{sec-radii}
The nuclear charge radii is an experimental observable and its analysis provides direct information on the presence of deformation in nuclei. In our case, we anticipate to obtain from such analysis some indication on the onset of deformation around neutron number $60$. Specifically, we expect to observe a kink in the isotope shift at this point. In this section, we will compare the theoretical values for radii and isotope shifts predicted by the IBM-CM with the experimental data  \cite{Ange13}.

The value of the nucleus' radius calculated using the IBM is closely associated with the matrix element of the $\hat{n}_d$ operator for the ground state. This value should be combined with a linear trend that depends on the number of bosons, which can be easily explained in terms of the liquid drop model. Additionally, in the case of the IBM-CM, it is necessary to consider both the regular and intruder configurations. In summary, in the IBM-CM the nuclear radius can be expressed as,
\begin{equation}
  r^2=r_c^2+ \hat{P}^{\dag}_{N}(\gamma_N \hat{N}+ \beta_N
  \hat{n}_d)\hat{P}_{N} + 
  \hat{P}^{\dag}_{N+2}(\gamma_{N+2} \hat{N}+ \beta_{N+2} \hat{n}_d) \hat{P}_{N+2}.
  \label{ibm-r2}
\end{equation}
%where $\hat{P}$ are projection operators on the regular (N) and the intruder (N+2) sectors. 
The appearing parameters are common for the entire chain of isotopes and are fixed to best reproduce the experimental data, which are referred to $^{108}$Mo and $^{104}$Ru, respectively. The resulting values for Mo are $\gamma_N=0.221$ fm$^2$, $\beta_N=-0.627$ fm$^2$, $\gamma_{N+2}=0.215$ fm$^2$, and $\beta_{N+2}=-0.024$ fm$^2$, while for Ru, the values are $\gamma_N=0.248$ fm$^2$ and $\beta_N=0.018$ fm$^2$. In the case of Ru, there is no dependence on the intruder part. This approach closely follows the methodology of a previous work  \cite{Zerg12}, considering only a single configuration.

In the case of Mo, the sudden increase in radius at $^{102}$Mo is correctly captured, along with the overall trend. However, the experimental and theoretical results do not coincide within the error bars, similar to what has been observed in other studies \cite{Garc11,Garc14b,Garc15,Garc19,Maya2022}. As a matter of fact, the onset of deformation is predicted more strongly than observed. This tiny discrepancy could be connected with the prediction of an equal degree of deformation for $^{104-108}$Ru while experimentally there is a more gradual variation. For Ru, where only the regular contribution is required, the linear trend of the radii is accurately reproduced, and no abrupt changes are observed. Note the different scale between panels (b) and (d).
\label{sec-s2n}
\begin{figure}[hbt]
  \centering
  \begin{tabular}{cc}
    \includegraphics[width=0.50\linewidth]{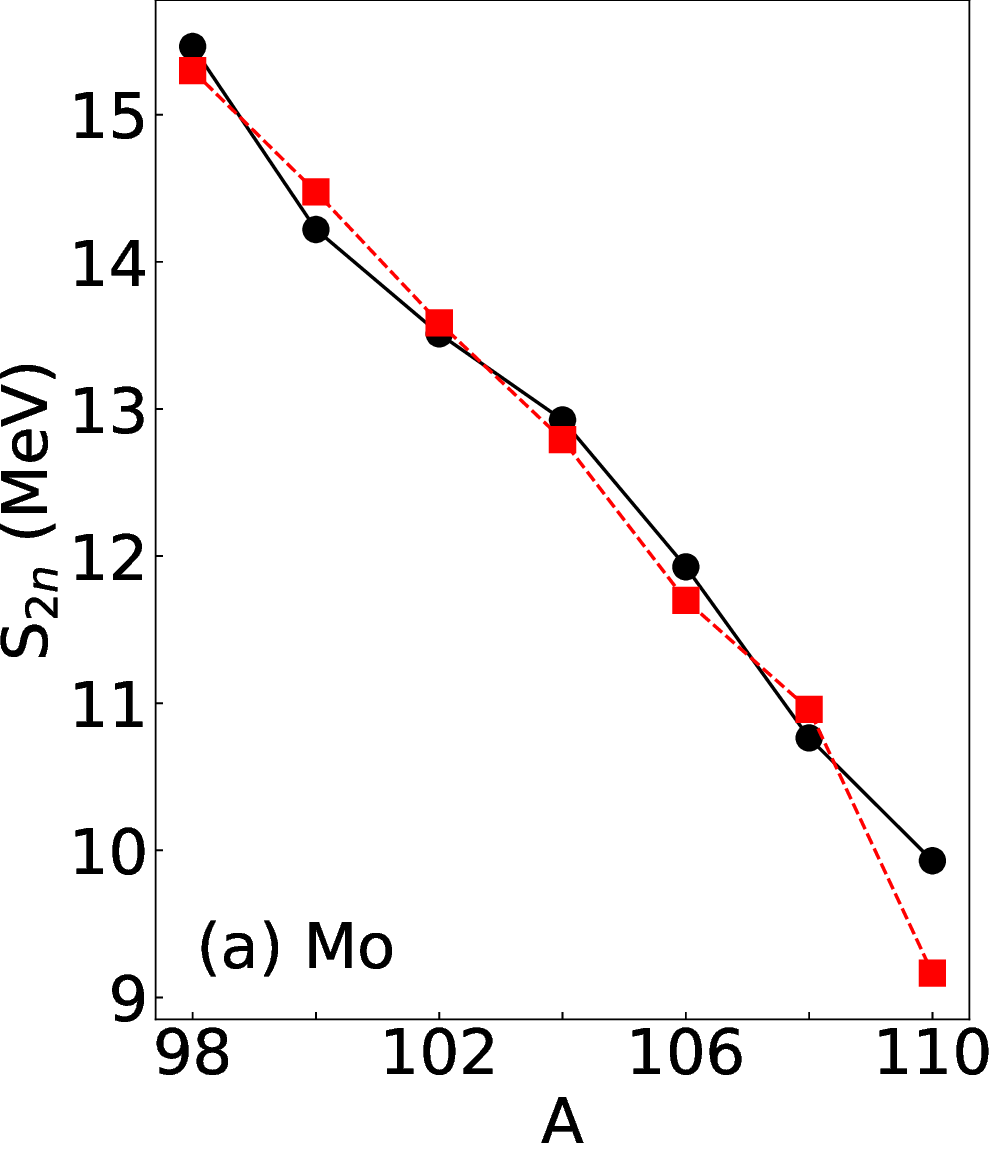}&
    \includegraphics[width=0.4665\linewidth]{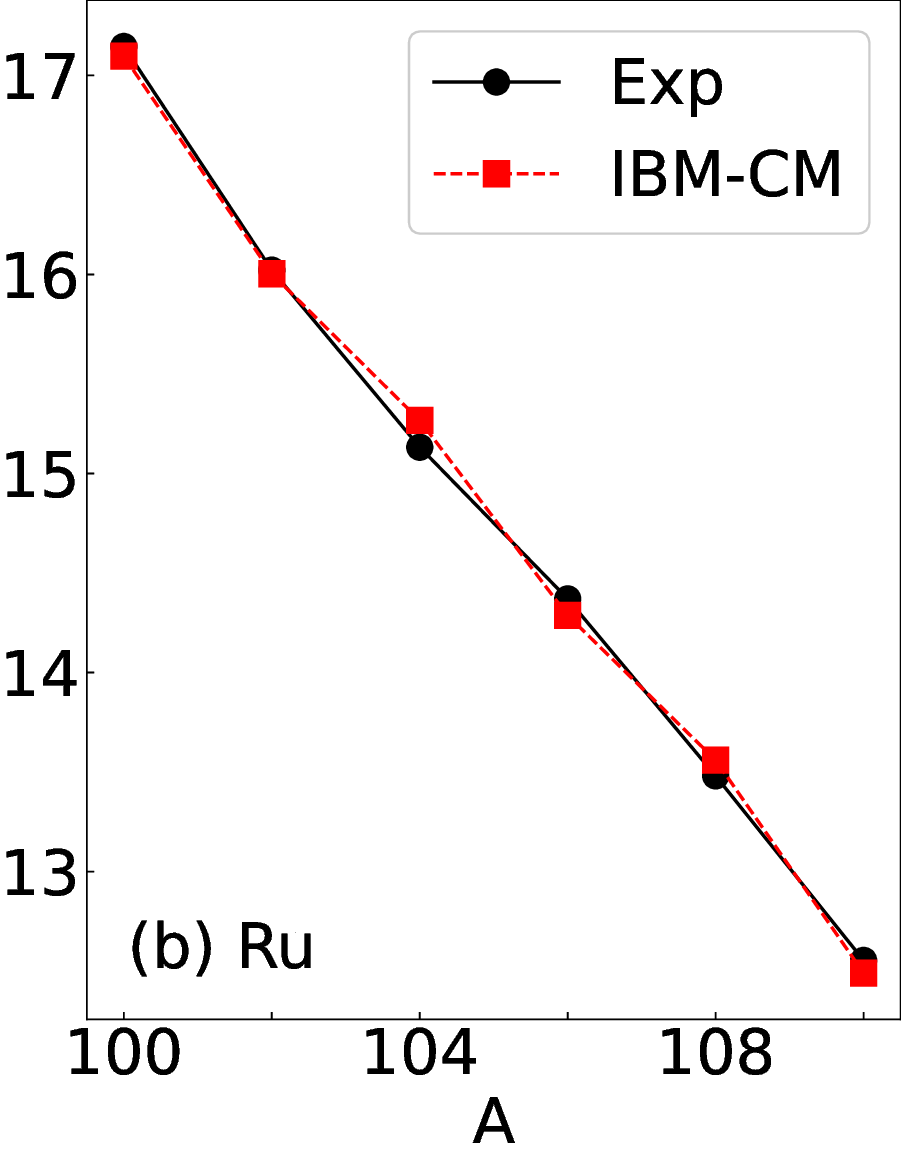}
  \end{tabular}
\caption{Comparison of experimental and theoretical two-neutron separation energies in Mo (panel (a)) and Ru (panel (b)) isotopes. }
\label{fig-s2n}
\end{figure}

\subsection{Two-neutron separation energies}
The definition of the S$_{2n}$ involves the value of the binding energy of two neighboring nuclei separated by two mass units, having the same value of Z, as expressed by the equation,
\begin{equation}
\label{s2n}
S_{2n}(A)=BE(A)-BE(A-2). 
\end{equation}
where $BE$ represents the binding energy of the nucleus, considered as positive. In the case of the IBM, an additional contribution depending on the number of neutrons and the square of the number of neutrons needs to be added to the calculated binding energy (see \cite{Foss02}). This introduces an extra linear term into the S$_{2n}$ value. Therefore, the S$_{2n}$ can be expressed as:
\begin{equation}
S_{2n}(A)={\cal A} +{\cal B} A+BE^{lo}(A)-BE^{lo}(A-2), 
\label{s2n-lin}
\end{equation}
where $BE^{lo}$ represents the ``local'' binding energy derived from the IBM Hamiltonian, and the coefficients ${\cal A}$ and ${\cal B}$ are assumed to be constant for an isotopic chain \cite{Foss02}. In the case of IBM-CM calculations, we anticipate that the effective number of bosons, or equivalently the mass number, for the ground state will be influenced by the presence of intruder states. To account for this effect, we propose as an \textit{ansatz},
\begin{equation}
S_{2n}(A)={\cal A} +{\cal B} (A+2(1-w))+BE^{lo}(A)-BE^{lo}(A-2),
\label{s2n-lin-new}
\end{equation}
where $w=w^1(0)$ (see Eq.\ (\ref{wk})). The values of ${\cal A}$ and ${\cal B}$ are determined, once the $BE^{lo}$'s are known, through a least-squares fit to the experimental values of S$_{2n}$, as explained in detail in \cite{Foss02,Garc14b,Garc19}. In our case, the obtained values are ${\cal A}=55.2$ MeV and ${\cal B}=-0.407$ MeV for Mo and ${\cal A}=65.7$ MeV and ${\cal B}=-0.499$ MeV for Ru. 

In Fig.\ \ref{fig-s2n}, the comparison between experimental and theoretical results is presented, highlighting the excellent agreement observed in both isotopic chains around neutron number $60$ ($A = 102$ in Mo and $A = 104$ in Ru). It is important to note that only the first portion of the neutron shell was utilized to determine the parameters ${\cal A}$ and ${\cal B}$. For the Ru isotopes, a clear linear trend is observed, which is accurately reproduced by the model. In the case of the Mo isotopes, a slight flattening is observed at $A = 102$, corresponding to neutron number $60$, although the IBM-CM calculation predominantly exhibits a linear trend. This specific point corresponds to the intersection of the intruder and regular configurations. It should be noted that in the Mo isotopes, the value $A = 108$ lies beyond the midpoint of the shell, even though the same linear portion as in the first half-shell is considered. The observed discrepancy in $^{110}$Mo arises from the discussion on the values of the linear coefficients in the first and second parts of the shell, as outlined in \cite{Foss02}.   
\begin{figure}
\includegraphics[width=.8\textwidth]{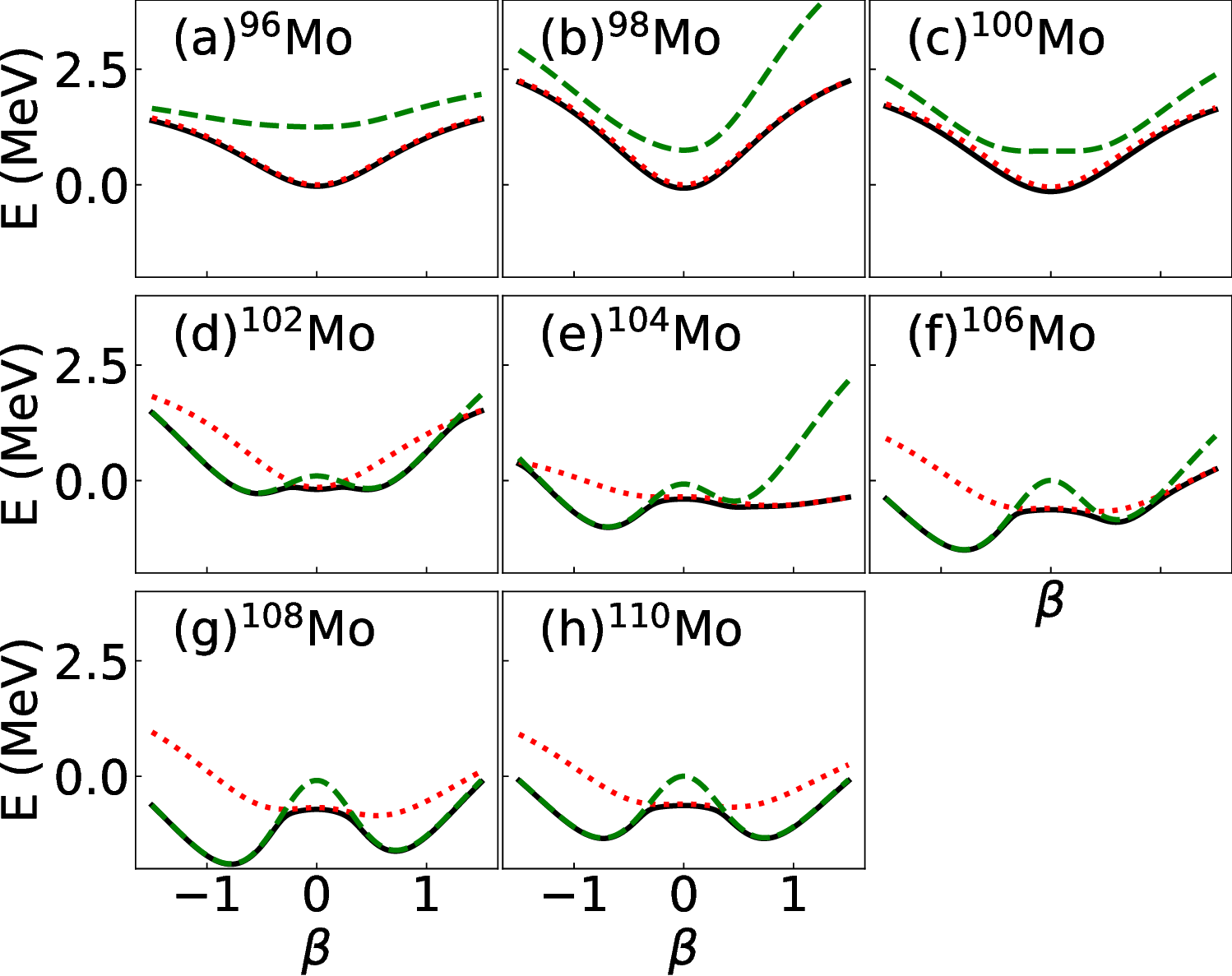}
\caption{Axial symmetry energy for $^{96-110}$Mo, corresponding to the IBM-CM Hamiltonian provided in Table \ref{tab-fit-par-mix-Mo}. The full configuration mixing calculation (full black line) is shown together with the unperturbed calculations for the regular sector (red dotted line) and for the intruder configuration (green dashed line).}
\label{fig_ibm_ener_axial-Mo}
\end{figure}

\begin{figure}
\includegraphics[width=.8\textwidth]{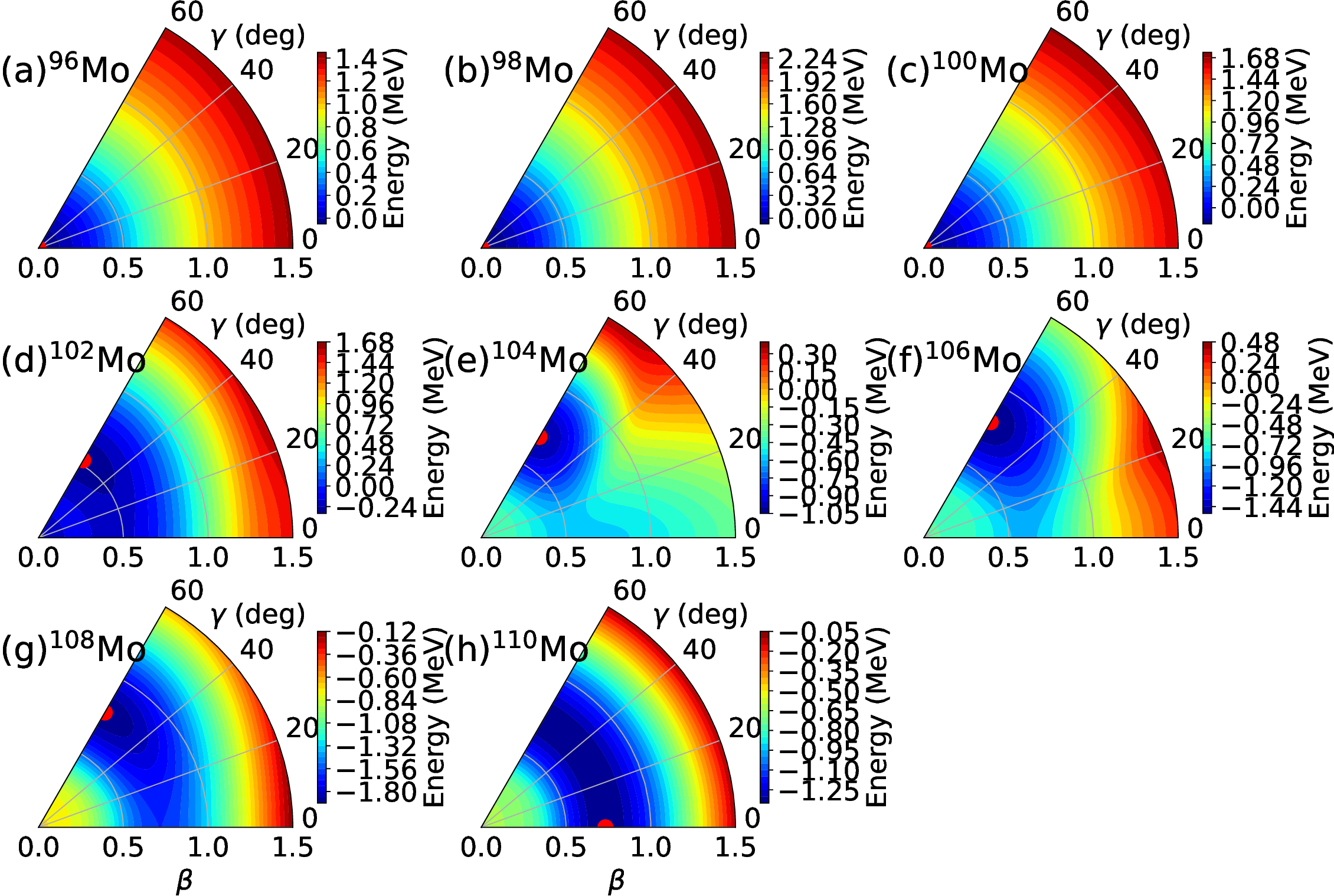}
\caption{Matrix coherent-state calculation in the $\beta-\gamma$ plane for $^{96-110}$Mo, corresponding to the IBM-CM Hamiltonian provided in Table \ref{tab-fit-par-mix-Mo}. The red dot marks the position of the absolute minimum.}
\label{fig_ibm_ener_surph-Mo}
\end{figure}

\section{Nuclear deformation and mean-field energy surfaces}
\label{sec-deformation}
One of the goals of this work is to analyze the onset of deformation around neutron number $60$. However, it is important to note that deformation itself is not directly an observable. Nevertheless, the IBM provides various approaches for calculating the deformation of a nucleus. The most common approach is the IBM mean-field method, which allows for the calculation of an energy functional based on deformation parameters \cite{gino80,diep80a,diep80b}. In the case of the IBM-CM, an expanded formalism was necessary to simultaneously describe both regular and intruder configurations, which involved the introduction of a matrix coherent-state method \cite{Frank02,Frank04,Mora08}. Detailed descriptions of the method and its application to Pt, Hg, Po, Zr, and Sr isotopes can be found in references ~\cite{Garc14a,Garc14b,Garc15,Garc19,Maya2022}.

Fig.\ \ref{fig_ibm_ener_axial-Mo} shows the axial energy surfaces and Fig.\ \ref{fig_ibm_ener_surph-Mo} shows the full $\beta-\gamma$ plane energy surfaces for Mo isotopes. In the axial case, the unperturbed calculations for regular (red dotted line) and intruder (green dashed line) configurations are presented, along with the full calculations (black solid line). It is evident that the intruder configuration evolves from a spherical shape to a flat surface in $^{100}$Mo, transitioning into an oblate deformed shape and eventually becoming nearly $\gamma$-unstable. This configuration gains correlation energy more rapidly than the regular configuration. On the other hand, the regular configuration evolves slowly from a spherical shape to a shallow prolate deformed energy surface. The full calculation clearly depicts the transition from a regular to an intruder ground state.

When analyzing the energy surfaces in the $\beta-\gamma$ plane (Fig. \ref{fig_ibm_ener_surph-Mo}), the transition from a spherical to an oblate shape can be observed in $^{102}$Mo. A secondary prolate minimum appears in $^{104}$Mo, and the nuclei become almost $\gamma$-unstable in $^{108-110}$Mo. Previous HFB calculations using a Gogny interaction \cite{Rodr10,Nomu16} exhibit similar energy surfaces to the present results, although there $^{100}$Mo is already prolate deformed. From $^{102}$Mo onwards, the shape is predominantly oblate, with triaxial minima and shallow valleys in the $\gamma$ direction. The coexistence of two minima, one oblate and the other prolate, is present in $^{104}$Mo. Note that the situation is similar to the case of Sr where the coexistence of a prolate and an oblate minimum exists \cite{Maya2022}, but differs from the Zr nuclei where an spherical and a prolate minimum coexist \cite{Garc20}.

\begin{figure}
\includegraphics[width=.8\textwidth]{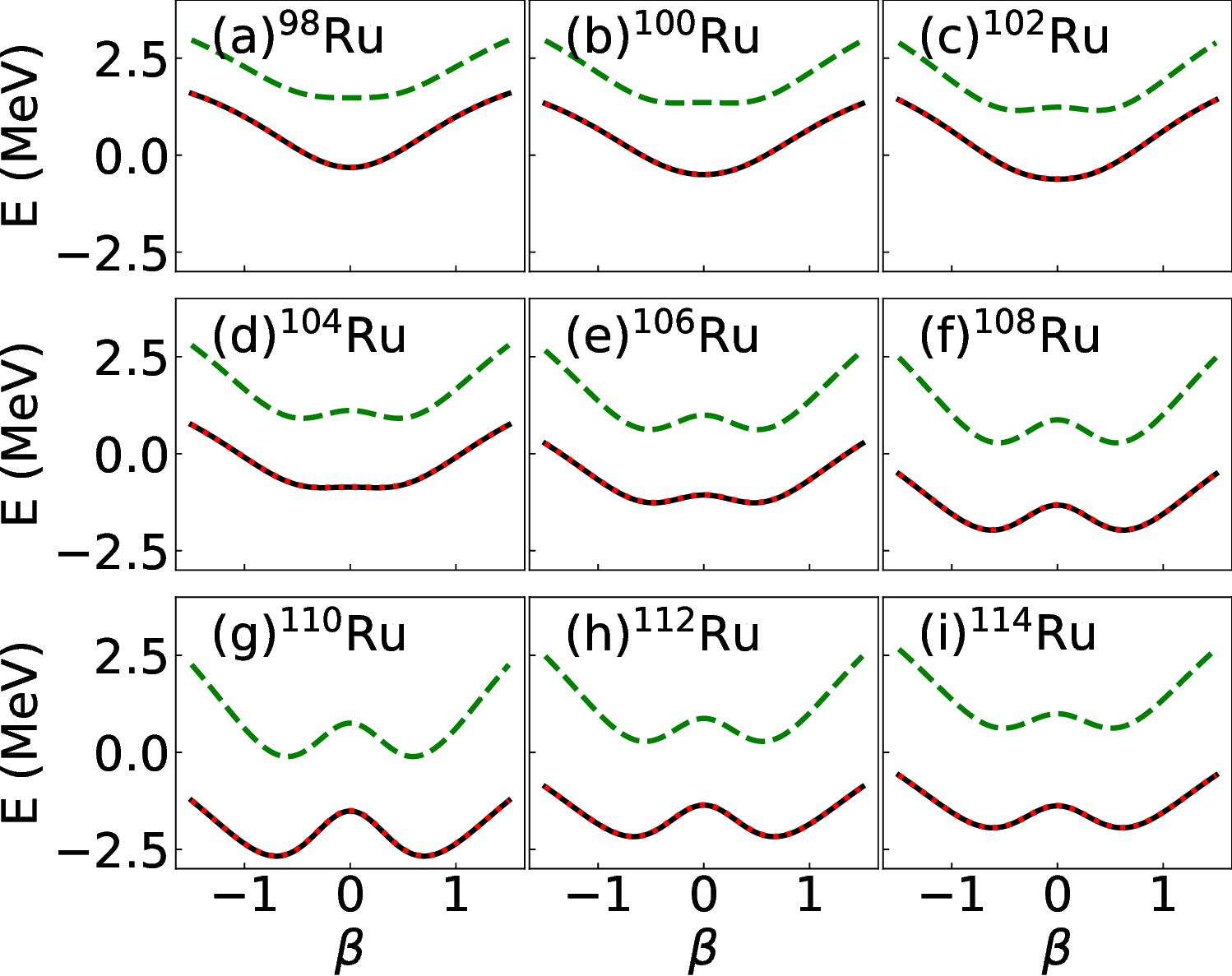}
\caption{Same as Fig.\ \ref{fig_ibm_ener_axial-Mo} but for $^{98-114}$Ru and Table \ref{tab-fit-par-mix-Ru}.}
\label{fig_ibm_ener_axial-Ru}
\end{figure}

\begin{figure}
\includegraphics[width=.8\textwidth]{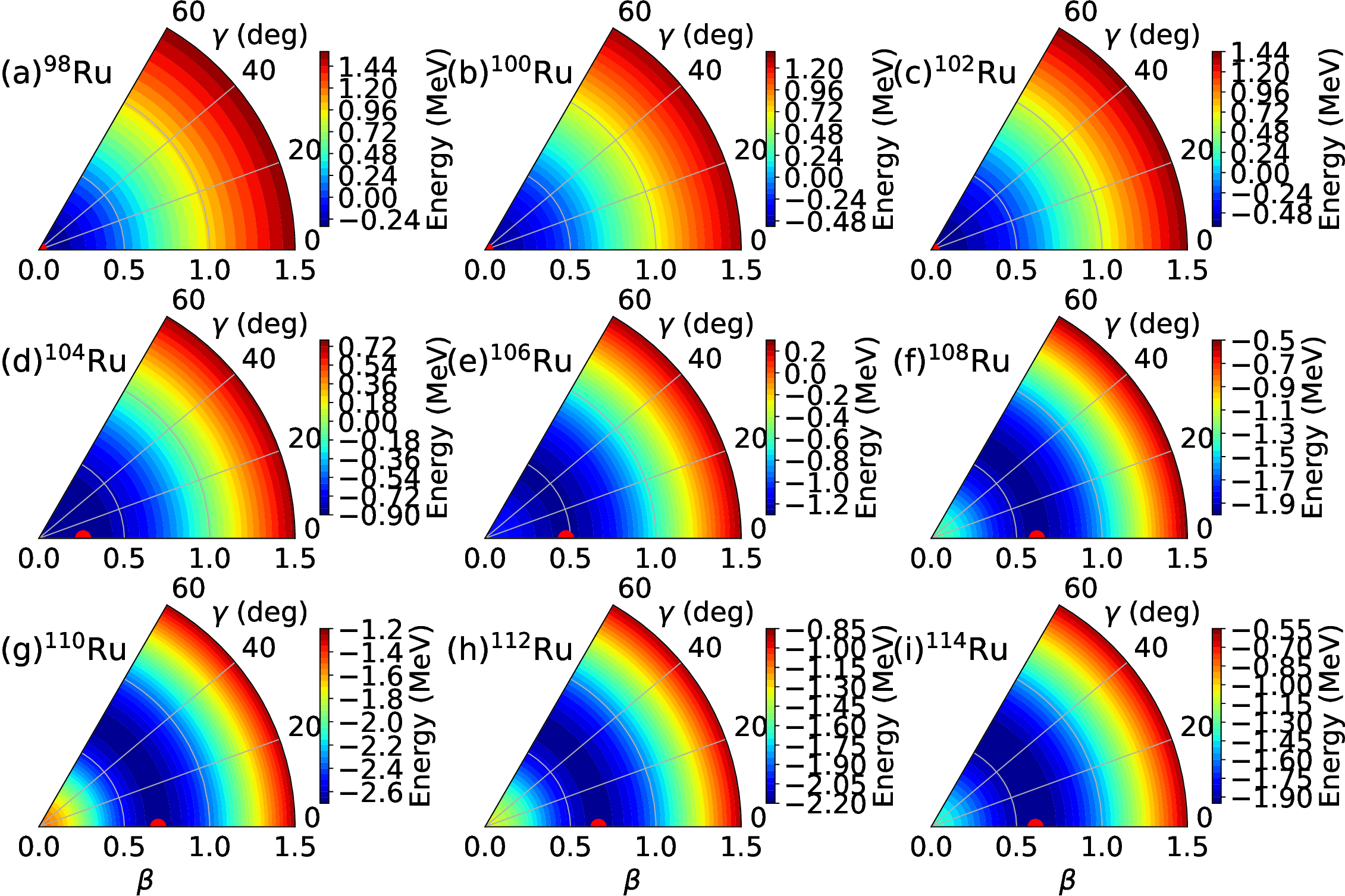}
\caption{Same as Fig.\ \ref{fig_ibm_ener_surph-Mo} but for $^{98-114}$Ru and Table \ref{tab-fit-par-mix-Ru}.}
\label{fig_ibm_ener_surph-Ru}
\end{figure}

In Fig.\ \ref{fig_ibm_ener_axial-Ru}, the axial energy surfaces for Ru isotopes are shown and the full $\beta-\gamma$ plane energy surfaces are displayed in Fig.\ \ref{fig_ibm_ener_surph-Ru}. In the axial case (Fig. \ref{fig_ibm_ener_axial-Ru}), both the unperturbed and full calculations are presented. It is noteworthy that the intruder configuration remains well separated from the regular configuration throughout, and this is consistent with the full calculation. The nuclei start out as spherical but gradually evolve into flatter shapes, reaching full flatness at $^{104}$Ru. Subsequently, they transform into $\gamma$-unstable shapes, with the deepest minimum occurring at $^{110}$Ru.

When examining the energy surfaces in the $\beta-\gamma$ plane (Fig. \ref{fig_ibm_ener_surph-Ru}), similar conclusions can be drawn. The flattest energy surface is observed at $^{104}$Ru, which serves as a boundary between the spherical and $\gamma$-unstable shapes. In a previous study \cite{Nomu16}, it was found that the lightest Ru isotopes exhibit a slightly prolate shape, while a very shallow triaxial minimum is observed in $^{104-106}$Ru. In $^{108-114}$Ru, oblate minima are obtained, albeit with a very flat $\gamma$ direction, and in certain cases, they become almost triaxial.

To gain a clearer understanding on the evolution of nuclear shape, it is useful to represent the value of $\beta$ corresponding to the minimum in the energy functional (indicated by red dots in Figs.\ \ref{fig_ibm_ener_surph-Mo} and \ref{fig_ibm_ener_surph-Ru}). It is important to note that the $\beta$ variable defined in the IBM does not directly correspond to the one used in the collective model. However, there exists a linear relationship \cite{gino80} connecting both variables. Therefore, the variables presented should be understood as being proportional to the Bohr-Mottelson ones. Note that positive values corresponds to prolate shapes while negative to oblate. 

\begin{figure}[hbt]
  \centering
  \begin{tabular}{cc}
  \includegraphics[width=0.50\linewidth]{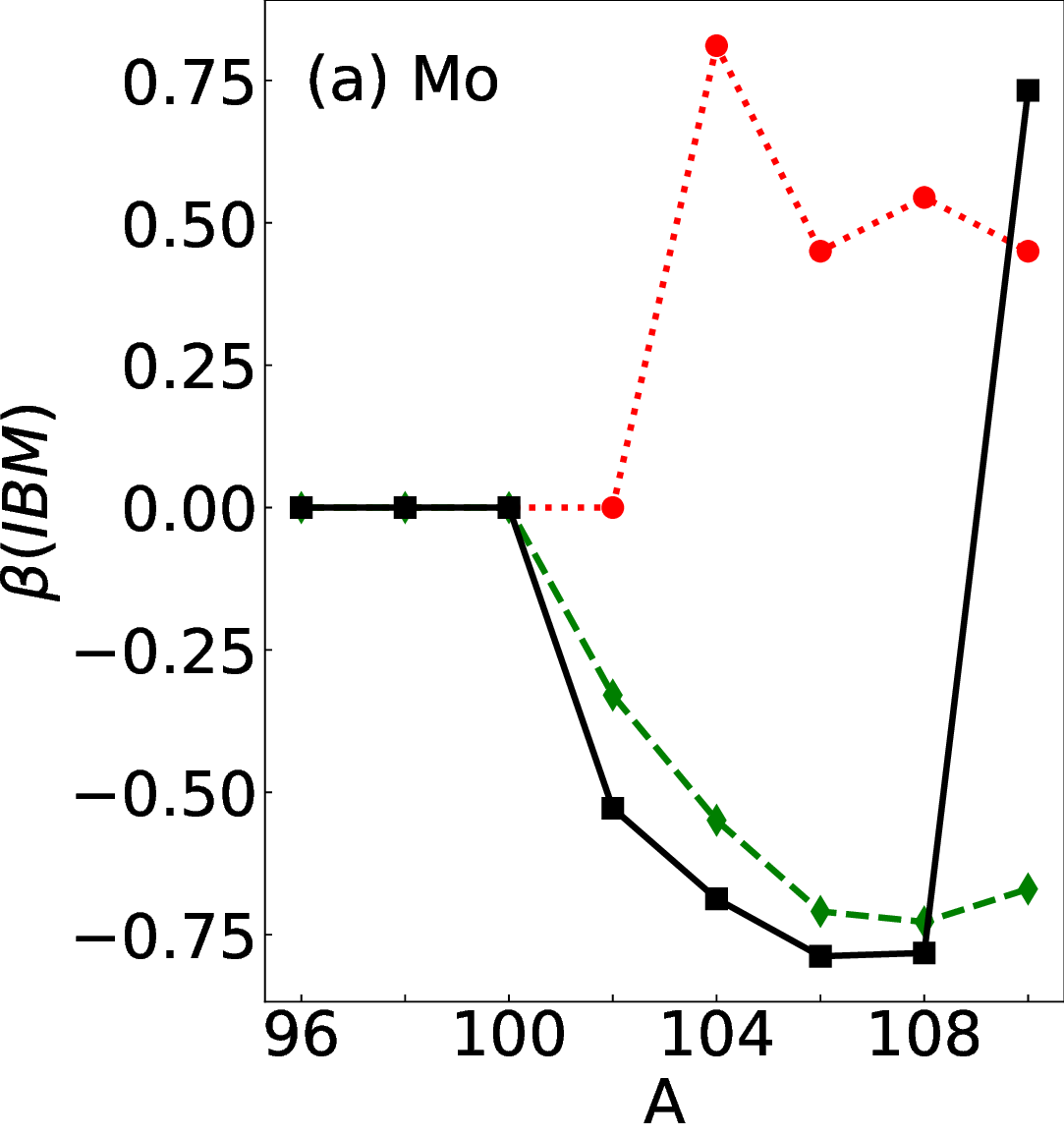}&
  \includegraphics[width=0.435\linewidth]{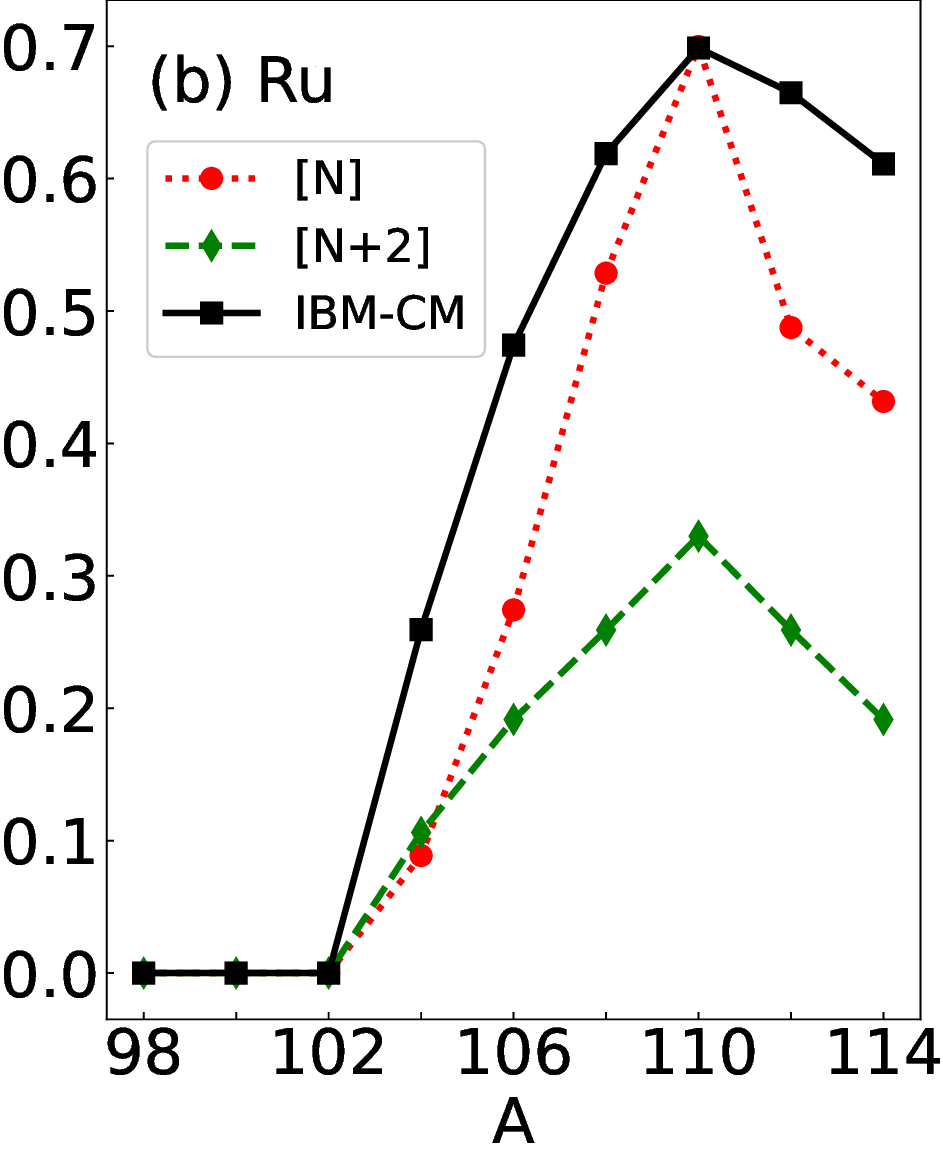}
  \end{tabular}
\caption{Value of $\beta$ extracted from the IBM-CM energy surface for Mo (panel (a)) and Ru (panel (b)) isotopes. $[N]$, $[N+2]$, and IBM-CM correspond to a pure regular configuration, a pure intruder configuration, and to the complete calculation.}
\label{fig-beta}
\end{figure}

Fig.\ \ref{fig-beta} illustrates the IBM $\beta$ values for Mo and Ru isotopes in panels (a) and (b), respectively. The values correspond to the full calculation (IBM-CM) as well as the regular ($[N]$) and intruder ($[N+2]$) configurations, assuming no interaction exists between the two configurations.

In the case of Mo isotopes (panel (a)), the regular configuration is spherical in the range $A=96-102$, transitioning to a prolate shape for $A=104-110$. On the other hand, the intruder configuration is spherical until $A=100$, but then it rapidly increases its value, becoming oblate. The full calculation exhibits a similar behavior to the intruder configuration, transitioning rapidly from spherical to prolate deformation around $A=102$. It is worth noting that although the obtained deformation for $A=110$ is oblate, the potential is almost $\gamma$-unstable. Therefore, the sign of $\beta$ for $^{110}$Mo is unimportant. In both configurations, the deformation develops quite rapidly.

Turning to the case of Ru isotopes (panel (b)), the regular configuration, as well as the full calculation, corresponds to spherical shapes for $^{98-102}$Ru. However, they undergo a sudden transition to $\gamma$-unstable deformation for $^{104-114}$Ru, with the maximum deformation occurring at the mid-shell. The intruder configuration exhibits a similar behavior, but it is important to remember that the intruder Hamiltonian remains fixed throughout the entire isotope chain.

Lastly, we will also calculate the value of the deformation in an almost model-independent manner using the work of Kumar, Cline, and their colleagues \cite{kumar72,Cline86}. This method allows the extraction of deformation using the experimental information coming from Coulomb excitation, which is a powerful tool for accessing information about the shape of a nucleus. The key concept is to utilize the notion of the ``equivalent ellipsoid'' for a given nucleus. This ellipsoid is defined as uniformly charged with the same charge, possessing the same $\left\langle r^{2}\right\rangle$ and $E2$ moments as the original nucleus characterized by a specific eigenstate.

By analyzing measured data from various transitions obtained through Coulomb excitation techniques, it is possible to extract the values of collective model variables, namely $\beta$ and $\gamma$, for a given state. This approach provides valuable experimental insights into the deformation of nuclei.
 
 \begin{table}
 \caption{Values of deformation parameters, $q^2$, $\beta$ and $\gamma$, extracted from the quadrupole shape invariants, together with the value of $w^k$ ([N] content, see Eq.\ (\ref{wk})), for Mo and Ru isotopes.}
\label{tab-q-invariant}
\begin{center}
 \begin{tabular}{cccccc|ccccc}
 \hline
 State & Iso & $q^{2}$ ($e^{2} b^{2}$) & ~~$\beta$~~ & $\gamma$ (deg) & $w^k$& ~Iso & $q^{2}$ ($e^{2} b^{2}$) & ~~$\beta$~~ & $\gamma$ (deg) & $w^k$\\
 \hline
   0$_{\text{1}}^+$ &$^{96}$Mo &0.28 & 0.17 &48 & 0.971 &$^{98}$Ru&0.40 &0.20 &30 &0.997\\ 
   0$_{\text{2}}^+$ &          &1.63 & 0.42 &50 & 0.158 &         &0.35 &0.18 &30 &0.988\\ 
   0$_{\text{3}}^+$ &          &0.62 & 0.26 &46 & 0.579 &         &2.05 &0.45 &30 &0.012\\  
 \hline                                  
   0$_{\text{1}}^+$ &$^{98}$Mo &0.37 & 0.20 &14 & 0.881 &$^{100}$Ru&0.50 &0.22 &30 &0.996\\ 
   0$_{\text{2}}^+$ &          &0.12 & 0.11 &14 & 0.122 &          &0.40 &0.19 &30 &0.995\\ 
   0$_{\text{3}}^+$ &          &0.49 & 0.23 &0  & 0.730 &          &2.76 &0.51 &30 &0.006\\  
 \hline                                  
   0$_{\text{1}}^+$ &$^{100}$Mo&0.53 & 0.23 &25 & 0.825 &$^{102}$Ru&0.64 &0.24 &30 &0.994\\ 
   0$_{\text{2}}^+$ &          &1.22 & 0.35 &31 & 0.241 &          &0.54 &0.22 &30 &0.992\\ 
   0$_{\text{3}}^+$ &          &0.64 & 0.26 &13 & 0.763 &          &3.60 &0.57 &30 &0.010\\  
 \hline                                  
   0$_{\text{1}}^+$ &$^{102}$Mo&1.16 & 0.34 &26 & 0.018 &$^{104}$Ru&0.84 &0.27 &30 &0.994\\ 
   0$_{\text{2}}^+$ &          &1.48 & 0.39 &18 & 0.968 &          &0.56 &0.22 &30 &0.996\\ 
   0$_{\text{3}}^+$ &          &0.90 & 0.30 &23 & 0.020 &          &0.73 &0.26 &30 &0.996\\  
 \hline                                  
   0$_{\text{1}}^+$ &$^{104}$Mo&1.45 & 0.38 &43 & 0.013 &$^{106}$Ru&1.00 &0.29 &30 &0.994\\ 
   0$_{\text{2}}^+$ &          &1.35 & 0.36 &55 & 0.946 &          &0.62 &0.23 &30 &0.995\\ 
   0$_{\text{3}}^+$ &          &1.01 & 0.32 &44 & 0.430 &          &0.86 &0.27 &30 &0.996\\  
 \hline                                  
   0$_{\text{1}}^+$ &$^{106}$Mo&1.61 & 0.39 &43 & 0.010 &$^{108}$Ru&0.96 &0.29 &30 &0.995\\ 
   0$_{\text{2}}^+$ &          &1.16 & 0.33 &30 & 0.297 &          &0.83 &0.27 &30 &0.997\\ 
   0$_{\text{3}}^+$ &          &0.99 & 0.31 &24 & 0.694 &          &0.61 &0.23 &30 &0.995\\  
 \hline                                  
   0$_{\text{1}}^+$ &$^{108}$Mo&2.23 & 0.46 &36 & 0.008 &$^{110}$Ru&1.05 &0.30 &30 &0.996\\ 
   0$_{\text{2}}^+$ &          &1.95 & 0.43 &28 & 0.016 &          &0.92 &0.28 &30 &0.997\\ 
   0$_{\text{3}}^+$ &          &1.29 & 0.35 &27 & 0.603 &          &0.71 &0.24 &30 &0.995\\  
 \hline                                  
   0$_{\text{1}}^+$ &$^{110}$Mo&1.91 & 0.42 &30 & 0.013 &$^{112}$Ru&1.14 &0.30 &30 &0.996\\ 
   0$_{\text{2}}^+$ &          &1.68 & 0.39 &30 & 0.012 &          &0.97 &0.28 &30 &0.998\\ 
   0$_{\text{3}}^+$ &          &0.96 & 0.30 &21 & 0.910 &          &0.73 &0.24 &30 &0.996\\  
 \hline                                  
   & & & & &                                           &$^{114}$Ru &0.98 &0.28 &30 &0.997\\ 
   & & & & &                                           &           &0.82 &0.25 &30 &0.998\\ 
   & & & & &                                           &           &0.60 &0.22 &30 &0.997\\  
 \hline                                                                        
 \end{tabular}
 \end{center}
 \end{table}
The procedure for calculating the deformation parameters involves the use of quadrupole shape invariants. Focusing specifically on the $0^+$ states, the following equations define these shape invariants (see \cite{Pove20} for an application of the method), 
\begin{eqnarray}
q_{2, i} &=&\sqrt{5}\left\langle 0_{i}^{+}\left|[\hat{Q} \times \hat{Q}]^{(0)}\right| 0_{i}^{+}\right\rangle, \label{eq11} \\
q_{3, i} &=&-\sqrt{\frac{35}{2}}\left\langle 0_{i}^{+}\mid[\hat{Q} \times \hat{Q} \times \hat{Q}]^{(0)} \mid 0_{i}^{+}\right\rangle. \label{eq12}
\end{eqnarray}
The deformation parameters are directly related to those of the triaxial rigid rotor, denoted as $q$ and $\delta$, respectively:
\begin{equation}
q=\sqrt{q_{2}}, 
\label{eq15}
\end{equation}

\begin{equation}
\delta=\frac{60}{\pi} \arccos \frac{q_{3}}{q_{2}^{3 / 2}},  
\label{eq16}
\end{equation}
where $\delta$ coincides with the parameter $\gamma$ of the Bohr-Mottelson model up to a first order approximation. 

The deformation parameter $\beta$ can also be obtained from the quadrupole shape invariant (\ref{eq11}) (see, e.g., references \cite{Srebrny06,clement07,Wrzo12}),
\begin{equation}
\beta = \dfrac{4\pi\sqrt{q_2}}{3 Z e r_0^2 A^{2/3}},
\label{beta}
\end{equation}
where $e$ is the proton charge and $r_0 = 1.2$ fm.

The theoretical values of $\beta$, $\gamma$, $q^2$, and the fraction of wave function belonging to the regular sector, $w^k$ (see Eq.\ (\ref{wk})), are presented in Table \ref{tab-q-invariant} for each $0_1^+$, $0_2^+$, and $0_3^+$ state across the entire chains of Mo and Ru isotopes. This table reveals the coexistence of different deformations within the same nucleus, with the regular states typically exhibiting less deformation compared to the intruder states.

In the case of Mo isotopes, the intruder states generally display a significant oblate deformation that increases with the mass number. However, an exception is observed in $^{98}$Mo, where a low deformation is observed. This is consistent with the ``collapse'' of the B(E2) values observed in its yrast band, possibly related to the closure of the neutron number $56$ subshell. The deformation of the first regular state steadily increases until $^{100}$Mo, but for $^{102}$Mo, there is a sudden increase, and the deformation remains relatively constant for the remaining isotopes. Regarding the value of $\gamma$, clear conclusions can only be drawn in a few cases, such as $^{96}$Mo and $^{98}$Mo, where the states are oblate and prolate, respectively. In other cases, the deformation is compatible with triaxiality.

Moving on to the Ru isotopes, the situation is relatively straightforward, with a steady increase in deformation for the regular states, with a fixed $\gamma$ value of $30$ degrees throughout. Only in two cases are intruder states observed, exhibiting a large deformation.
\begin{figure}[hbt]
  \centering
  \begin{tabular}{ccc}
    \includegraphics[width=.33\linewidth]{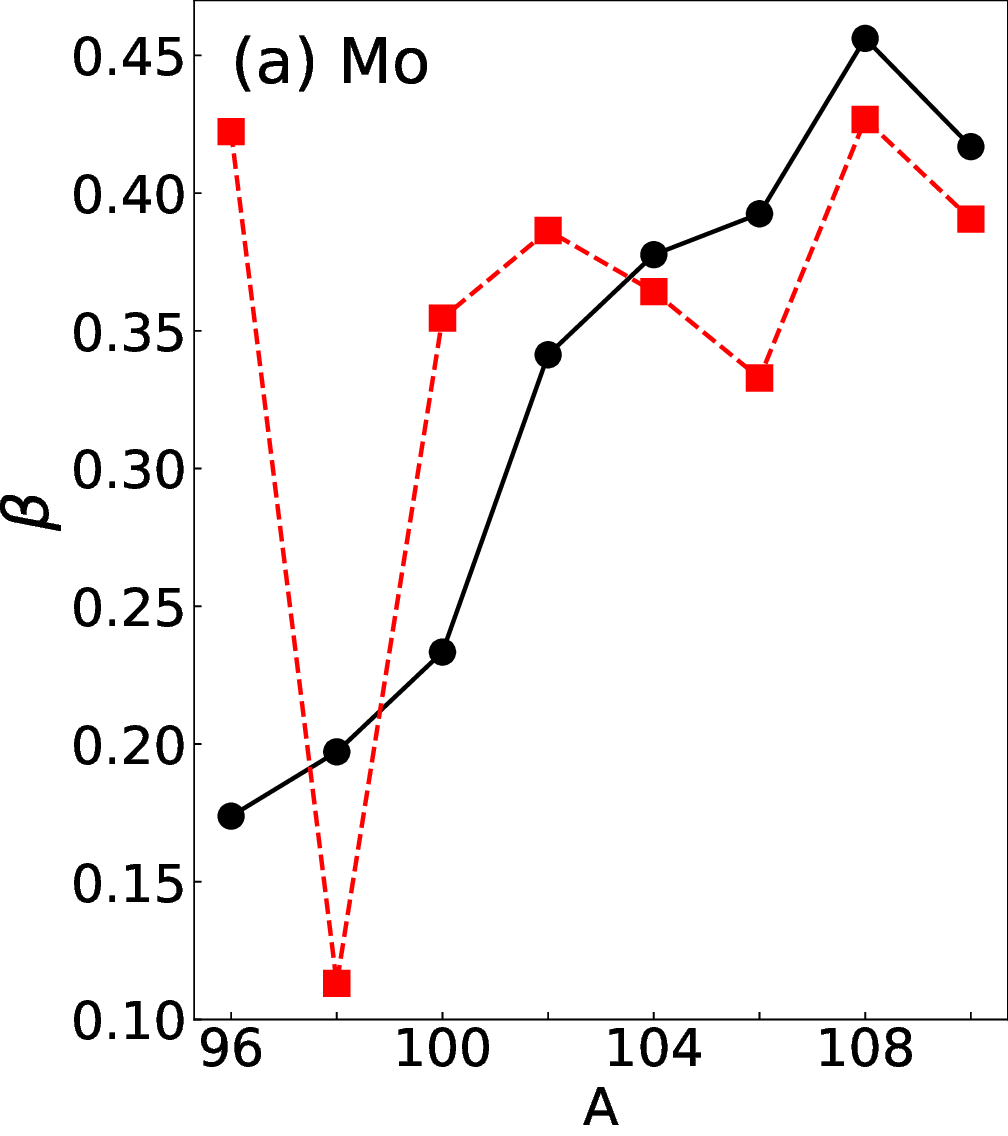}&
    \includegraphics[width=.305\linewidth]{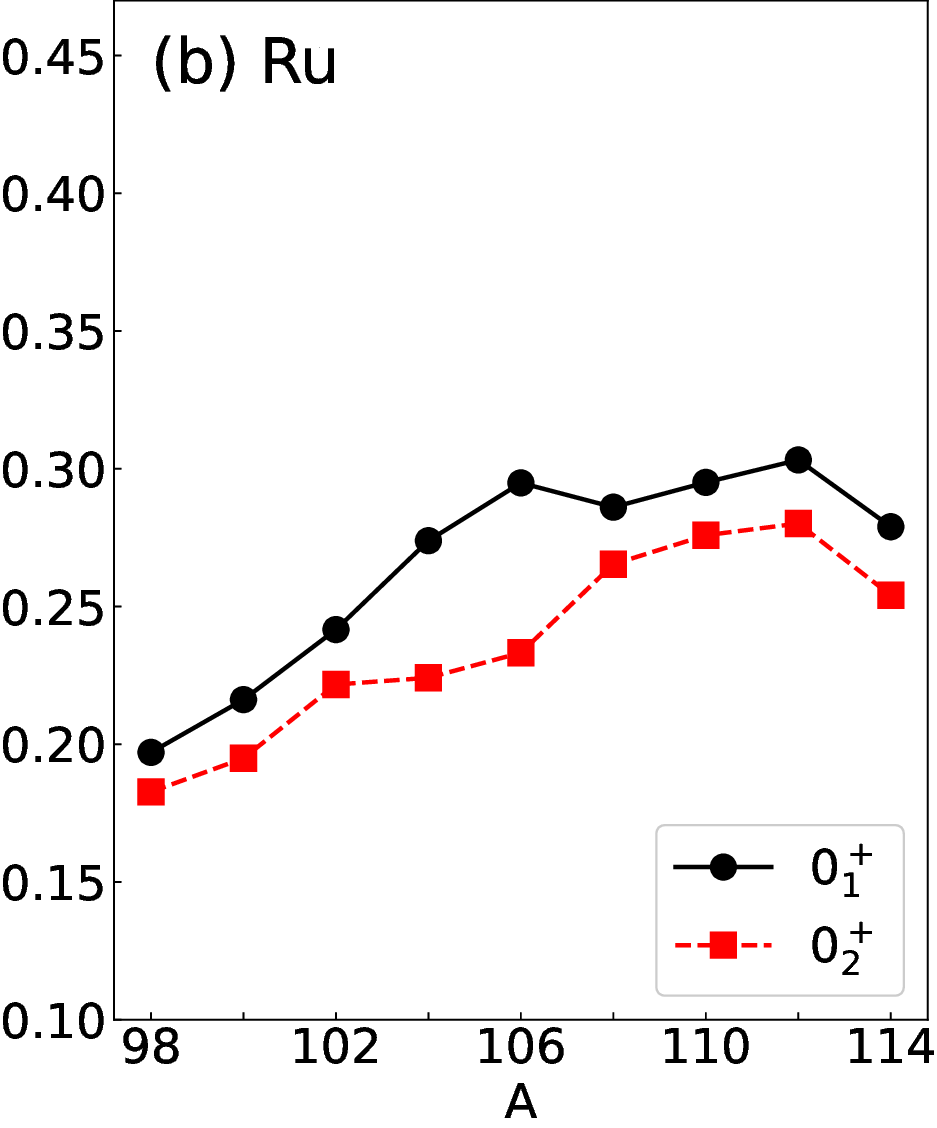}
    \includegraphics[width=.305\linewidth]{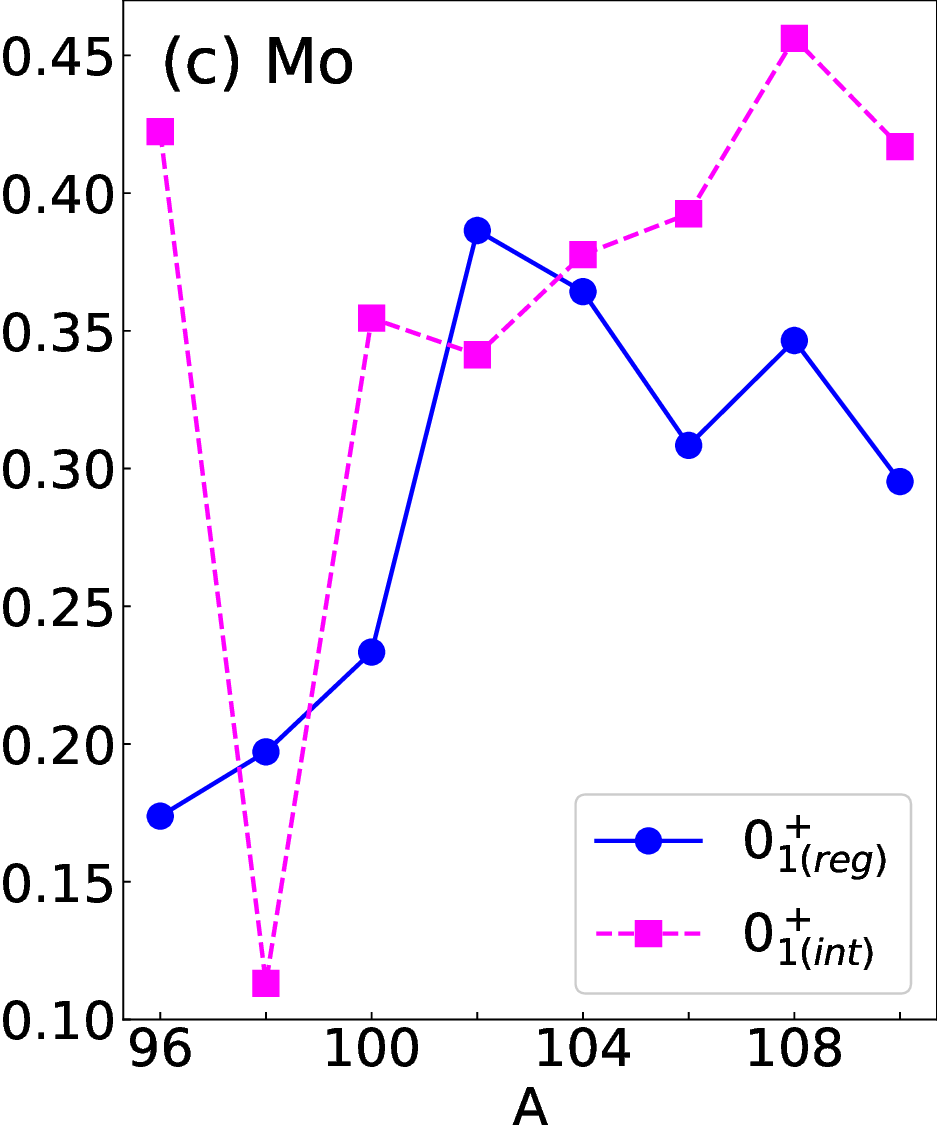}
  \end{tabular}
  \caption{Value of the deformation, $\beta$, extracted from the value of the quadrupole shape invariants for Mo (panel(a)) and Ru (panel (b)). Panel (c) is also for Mo but here the $\beta$ values for the first regular and the first intruder $0^+$ states are shown.}
  \label{fig-beta-fromQ2}
\end{figure}

As a complement to Table \ref{tab-q-invariant}, the trend of the value of $\beta$ for the first two $0^+$ states in Mo and Ru isotopes is depicted in Fig.\ \ref{fig-beta-fromQ2}.
In Ru isotopes (panel (b)), the behavior is relatively straightforward, with a constant increase in $\beta$ up to around the mid-shell region, reaching a value of approximately $0.30$. This trend is similar to the one shown in Fig.\ \ref{fig-beta}. Both states belong to the regular sector.
In the case of Mo isotopes (panel (a)), the situation is more complex, especially for the $0_2^+$ state. The $0_1^+$ state exhibits a rapid increase in $\beta$ at $A=102$, rising from $0.23$ to $0.34$, and then showing a more gradual increase up to the mid-shell region where it reaches its maximum value of $0.42$.
In addition, it is also of interest to plot the value of $\beta$ for the first $0^+$ state belonging to the regular sector ($0^+{\text{reg}}$) and the first $0^+$ state belonging to the intruder sector ($0^+{\text{int}}$) in Mo isotopes, as shown in panel (c) (note that in Ru only regular states are observed). It can be observed that the regular state exhibits a rapid increase in deformation at $A=102$ and subsequently shows a steady decrease. On the other hand, the intruder state maintains a relatively high and constant value of $\beta$, except for the aforementioned exception observed at $A=98$.

\section{The quest of a quantum phase transition}
\label{sec-quest}
\begin{figure}[hbt]
  \centering
    \includegraphics[width=0.5\linewidth]{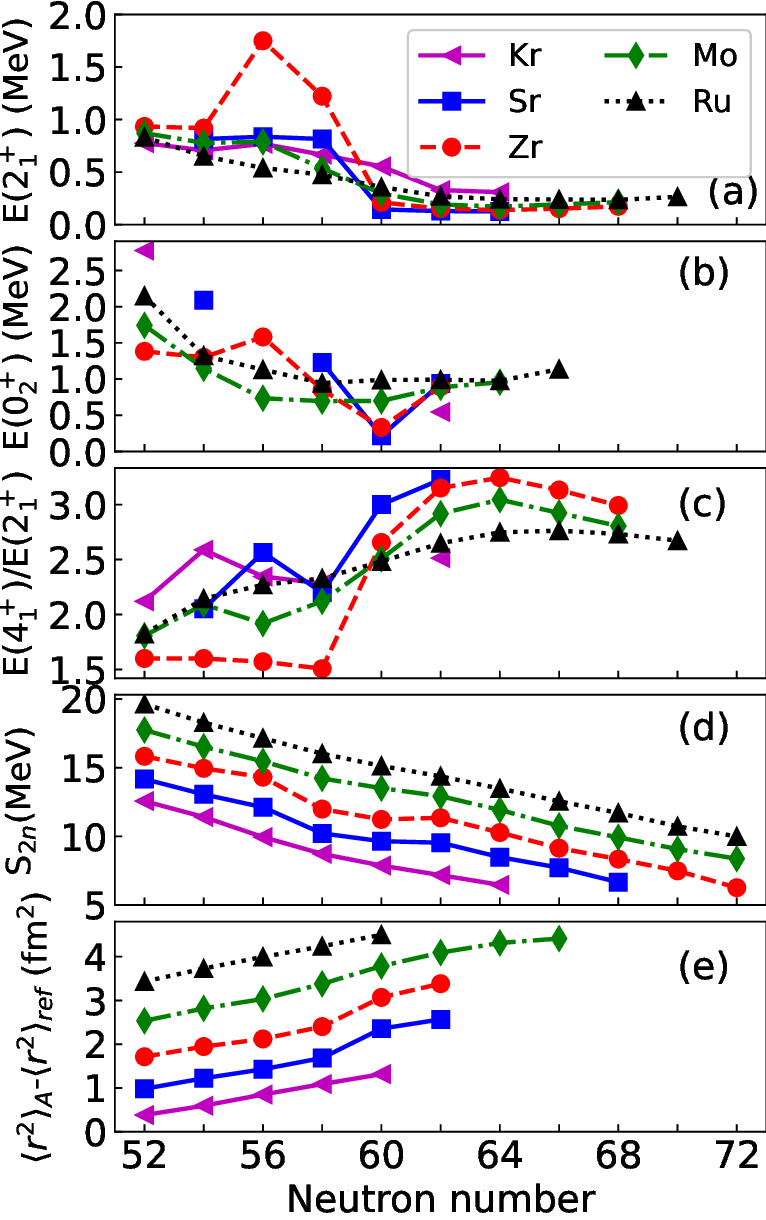}
  \caption{Values for key QPT/shape coexistence observables for Kr, Sr, Zr, Mo, and Ru isotopes as a function of neutron number. (a) $E(2_1^+)$, (b) $E(0_2^+)$, (c) $E(4_1^+)/E(2_1^+)$, (d) S$_{2n}$, and (e) $\langle r^2 \rangle- \langle r^2_{ref} \rangle$ (the reference values are different for each isotope chain to better distinguish the data).}
  \label{fig-qpt-evidences}
\end{figure}
The potential relationship between the phenomenon of shape coexistence and the existence of Quantum Phase Transitions (QPTs) has been the subject of recent investigations \cite{Garc20,Gavr19,Gavr19b,Gavr22}. In brief, a QPT occurs in systems where the ground state's structure undergoes a sudden change when a control parameter varies slightly around a specific value \cite{Sach11}. It is important to note that this phenomenon occurs at absolute zero temperature. The presence of a QPT is generally associated with a combination of Hamiltonians possessing different symmetries. Specifically, we can consider a scenario where two Hamiltonians, each associated with a particular symmetry (A or B), are combined. Consequently, the Hamiltonian of the system can be expressed as follows:
\begin{equation}
\label{eq:H_QPT}
\hat H= (1-x) \hat H_A+ x\hat H_B.
\end{equation}
This formulation allows us to investigate the interplay between different symmetries, A and B, by adjusting the parameter $x$, which determines the relative weight or contribution of each symmetry to the overall Hamiltonian.

In most cases, the symmetry involved in the QPT can be considered as a dynamical symmetry \cite{Iach98}. The QPT occurs at a critical value $x=x_c$, where the wave function undergoes an abrupt transition from having symmetry A to having symmetry B, even though the full Hamiltonian does not possess any specific symmetry except at $x=0$ or $1$. This phenomenon is closely connected to the concept of quasidynamical symmetry proposed by D. Rowe in \cite{Rowe04}. The existence of a QPT also implies a sudden change in the so-called order parameter, which vanishes in the symmetric phase and takes a nonzero value in the broken phase \cite{Sach11}. Thus, the order parameter carries information about the symmetry of the system's ground state.

QPTs can be classified in a manner similar to the phase transitions that occur in macroscopic systems at non-zero temperature, employing the Erhenfest classification \cite{Land69}. Based on this classical classification, QPTs can be categorized as first-order and second-order (or continuous) QPTs \cite{Land69}. In a first-order QPT (where the first derivative of the ground state energy with respect to the control parameter exhibits a discontinuity), there exists a narrow region around $x_c$ where both states with different symmetries, A and B, coexist. In the case of a second-order QPT (according to the Erhenfest classification, where the second derivative of the ground state energy with respect to the control parameter displays a discontinuity), there is no coexistence of symmetries around the critical region $x_c$.

A crucial characteristic of a QPT is that it leads to the degeneracy of a set of states and the compression of the spectra \cite{Garc01,Garc20}. In fact, in the thermodynamic limit, the excitation energy of the first excited state approaches zero at the critical point.

When studying QPTs in nuclear systems, one encounters challenges related to the finite size of the system and the approximate nature of the control parameter, often identified with the nuclear mass or neutron number. As a result, all the main characteristics of a QPT can only be observed in an approximate manner within a nucleus, and the abrupt changes are typically smoothed out \cite{Iach98}. In Figure 
\ref{fig-qpt-evidences}, we present experimental data for Kr, Zr, Sr, Mo, and Ru, including the values of $E(2_1^+)$, $E(0_2^+)$, $E(4_1^+)/E(2_1^+)$, S$_{2n}$, and $\langle r^2 \rangle$.%- \langle r^2 \rangle_\text{ref}$. 
These quantities serve as indicators for the presence of a QPT or the existence of shape coexistence. These nuclei are located near the subshell closure at $Z=40$ and are close to the neutron number $60$, where a rapid onset of deformation is observed.

In panel (a), it is evident that in all nuclei, the energy of the $2_1^+$ state decreases, indicating the appearance of deformation, particularly from neutron number $60$ onwards. However, notable differences exist between Sr and Zr, where the drop is very abrupt, Mo, which exhibits a slower decrease, and Ru or Kr, where the transition is smoother. The behavior in Sr and Zr has been interpreted in terms of the crossing of two different particle-hole configurations \cite{Garc19,Maya2022} or as a first-order QPT \cite{Garc20,Garc05}.

Moving to panel (b) and considering the energy systematics of the $0_2^+$ state, a distinct minimum is observed for Sr and Zr at neutron number $60$, while Ru and Mo exhibit a relatively flat energy trend around this neutron number. In Kr, it is not possible to extract a clean conclusion. The deep minimum in Sr and Zr again suggests the crossing of two configurations or the existence of a QPT, while the smoother behavior in Mo and Ru implies a slower evolution.

Panel (c) depicts the ratio $E(4_1^+)/E(2_1^+)$, which serves as a clear indicator of the onset of deformation. A value around $2$ corresponds to a spherical nucleus, $2.5$ to a $\gamma$-unstable rotor, and $3.3$ to a rigid rotor. Sr and Zr clearly exhibit an evolution from sphericity to a rigid rotor, Ru and Kr indicate an evolution into a $\gamma$-unstable rotor, and Mo lies in between both cases. This observable can also be considered as an approximate order parameter, behaving as a first-order QPT in Sr and Zr, and potentially as a second-order QPT in Mo (less clear in Ru).

In panel (d), the two-neutron separation energy, S$_{2n}$, is displayed. This observable, which can be understood as the derivative of the binding energy, is a smoking gun for the presence of a QPT. The sudden change in its slope around neutron number $60$ in Sr and Zr suggests the existence of a first-order QPT, while Mo exhibits a small perturbation indicating a possible second-order QPT. No departure from the linear trend is observed in Ru and Kr.

Finally, panel (e) presents the mean-square radii. Note that the origin has been shifted differently for each isotope for clarity. Once again, a sudden increase in the radius is observed in Sr and Zr around neutron number $60$, while Mo shows a less drastic increase and Ru and Kr exhibit a linear trend.

Based on the information presented in Fig.\ \ref{fig-qpt-evidences}, it is reasonable to assume that Sr and Zr undergo a first-order QPT around neutron number $60$, Mo undergoes a second-order QPT, while Ru exhibits a smooth evolution without any abrupt changes. In Sr, Zr, and Mo, the ratio $E(4_1^+)/E(2_1^+)$ can be considered as an order parameter, indicating the presence of a QPT, while S$_{2n}$ points towards the existence of a discontinuity in the first or second derivative of the binding energy.

Alternatively, all the observed features in Figure \ref{fig-qpt-evidences} can also be explained in terms of the crossing of two configurations. The increase in $E(4_1^+)/E(2_1^+)$ and the decrease in $E(2_1^+)$ can be easily explained by the crossing of a spherical and a deformed configuration. The drop in the energy of the $0_2^+$ state can be attributed to the presence of an intruder configuration that gains correlation energy more rapidly than the regular configuration. Finally, the deviation from the linear trend in S$_{2n}$ and mean-square radius can also be explained by the crossing of two configurations.

Therefore, both the QPTs and the crossing of two configurations can provide explanations for the observed phenomena in the figure.

In previous works, the observed QPT in Zr has been explained in terms of the crossing of two weakly interacting configurations \cite{Gavr19b, Garc20, Gavr22}, providing a straightforward explanation for the behavior of S$_{2n}$. A recent analysis has also been conducted for Sr nuclei \cite{Maya2023u} with similar conclusions. Moreover, in Zr, the different configurations, especially the intruder one, undergo their own QPT. The concept of ``Type I'' and ``Type II'' QPTs was introduced in \cite{Gavr19b, Gavr22}, referring to QPTs occurring within a single configuration or involving two configurations, respectively. In the case of Mo, it has been shown in Sec.\ \ref{sec-corr_energy} that two configurations indeed cross, although the observed changes in S$_{2n}$ are less abrupt compared to Zr. This situation is reminiscent of the behavior observed in Pt \cite{Garc09}, where a crossing of two configurations occurs without inducing a QPT. It is worth mentioning that the interaction between configurations was quite strong in Pt, which resulted in the suppression of the QPT. The mixing between configurations is controlled by the parameter $\omega$, with values of 15 keV, 15 keV, and 50 keV for Zr, Mo, and Pt, respectively. However, the interaction between configurations depends on the matrix element $\langle 0^+_{\rm reg}|\hat{V}_{\rm mix}| 0^+_{\rm int}\rangle$, which has values around the region where the configurations cross, namely 80 keV, 250 keV, and 250 keV for Zr, Mo, and Pt, respectively. Therefore, it is evident that the case of Mo resembles Pt, where two configurations cross but a strong interaction between them hinders the presence of abrupt changes in the spectrum. Nevertheless, Mo still exhibits some key elements of a QPT (see Fig. \ref{fig-qpt-evidences}). When considering the unperturbed configurations, both rapidly transition from a spherical to a deformed shape, either oblate (regular) or prolate (intruder). According to Figs. \ref{fig_ibm_ener_axial-Mo} and \ref{fig-beta}, the regular and intruder configurations undergo a ``Type I'' QPT, while the ground state undergoes a ``Type II'' QPT around $A=102$.

The situation in Ru isotopes is indeed different, as the evolution of the ground state is fully determined by a single configuration. According to Figs. \ref{fig_ibm_ener_axial-Ru} and \ref{fig_ibm_ener_surph-Ru}, the energy surface of Ru isotopes is initially spherical for the lighter ones, but it starts to flatten and becomes fully flat at $A=104$ (neutron number $60$). From this point onwards, a $\gamma$-unstable deformation develops. Based on this behavior, Ru isotopes undergo a ``Type I'' QPT of second order, as indicated by Fig. \ref{fig-qpt-evidences} and as proposed in \cite{Fran2001}.

\section{Summary and conclusions}
\label{sec-conclu}
This work has focused on analyzing the even-even $^{96-110}_{~~~~~42}$Mo and $^{98-114}_{~~~~~44}$Ru isotopes using the interacting boson model with configuration mixing (IBM-CM). Initially, the parameters of the Hamiltonian and the $\hat{T}(E2)$ operators were determined through a least-squares fitting process using available experimental excitation energies and $B(E2)$ values. Additionally, radii and two-neutron separation energies were computed and compared to experimental data. Subsequently, a detailed analysis was conducted on the wave function of the states, their deformation, and their mean-field energy surfaces. Finally, the potential interplay between shape coexistence and quantum phase transition was explored.

The primary objective of this study is to determine the extent of the shape coexistence region around $Z=40$. In the case of Sr and Zr isotopes, the influence of intruder states becomes evident at neutron number $60$, where there is a crossing between the regular and intruder configurations in the ground state. Therefore, it is natural to investigate whether this phenomenon persists up to $Z=42-44$. Previous research has already examined the case of Kr ($Z=36$), concluding that the intruder configuration no longer plays a significant role in these nuclei and a rather smooth systematics is observed \cite{Marg2009, Albe2012}, although in \cite{Flav2017} the authors claim that in $^{98-100}$Kr, intruder states can play a significant role. However, the situation in the other direction remains unclear.

One of the main conclusions drawn from this study is that shape coexistence plays a significant role in Mo isotopes, where a crossing between the intruder and regular configurations occurs at neutron number $60$ in the ground state. However, in Ru isotopes, the intruder states appear at much higher energies and do not affect the behavior of the low-lying states. Therefore, it seems that the boundary of the shape coexistence phenomenon lies within the Mo isotopes.

The region under investigation is also known for the presence of quantum phase transitions (QPTs). Based on the experimental information, it is challenging to assert the existence of QPTs in Mo and Ru nuclei. However, other indirect features indicate the presence of QPTs in both nuclei, albeit of different nature. In Mo isotopes, the existence of a configuration crossing suggests a first-order QPT (``Type II'' QPT) similar to that observed in Sr and Zr isotopes. However, the strong interaction between configurations smooths out the systematics, resembling the situation in Pt nuclei. Hence, it can be stated that in Mo isotopes, a ``Type II'' QPT exists, but it is not of first order, because the observed changes in the key observables shown in Fig.\ \ref{fig-qpt-evidences} are rather smooth. In Ru isotopes, all the analyzed observables (refer to Fig.\ \ref{fig-qpt-evidences}) indicate a gradual trend. However, the mean-field analysis of the nuclei suggests the presence of a ``Type I'' second-order QPT at neutron number $60$, as proposed in \cite{Fran2001}. Thus, neighboring nuclei exhibit QPTs, but with different origins. In Mo isotopes, the QPT is induced by shape coexistence (``Type II'' QPT), whereas in Ru isotopes, it involves a single configuration (``Type I'' QPT). In both cases, the QPT is of second order.

It is important to acknowledge the limitations of the obtained results, primarily stemming from the lack of experimental data on transition rates involving states of the regular and intruder configurations. This absence of information makes it challenging to precisely determine certain parameters in the model operators. Fortunately, the study of isotopes with $Z=40$ is currently an active area of experimental research, which holds promise for obtaining additional data in the future.

A potential future extension of this work would involve investigating odd-even nuclei, as demonstrated in a compelling manner for Nb isotopes in the reference \cite{Gavr22b}. This would represent the next step in configuration mixing studies and could provide further insights and understanding of the phenomena of shape coexistence and QPTs.

\section{Acknowledgment}
One of the authors (JEGR) wants to thank K.\ Heyde for his continuous advice and support. This work was partially supported by the Consejer\'{\i}a de Econom\'{\i}a, Conocimiento, Empresas y Universidad de la Junta de Andaluc\'{\i}a (Spain) under Groups FQM-160 and FQM-370, and under projects P20-00617, P20-01247 and US-1380840; also through the projects PID2019-104002GB-C21, PID2019-104002GB-C22, and PID2020-114687GB-I00 funded by MCIN/AEI/10.13039/50110001103 and ``ERDF A way of making Europe''. 
%This work was supported by the grant number PID2019-104002GB-C21 funded by MCIN/AEI/10.13039/501100011033 and FEDER ``A way of making Europe''. 
Resources supporting this work were provided by the CEAFMC and the Universidad de Huelva High Performance Computer (HPC@UHU) funded by ERDF/MINECO project UNHU-15CE-2848.   

\bibliography{references-IBM-CM,references-QPT}

\begin{thebibliography}{110}
\bibitem [{\citenamefont {Maya-Barbecho}\ and\ \citenamefont
  {Garc\'{\i}a-Ramos}()}]{Maya2023u}%
  \BibitemOpen
  \bibfield  {author} {\bibinfo {author} {\bibfnamefont {E.}~\bibnamefont
  {Maya-Barbecho}}\ and\ \bibinfo {author} {\bibfnamefont {J.~E.}\ \bibnamefont
  {Garc\'{\i}a-Ramos}},\ }\bibfield  {title} {\enquote {\bibinfo {title} {{Sr
  isotopes: the interplay between shape coexistence and quantum phase
  transitions}},}\ }\href@noop {} {\bibinfo  {journal} {EPJ Web Conf. (to be
  published)}\ }\BibitemShut {NoStop}%
\end{thebibliography}

\end{document}